\documentclass[11pt]{article} 

\usepackage[showcomments]{ajtex}

\DeclareRobustCommand{\DIEP}{\ensuremath{%
    \mathchoice{\includegraphics[height=2ex]{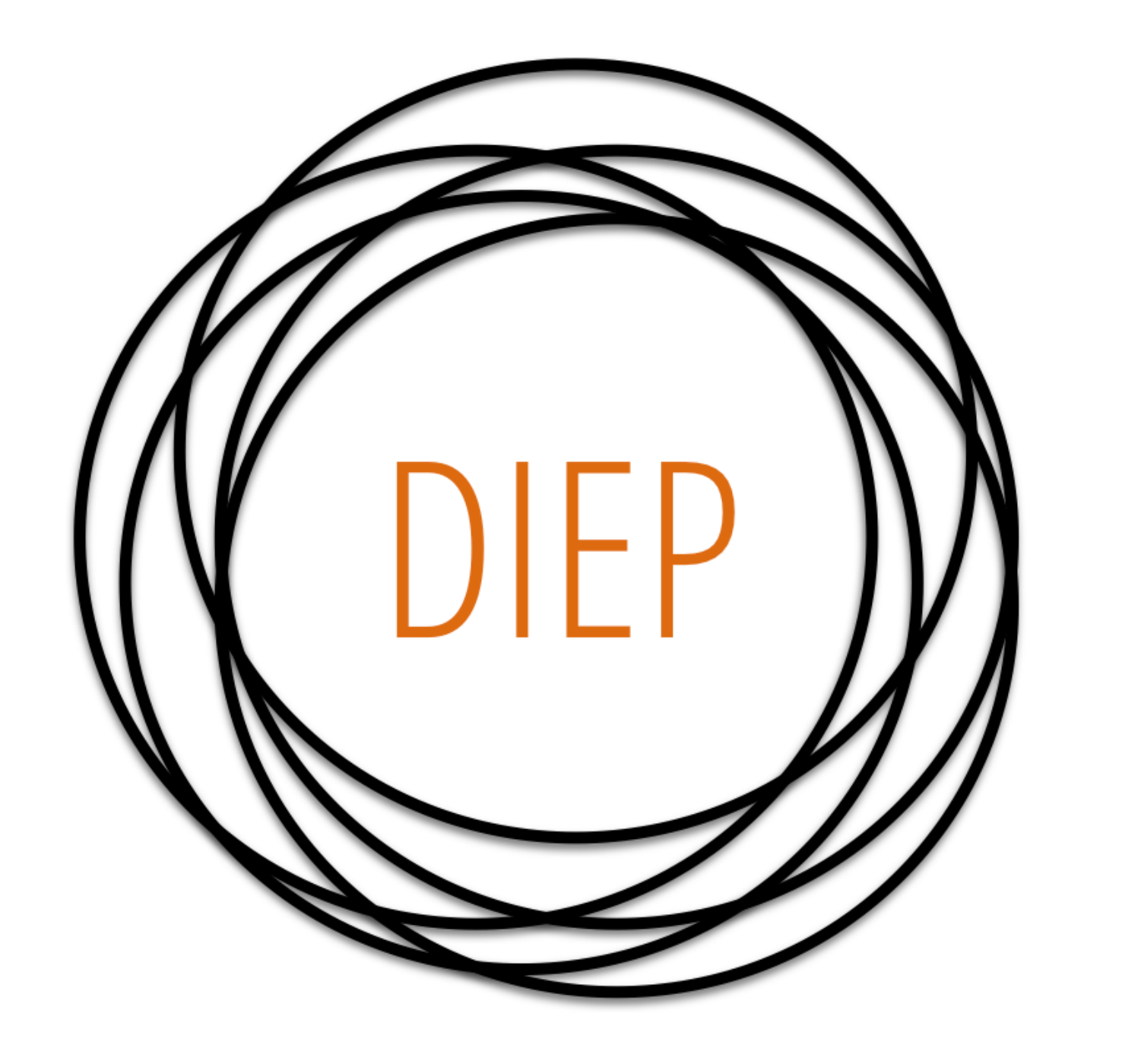}}
    {\includegraphics[height=2ex]{DIEPs.pdf}}
    {\includegraphics[height=1.5ex]{DIEPs.pdf}}
    {\includegraphics[height=1ex]{DIEPs.pdf}}
  }}

\title{Effective field theory for hydrodynamics\\ without boosts}

\author[a,\DIEP]{Jay Armas}\email{j.armas@uva.nl} 
\author[b]{Akash Jain}\email{ajain@uvic.ca}

\affiliation[a]{Institute for Theoretical Physics, University of Amsterdam, 1090 GL Amsterdam, The Netherlands} 
  
\affiliation[\DIEP]{Dutch Institute for Emergent Phenomena (DIEP), 1090 GL Amsterdam, The Netherlands} 

\affiliation[b]{Department of Physics \& Astronomy, University of Victoria, PO
  Box 1700 STN CSC, Victoria, BC, V8W 2Y2, Canada}

\abstract{We formulate the Schwinger-Keldysh effective field theory of
  hydrodynamics without boost symmetry. This includes a spacetime covariant
  formulation of classical hydrodynamics without boosts with an additional
  conserved particle/charge current coupled to Aristotelian background
  sources. We find that, up to first order in derivatives, the theory is
  characterised by the thermodynamic equation of state and a total of 29
  independent transport coefficients, in particular, 3 hydrostatic, 9
  non-hydrostatic non-dissipative, and 17 dissipative. Furthermore, we study the
  spectrum of linearised fluctuations around anisotropic equilibrium states with
  non-vanishing fluid velocity. This analysis reveals a pair of sound modes that
  propagate at different speeds along and opposite to the fluid flow, one charge
  diffusion mode, and two distinct shear modes along and perpendicular to the
  fluid velocity. We present these results in a new hydrodynamic frame that is
  linearly stable irrespective of the boost symmetry in place. This provides a
  unified covariant stable approach for simultaneously treating Lorentzian,
  Galilean, and Lifshitz fluids within an effective field theory framework and
  sets the stage for future studies of non-relativistic intertwined patterns of
  symmetry breaking. }

\usepackage[bbgreekl]{mathbbol}

\def\bbn{\mathbb{n}}
\def\bbh{\mathbb{h}}

\def\bbPhi{\mathbb{\Phi}}

\def\rel{\text{rel}}
\def\gal{\text{gal}}
\def\kB{k_{\text B}}
\def\hs{\text{hs}}
\def\nhs{\text{nhs}}

\def\SK{Schwinger-Keldysh\xspace}

\setlist[itemize]{itemsep=0pt,topsep=5pt}
\setlist[enumerate]{itemsep=0pt,topsep=5pt}

\usepackage{tikz}
\usetikzlibrary{decorations.markings,decorations.pathmorphing,arrows.meta}

\tikzset{aux/.style={decorate,decoration={snake,segment length=1.5mm,
      amplitude=0.4mm}}}
\tikzset{left/.style={arrows={Stealth[scale length=0.5, scale width=1.5]-}}}
\tikzset{right/.style={arrows={-[sep=-2pt]Stealth[scale length=0.5, scale
      width=1.5]}}}
\tikzset{leftright/.style={arrows={Stealth[scale length=0.5, scale
      width=1.5]-Stealth[scale length=0.5, scale width=1.5]}}}

\definecolor{lightblue}{RGB}{200,225,255}
\definecolor{lightblueborder}{RGB}{30,100,255}

\def\arraystretch{1.2}

\begin{document}

\maketitle 

\section{Introduction}

Hydrodynamics describes the long-wavelength collective behaviour of low-energy
excitations in a broad range of physical systems. In this regime, the dynamics
is insensitive to most microscopic details and is universally captured by a set
of conservation laws. The range of applicability of hydrodynamics spans widely
separated scales, in particular those of quantum gravity
\cite{Bhattacharyya:2008jc, Armas:2016mes}, viscous electron flows
\cite{2016NatPh..12..672L, 2017NatPh..13.1182K}, biological fluids
\cite{2019PhRvF...4k0506M}, and the dynamics of black hole accretion disks
\cite{Font:2008fka}, to mention only a few. Traditionally, hydrodynamics has
been a phenomenological field of study. One specifies the symmetry-breaking
pattern; postulates a set of currents with associated conservation laws; invokes
the second law of thermodynamics (through the positivity of the divergence of an
entropy current) together with Onsager’s relations, and determines the
constitutive relations in a gradient expansion~\cite{Kovtun:2012rj}. While this
classical approach has been extremely successful, one expects that symmetry
alone should be sufficient to characterise the hydrodynamic regime. 

Treating hydrodynamics as a \emph{bona fide} thermal field theory, a more
fundamental approach has been developed in the context of relativistic fluids in
the past few years~\cite{Crossley:2015evo, Haehl:2015foa, Jensen:2017kzi}, and
has recently been adapted to Galilean-invariant fluids as
well~\cite{Jain:2020vgc}. This formulation is based on the Schwinger-Keldysh
effective field theory (EFT) framework for non-equilibrium thermal systems;
see~\cite{Glorioso:2018wxw} for a review. The starting point of this EFT
framework is a generating functional from which correlation functions of
hydrodynamic operators and hydrodynamic equations of motion can be derived. In
addition, the framework systematically accounts for the effects of
stochastic/thermal noise on the hydrodynamic evolution via stochastic
interactions. In order to describe out-of-equilibrium thermal systems, the EFT
generating functional must satisfy certain requirements, such as KMS
symmetry, which lead to an emergent second law of thermodynamics and
implementation of the Onsager's relations at the classical level, in addition to
fluctuation-dissipation constraints on the correlation functions. The main goal of this work is to develop a Schwinger-Keldysh EFT for the
hydrodynamic description of physical systems that lack any boost symmetry,
Lorentzian or Galilean, to begin with. In particular, this covers systems which
have their boost symmetry explicitly broken due to the presence of a background
medium\footnote{In \cite{Nicolis:2015sra, Alberte:2020eil} the case of
  spontaneous breaking of Lorentz boost symmetry was considered. This is
  distinct from the setup considered in this paper, where the boost symmetry is
  explicitly broken and the respective Ward identity is absent to begin
  with. This can be thought of as being accomplished by ``integrating out'' the
  medium through which the fluid is moving.} or systems that do respect a boost
symmetry but is not explicitly manifest at the macroscopic level. Systems
without a boost symmetry are ubiquitous in non-Fermi liquid phases of matter
such as metallic quantum critical systems~\cite{Hertz:1976zz}. Fermi-liquids can 
also exhibit phases characterised by the absence of a boost symmetry, the prime
example being liquid helium-3 at sufficiently low
temperatures~\cite{Alberte:2020eil}.\footnote{In the context of quantum matter,
  it has also been argued that electron flows in graphene may break Lorentz
  invariance due to the presence of long-range Coulomb
  interactions~\cite{Lucas:2017idv}.} In the realm of classical physics, various
many-body systems in soft matter physics and
biophysics~\cite{2013RvMP...85.1143M} do not respect Galilean boost
symmetry. Common examples include models with self-propelled agents such as
flocks of birds and colonies of bacteria swimming in a medium.

In general, however, physical systems that break boost symmetry also exhibit
other patterns of symmetry breaking. In the context of quantum matter, spatial
translations are usually also spontaneously broken, due to the presence of the
ionic lattice, or explicitly broken due to the presence of
impurities~\cite{1995pcmp.book.....C}. Charge density wave phases are one such
example; see also~\cite{Delacretaz:2017zxd,Armas:2020bmo}. In the setting of
classical fluids, self-propelled agents break spacetime translations explicitly
due to the presence of driving forces~\cite{PhysRevE.58.4828}, while active
liquid crystal phases can break translations and rotations
spontaneously~\cite{2018RPPh...81g6601J}.  Such situations compromise the
gradient expansion of hydrodynamics and thus it is important to move away from
traditional treatments and understand what are \emph{the rules of the game} for
building hydrodynamics models with intertwined patterns of symmetry breaking.

Schwinger-Keldysh EFT provides a controlled framework for developing such
hydrodynamic theories and studying stochastic corrections to classical
hydrodynamics. In particular, it was shown recently in the context of isotropic
relativistic fluids that stochastic corrections break the hydrodynamic
derivative expansion at third derivative order, leading to non-classical
contributions to hydrodynamic correlation
functions~\cite{Jain:2020fsm}. However, such effects may possibly appear earlier
in the derivative expansion in systems with specific kinds of broken
symmetries. The work presented here considers only the case of broken boost
symmetry and has a three-fold purpose: (1) to accurately classify the transport
properties of hydrodynamics without boosts in the presence of a conserved U(1)
particle-number/charge current; (2) to provide a unified field theoretic
framework that can simultaneously describe Lorentzian as well as Galilean and
Lifshitz fluids,\footnote{This framework could also potentially describe
  Carrollian-boost invariant fluids \cite{Ciambelli:2018wre, Ciambelli:2018xat,
    deBoer:2017ing}, but we have not explored this possibility here.} which can
be obtained by restoring different types of boost symmetries or taking different
scaling limits; and (3) to provide the necessary foundations for future
explorations addressing more complicated patterns of symmetry breaking and
associated stochastic contributions.

Several recent works have motivated this study. Earlier literature established a
classical covariant approach to ideal fluids without boost symmetry, by coupling
the fluid to Aristotelian geometry~\cite{deBoer:2017ing}. A full treatment of
one-derivative corrections was carried out in~\cite{deBoer:2020xlc}, but the
U(1) current responsible for particle-number/charge conservation was not
introduced. A linearised analysis of fluctuating isotropic and homogeneous
configurations in charged hydrodynamics without boosts was done
in~\cite{deBoer:2017abi}. A classification of first order transport in flat
spacetime with the additional U(1) current was undertaken
in~\cite{Novak:2019wqg}, but a complete analysis of the second law constraints
was not carried out. In this work, we develop further all of these lines of
research by providing a complete covariant treatment and classification of
transport in hydrodynamics without boosts within a field theoretic framework,
including the presence of a U(1) current, and consider the most general
fluctuation analysis around equilibrium states that are inherently anisotropic.

In contrast with all the previous literature, we present our results in a new
hydrodynamic frame, which we call \emph{density frame}, that is linearly stable
(in the sense of \cite{Kovtun:2019hdm,Poovuttikul:2019ckt, Hoult:2020eho})
irrespective of the boost symmetry in place (Galilean or Lorentzian), or absence
thereof, and is thus better suited for potential numerical simulations. This
frame choice aligns the fluid velocity with the flow of momentum, rather than
the flow of internal energy (as in the Landau frame) or charge/particle-number
(as in the Eckart frame). Note that momentum is a reference-frame dependent
quantity. Therefore, when employed in Galilean or relativistic hydrodynamics,
the density frame will lead to a manifestly non-covariant representation of the
respective constitutive relations. However, the equations of motion are still
manifestly covariant and boost-invariant (up to second derivative
corrections). We emphasise that hydrodynamic models, irrespective of the
hydrodynamic frame utilised to represent the constitutive relations, are not
suitable to make reliable universal predictions about gapped ``high-energy''
modes. In this sense, the aforementioned linear stability (i.e. the absence of
unstable gapped modes) in the density frame is \emph{not} a physical prediction
of the model. It is rather a technical characteristic of the model that makes it
``more suitable'' for setting up initial-value problems aimed at exploring the
low-energy long-wavelength physics of fluids.

This paper is organised as follows. In \cref{sec:essentials}, we introduce
classical aspects of hydrodynamics without boost symmetry, conservation laws as
well as entropy production, and the basics of Aristotelian geometry to which
these fluids couple to. In \cref{sec:EFT}, we use these considerations in order
to formulate the Schwinger-Keldysh effective field theory for these systems. In
\cref{sec:1der}, we write down the specific Lagrangian that includes all the
dissipative and non-dissipative transport coefficients that characterise the
effective theory up to first order in a gradient expansion. In
\cref{sec:gal-rel}, we examine special limits where one recovers Lorentzian,
Galilean, and Lifshitz fluids. In section \ref{sec:linearised}, we study
fluctuations around generic anisotropic equilibrium configurations and obtain
explicit expressions for sound, shear, and charge diffusion modes in a linearly
stable hydrodynamic frame. Finally, we conclude with some discussion in
\cref{sec:discussion}. \Cref{app:frame} is dedicated to expressing our results
in the Landau frame in order to compare with the previous
literature. \Cref{sec:interactions} provides the interaction Lagrangian for the
linearised effective field theory of hydrodynamics without boosts, which can be
used for studying stochastic contributions to hydrodynamic correlation
functions.

\section{Classical boost-agnostic hydrodynamics}
\label{sec:essentials}

In this section, we review various aspects of classical boost-agnostic
hydrodynamics. We start with the energy, momentum, and charge/particle-number
conservation equations and use the second law of thermodynamics to derive the
constitutive relations of an ideal fluid without boost symmetry. We discuss how to
introduce curved background sources into these equations coupled to various
hydrodynamic observables, which will be crucial for our subsequent discussion of
the EFT framework. Following this, we outline a generic procedure to implement
the second law constraints at arbitrarily high orders in the derivative
expansion using the adiabaticity equation. A more concrete construction of the
allowed one-derivative corrections is presented later in \cref{sec:1der}.

\subsection{Ideal hydrodynamics on flat background}

\subsubsection{Symmetries and conservation laws}
\label{sec:conservation-eqs}

Hydrodynamics is a theory of locally conserved quantities. One starts by
outlining the complete set of Noether currents associated with any global
symmetries that the system might enjoy, and expresses various ``fluxes'' in
terms of the conserved ``densities'', arranged in a perturbative expansion in
derivatives. For a given set of such ``constitutive relations'', the time
evolution of the conserved densities is determined by their respective
conservation equations. In typical hydrodynamic systems, these conserved
densities are the energy $\epsilon$, momentum $\pi^i$, and particle number
density $n$ of the fluid, associated with time and space translational
invariance and an abstract internal U(1) phase shift invariance of the
theory. The associated fluxes are the energy flux $\epsilon^i$, stress tensor
$\tau^{ij}$, and mass/particle number flux $j^i$, with conservation equations
\begin{align}
  \text{Energy conservation:}
  &\qquad
    \dow_t \epsilon
    + \dow_i \epsilon^i = 0~, \nn\\
  \text{Momentum conservation:}
  &\qquad
    \dow_t \pi^j + \dow_i \tau^{ij} = 0~, \nn\\
  \text{Continuity equation:}
  &\qquad
    \dow_t n + \dow_i j^i = 0~.
    \label{eq:conservation}%
\end{align}%

In addition, hydrodynamic systems usually feature rotational invariance and some
kind of boost invariance. Provided that the fluid does not carry an intrinsic
spin density, rotational invariance requires the orbital angular momentum
density to be conserved
\begin{equation}
  \text{Angular-momentum conservation:}\quad
  \dow_t \lb \pi^{i} x^{j} - \pi^j x^i \rb
  + \dow_k \lb \tau^{ki}x^{j} - \tau^{kj}x^i \rb
  = \tau^{ji} - \tau^{ij}~,
\end{equation}
ensuring the stress tensor to be symmetric. If the theory is required to be
invariant under Galilean boosts, the center of inertia will need to be conserved
\begin{subequations}
  \begin{equation}
    \text{Center-of-mass conservation:}\quad
    \dow_t \lb m\,n\, x^{i} - \pi^i t \rb
    + \dow_k \lb m\, j^k x^{i} - \tau^{ki} t \rb
    = m\, j^i - \pi^i~,
  \end{equation}
  where $m$ is the constant mass per particle. This leads to the momentum
  density being aligned with the mass flux $\pi^i = m\,j^i$. Similarly, we have
  a conserved center-of-energy in the relativistic case
  \begin{equation}
    \text{Center-of-energy conservation:}\quad
    \dow_t \lb \frac{1}{c^2} \epsilon\, x^{i} - \pi^i t \rb
    + \dow_k \lb \frac{1}{c^2} \epsilon^k x^{i} - \tau^{ki} t \rb
    = \frac{1}{c^2} \epsilon^i - \pi^i~,
  \end{equation}
\end{subequations}
where $c$ is the speed of light, equating the momentum density to the
energy-flux $\pi^i = \epsilon^i/c^2$ instead.\footnote{Note that in the
  relativistic theory, we have a conserved energy momentum tensor
  $T^{\mu\nu}$. Momentum density equalling the energy-flux is merely the
  statement that $T^{ti} = T^{it}$.} The paradigm of the present work is to
study systems which might not necessarily respect a boost symmetry. In this
sense, we do not tie $\pi^i$ to either $j^i$ or $\epsilon^i$. We will still
focus on systems respecting rotational invariance (on hydrodynamic length
scales), so $\tau^{ij}$ is assumed to be symmetric.

\subsubsection{Constitutive relations and second law}
\label{sec:flat-ideal-hydro}

The starting point of hydrodynamics is the assumption that the low-energy dynamics
of the system near thermal equilibrium is entirely governed by its conserved
operators: density $n$, energy density $\epsilon$, and momentum density $\pi^i$.
Hydrodynamics is then characterised by the most generic expressions for the
fluxes $j^i$, $\epsilon^i$, and $\tau^{ij}$, written in terms of the chosen
variables and their spatial derivatives, i.e.
\begin{equation}
  j^i[n,\varepsilon,\pi^i,\dow_i]~, \qquad
  \epsilon^i[n,\varepsilon,\pi^i,\dow_i]~, \qquad
  \tau^{ij}[n,\varepsilon,\pi^i,\dow_i]~.
\end{equation}
These are known as the \emph{hydrodynamic constitutive relations}. Note that the
temporal derivatives of various quantities are determined by
\cref{eq:conservation} and hence are not independent. Our assumption of
\emph{near-equilibrium} allows us to arrange the constitutive relations in a
\emph{derivative expansion}, truncated at a given order in derivatives according
to the phenomenological sensitivity required. At any given order in the
derivative expansion, the constitutive relations contain all the possible tensor
structures made out of derivatives of $n$, $\epsilon$, and $\pi^i$, consistent
with symmetries, appended with arbitrary \emph{transport coefficients} as a
functions of $n$, $\epsilon$, $\vec \pi^2$.

The hydrodynamic constitutive relations are required to respect certain
phenomenological constraints. Most important of these is the ``local second law
of thermodynamics'' that requires that there must exist an entropy density $s^t$
and an associated flux $s^i$ such that
\begin{equation}
  \dow_t s^t + \dow_i s^i \geq 0~.
  \label{eq:second-law-flat}
\end{equation}
At the leading order in the derivative expansion, entropy density is merely
given by an arbitrary function of $n$, $\epsilon$, and $\vec\pi^2$, i.e.
$s^t = s(\epsilon,n,\vec\pi^2)$. Let us define intensive parameters: temperature
$T(\epsilon,n,\vec\pi^2)$, chemical potential $\mu(\epsilon,n,\vec\pi^2)$,
velocity $u^i(\epsilon,n,\vec\pi^2)$, and pressure $p(\epsilon,n,\vec\pi^2)$ via
the thermodynamic relations: local first law of thermodynamics and Euler
relation respectively
\begin{equation}
  T\df s = \df \epsilon - \mu\, \df n - u^i \df\pi_i~,  \qquad
  p = Ts+\mu n + u^i\pi_i  - \epsilon~.
  \label{eq:thermodynamics-micro}
\end{equation}
Due to rotational invariance, the velocity must be aligned with momentum, i.e.
$u^i = \pi^i/\rho$, where $\rho$ is the momentum susceptibility. It is easy to
check that
\begin{align}
  \partial_t s^t + \dow_i s^i
  &= 
  - \frac{1}{T^2} \lb \epsilon^i - (\epsilon+p) u^i \rb \dow_i T
  - \lb j^i - n\, u^i\rb \dow_i \frac{\mu}{T}
  - \lb \tau^{ij} - \rho\, u^i u^j - p\, \delta^{ij} \rb \dow_i \frac{u_j}{T}~,
  \label{eq:2ndLaw-non-cov}
\end{align}
where we have identified the entropy flux as
\begin{equation}
  T\,s^i = p\, u^i + \epsilon^i - \mu\, j^i
  - \tau^{ij} u_j + \mathcal{O}(\dow)~.
\end{equation}
We need to require that the RHS of \cref{eq:2ndLaw-non-cov} is positive
semi-definite for arbitrary fluid configurations. At the leading order in
derivatives, this leads to the ideal fluid constitutive relations
\begin{gather}
  \epsilon^i = (\epsilon+p)u^i + \mathcal{O}(\dow)~, \qquad
  \tau^{ij} = \rho\,u^i u^j + p\,\delta^{ij} + \mathcal{O}(\dow)~, \qquad
  j^i = n\, u^i + \mathcal{O}(\dow)~, \nn\\
  s^i = s\, u^i + \mathcal{O}(\dow)~.
  \label{eq:flat-consti}
\end{gather}
The respective dynamics is given by substituting these into the conservation
equations \eqref{eq:conservation}. Note that entropy is conserved at ideal
order. We can, in principle, extend this analysis to higher orders in the
derivative expansion (see~\cite{deBoer:2020xlc}). We shall return to this when
equipped with more tools.

The relation $s = s(\epsilon,n,\vec\pi^2)$, or equivalently
$\epsilon = \epsilon(s,n,\vec\pi^2)$, can be understood as the micro-canonical
equation of state of the fluid and completely characterises its constitutive
relations at ideal order through the thermodynamic relations
\eqref{eq:thermodynamics-micro}. We note, however, that hydrodynamics as a
physical system is better defined in the grand canonical ensemble, because a
fluid element is allowed to freely exchange particles, energy, and momentum with
its surroundings. Keeping this in mind, we can take the fundamental dynamical
fields to be $T$, $\mu$, $u^i$ instead of $\epsilon$, $n$, and $\pi_i$. In this
case, the equation of state is given in terms of $p(T,\mu,\vec u^2)$ instead of
$\epsilon(s,n,\vec\pi^2)$ with the thermodynamic relations: Gibbs-Duhem relation
and Euler relation respectively
\begin{equation}
  \df p = s\, \df T + n\, \df\mu + \pi_i \df u^i, \qquad
  \epsilon = Ts + \mu n + u^i\pi_i - p.
  \label{eq:thermodynamics}
\end{equation}
Recall that $\pi_i = \rho\, u_i$. These relations define $\epsilon$, $n$, and
$\pi_i$ in terms of $T$, $\mu$, and $u^i$.

\subsection{Coupling to background sources}

\subsubsection{Aristotelian background sources}
\label{sec:sources}

We would like to introduce a set of Aristotelian background sources to which
fluids without boost symmetry couple to, as discussed in~\cite{deBoer:2020xlc}. These are similar to the
Newton-Cartan sources prevalent in Galilean hydrodynamics, but with no Milne
boost symmetry. The absence of this symmetry, in fact, makes these sources
easier to implement in an effective theory. These are given by
\begin{equation}
  \text{Clock-form:}\quad n_\mu~, \qquad
  \text{Spatial metric:}\quad h_{\mu\nu}~, \qquad
  \text{Gauge field:}\quad A_\mu~,
  \label{eq:sources}
\end{equation}
where $h_{\mu\nu}$ is a symmetric matrix of signature
$(0,1,1,1,\ldots)$. $n_\mu$ and $h_{\mu\nu}$ can be vaguely understood as the
time and space components of the relativistic spacetime metric $g_{\mu\nu}$, now
being treated independently due to the lack of any boost symmetry. Since
$h_{\mu\nu}$ is a degenerate matrix, it admits a zero-eigenvector $v^\mu$,
normalised as $v^\mu n_\mu = 1$, such that $v^\mu h_{\mu\nu} = 0$. Its spatial
components $v^i$ can be identified as the velocity of a lab frame observer. This
can be used to define an ``inverse spatial metric'' $h^{\mu\nu}$ via the
relations $h^{\mu\nu}n_\nu = 0$ and
$h^{\mu\nu} h_{\nu\lambda} + v^\mu n_\lambda = \delta^\mu_\lambda$. Note that
$h^{\mu\nu}$ is \emph{not} the inverse of $h_{\mu\nu}$. Together,
\begin{equation}
  \text{Frame velocity:}\quad v^\mu~, \qquad
  \text{Inverse spatial metric:}\quad h^{\mu\nu}~,
\end{equation}
should be understood along the same lines as the inverse metric $g^{\mu\nu}$ in
relativistic field theories. They are entirely fixed by the sources
\eqref{eq:sources} via the conditions
\begin{equation}
  v^\mu n_\mu = 1~, \qquad
  v^\mu h_{\mu\nu} = 0~, \qquad
  h^{\mu\nu}n_\nu = 0~, \qquad
  h^{\mu\nu} h_{\nu\lambda} + v^\mu n_\lambda = \delta^\mu_\lambda~.
  \label{eq:NC-normalisations}
\end{equation}
The flat background limit is given as $n_\mu = \delta_\mu^t$,
$h_{\mu\nu} = \delta^i_\mu \delta_{i\nu}$, $A_\mu = 0$, $v^\mu = \delta^\mu_t$,
and $h^{\mu\nu} = \delta^{i\mu}\delta_i^\nu$.

Let $d$ be the number of spatial dimensions. The clock-form $n_\mu$ couples to
the energy density and flux $\epsilon^\mu$, the $d$ independent components of
the frame velocity $v^\mu$ couple to the momentum density $\pi_\mu$ (normalised
as $v^\mu \pi_\mu = 0$), while the remaining $d(d+1)/2$ independent components
of the spatial metric $h_{\mu\nu}$ couple to the stress tensor $\tau^{\mu\nu}$
(satisfying $\tau^{\mu\nu} n_\nu = 0$ and $\tau^{\mu\nu} = \tau^{\nu\mu}$), and
finally the gauge field $A_\mu$ couples to the particle number current
$j^\mu$. In terms of the non-covariant densities and fluxes we have
\begin{equation}
  \epsilon^\mu =
  \begin{pmatrix}
    \epsilon \\ \epsilon^i
  \end{pmatrix}, \quad
  \pi_\mu =
  \begin{pmatrix}
    - v^k \pi_k/v^t \\ \pi_i
  \end{pmatrix}, \quad
  \tau^{\mu\nu} =
  \begin{pmatrix}
   n_k n_l\tau^{kl}/n_t^2 & - n_k \tau^{kj}/n_t \\
   - n_k \tau^{ki}/n_t & \tau^{ij}
 \end{pmatrix}, \quad
 j^\mu =
 \begin{pmatrix}
   n \\ j^i
 \end{pmatrix}~.
 \label{eq:cov-operators}
\end{equation}
We will often use $\pi^\mu = h^{\mu\nu} \pi_\nu$ satisfying $\pi^\mu n_\mu = 0$.
The coupling can be denoted in terms of the variation of an (equilibrium)
effective action $S$ describing the theory as
\begin{align}
  \delta S
  &= \int \df t\df^d x \sqrt{\gamma}\,
  \lb j^\mu \delta A_\mu - \epsilon^\mu \delta n_\mu
  - \pi_\mu \delta v^\mu
    + \half \tau^{\mu\nu} \delta h_{\mu\nu} \rb \nn\\
  &= \int \df t\df^d x \sqrt{\gamma}\,
  \lb j^\mu \delta A_\mu - \epsilon^\mu \delta n_\mu
  + \lb v^\mu \pi^\nu + \half \tau^{\mu\nu} \rb \delta h_{\mu\nu} \rb,
  \label{eq:basic-delS}
\end{align}
where $\gamma = \det(h_{\mu\nu}+n_\mu n_\nu)$. The second line is to highlight
that $v^\mu$ is not an independent source.

We require the theory to be invariant under local diffeomorphisms parametrised
by arbitrary invertible maps $x'^\mu(x)$ and U(1) gauge transformation
parametrised by $\Lambda(x)$. We have collectively denoted the spacetime
coordinates $x^\mu = (t,x^i)$. Their action on the background fields is defined
as usual
\begin{gather}
  n_\mu(x)
  \to n'_\mu(x') = \frac{\dow x^\nu}{\dow x'^\mu} n_\nu(x)~, \qquad
  h_{\mu\nu}(x)
  \to h'_{\mu\nu}(x')
  = \frac{\dow x^\rho}{\dow x'^\mu}\frac{\dow x^\sigma}{\dow x'^\nu}
  h_{\rho\sigma}(x)~, \nn\\
  A_\mu(x)
  \to A'_\mu(x') = \frac{\dow x^\nu}{\dow x'^\mu} \lb A_\nu(x)
  + \dow_\nu\Lambda(x) \rb~,
  \label{eq:NC.diffeo}
\end{gather}
and similarly for $v^\mu$ and $h^{\mu\nu}$. Under infinitesimal version of these
transformations, with $x'^\mu(x) = x^\mu + \xi^\mu(x)$, the diffeomorphisms
merely act as Lie derivatives
\begin{equation}
  n_\mu \to n_\mu + \lie_\xi n_\mu~,  \quad
  h_{\mu\nu} \to h_{\mu\nu} + \lie_\xi h_{\mu\nu}~, \quad
  A_\mu\to A_\mu +\lie_\xi A_\mu +\dow_\mu\Lambda~.
\end{equation}
Implementing this on the action variation in \cref{eq:basic-delS}, we can work
out the covariant conservation equations
\begin{align}
  \text{Energy conservation:}\qquad
  &\lb \nabla_\mu + F^n_{\mu\lambda} v^\lambda \rb \epsilon^\mu
    = - v^\nu \lb F_{\nu\mu} j^\mu - F^n_{\nu\mu} \epsilon^\mu \rb
    - \tau^{\mu\lambda} h_{\lambda\nu} \nabla_\mu v^\nu~, \nn\\
  \text{Momentum conservation:}\qquad
    &\lb \nabla_\mu + F^n_{\mu\lambda} v^\lambda \rb
      \lb v^\mu \pi^\nu + \tau^{\mu\nu} \rb
      = h^{\nu\lambda} \lb F_{\lambda\mu} j^\mu - F^n_{\lambda\mu} \epsilon^\mu \rb
      - \pi^\mu \nabla_\mu v^\nu~, \nn\\
  \text{Continuity equation:}\qquad
  &\lb \nabla_\mu + F^n_{\mu\lambda} v^\lambda \rb j^\mu
    = 0~.
    \label{eq:NC.Conservation}
\end{align}
Here $F_{\mu\nu} = 2\dow_{[\mu}A_{\nu]}$ and
$F^n_{\mu\nu} = 2\dow_{[\mu}n_{\nu]}$ are the field strengths associated with
$A_\mu$ and $n_\mu$, and $\nabla_\mu$ is the covariant derivative operator
associated with the connection\footnote{In Galilean theories, it is often
  convenient to work with a different connection, namely
  \begin{equation}\nonumber
    \tilde\Gamma^\lambda_{\mu\nu}
    = v^{\lambda} \dow_\mu n_{\nu}
    + \half  h^{\lambda\rho}
    \lb \dow_\mu h_{\nu\rho} + \dow_\nu h_{\mu\rho} - \dow_\rho h_{\mu\nu} \rb
    + n_{(\mu} F_{\nu)\rho} h^{\rho\lambda},
  \end{equation}
  which is Milne boost-invariant on backgrounds with $F^n_{\mu\nu} =0$. Since we
  do not have any boost invariance, we choose to work with the simpler
  connection. The choice of connection has no bearing on the physical results.}
\begin{align}
  \Gamma^\lambda_{\mu\nu}
  = v^{\lambda} \dow_\mu n_{\nu}
  + \half  h^{\lambda\rho}
  \lb \dow_\mu h_{\nu\rho} + \dow_\nu h_{\mu\rho} - \dow_\rho h_{\mu\nu} \rb~.
  \label{eq:NC-connection}
\end{align}
This connection satisfies
\begin{gather}
  \nabla_\mu n_\nu = \nabla_\mu h^{\nu\lambda} = 0~, \qquad
  \nabla_\lambda h_{\mu\nu}
  = - n_{(\mu} \lie_v h_{\nu)\lambda}~, \qquad
  h_{\nu\lambda}\nabla_\mu v^\lambda
  = \half \lie_v h_{\mu\nu}, \nn\\
  \Gamma^\mu_{\mu\nu} + F^n_{\nu\mu} v^\mu
  = \frac{1}{\sqrt{\gamma}} \dow_\nu \sqrt{\gamma}~, \qquad
  2\Gamma^\lambda_{[\mu\nu]} = v^\lambda F^n_{\mu\nu}~.
\end{gather}
Note that this connection is torsional. In addition to torsional contributions
on the left, the conservation of energy and momentum in
\cref{eq:NC.Conservation} is sourced by Lorentz force-like terms coupled to the
field strengths $F_{\mu\nu}$ and $F_{\mu\nu}^n$, and pseudo-force terms coupled
to the covariant derivative of the frame velocity $\nabla_\mu v^\nu$.

\subsubsection{Hydrodynamics on curved background}

The ideal order hydrodynamic constitutive relations \eqref{eq:flat-consti} can
be coupled to background sources naturally as
\begin{gather}
  \epsilon^\mu = \epsilon\,u^\mu + p\,\vec u^\mu + \mathcal{O}(\dow)~, \qquad
  \pi^\mu = \rho\, \vec u^\mu + \mathcal{O}(\dow)~, \qquad
  \tau^{\mu\nu} = \rho\,\vec u^\mu \vec u^\nu + p\, h^{\mu\nu}
  + \mathcal{O}(\dow)~, \nn\\
  j^\mu = n\, u^\mu + \mathcal{O}(\dow)~, \qquad
  s^\mu = s\, u^\mu + \mathcal{O}(\dow)~,
  \label{eq:consti-ideal}
\end{gather}
where the covariant version of various hydrodynamic observables are defined in
\cref{eq:cov-operators}. We have also taken $u^t = (1- u^i n_i)/n_t$, which is
just equal to $1$ on a flat background, so that the covariant fluid velocity
$u^\mu$ satisfies the normalisation condition $u^\mu n_\mu = 1$. The velocity of
the fluid with respect to the Galilean frame is defined as
$\vec u^\mu = u^\mu - v^\mu$,
$\vec u_\mu = h_{\mu\nu} u^\nu = h_{\mu\nu} \vec u^\nu$, satisfying
$\vec u^\mu n_\mu = 0$, $\vec u_\mu v^\mu = 0$. These constitutive relations
should be understood as written in the grand canonical ensemble characterised by
a function $p(T,\mu,\vec u^2)$, where
$\vec u^2 = u^\mu u^\nu h_{\mu\nu} = \vec u^\mu \vec u^\nu h_{\mu\nu}$, and the
thermodynamic relations \eqref{eq:thermodynamics}. The structure of the
constitutive relations is fixed by the second law of thermodynamics; it can be
checked that \cref{eq:consti-ideal} represents the most generic leading
derivative order constitutive relations satisfying
\begin{equation}
  \lb \nabla_\mu + F^n_{\mu\lambda} v^\lambda \rb s^\mu
  = \frac{1}{\sqrt{\gamma}} \dow_\mu \lb \sqrt{\gamma}\, s^\mu \rb
  \geq 0~, \qquad
  s^\mu = s\, u^\mu + \mathcal{O}(\dow)~.
\end{equation}

The equations of motion of hydrodynamics are obtained by substituting the
constitutive relations \eqref{eq:consti-ideal} into the conservation equations
\eqref{eq:NC.Conservation}. We obtain
\begin{gather}
  \frac{\vec u^\nu/T^2}{\sqrt{\gamma}}
  \delta_\scB (\sqrt{\gamma}\, T^2\rho)
  - (\epsilon+p) h^{\mu\nu} \delta_\scB n_\mu
  + n h^{\mu\nu} \delta_\scB A_\mu 
  + \rho\, u^\sigma  h^{\nu\rho} \delta_\scB h_{\sigma\rho}
  = \mathcal{O}(\dow^2)~, \nn\\
  \frac{1}{\sqrt{\gamma}} \delta_\scB (\sqrt{\gamma}\, Tn)
  = \mathcal{O}(\dow^2)~, \qquad
    \frac{1}{\sqrt{\gamma}} \delta_\scB \lb \sqrt{\gamma}\, Ts\rb
    = \mathcal{O}(\dow^2)~.
    \label{eq:delB-EOM}
\end{gather}
Here $\delta_\scB$ denotes a Lie derivative along $u^\mu/T$ combined with a
gauge-shift along $(\mu - u^\mu A_\mu)/T$. Explicitly, we find
\begin{gather}
  \delta_\scB n_\mu
  = - \frac{1}{T^2} \dow_\mu T - \frac{1}{T} F^n_{\mu\nu} u^\nu, \qquad
  \delta_\scB h_{\mu\nu}
  =
  2h_{\lambda(\mu} \nabla_{\nu)} \frac{u^\lambda}{T}
  + \frac{u^\lambda}{T} \nabla_\lambda h_{\mu\nu}~, \nn\\
  \delta_\scB A_\mu
  = \dow_\mu \frac{\mu}{T} - \frac{1}{T} F_{\mu\nu} u^\nu~.
\end{gather}
Note the identity for arbitrary function $f(T,\mu,\vec u^2)$, 
\begin{align}
  \frac{1}{\sqrt{\gamma}}
  \delta_\scB \lb \sqrt{\gamma}\, f \rb
  &= 
  - \lB \lb T \frac{\dow f}{\dow T}
  + \mu \frac{\dow f}{\dow \mu}
  + 2\vec u^2\frac{\dow f}{\dow \vec u^2}
  \rb u^\mu
    - f\, v^\mu \rB \delta_\scB n_\mu \nn\\
  &\qquad
  + \frac{\dow f}{\dow \mu} u^\mu \delta_\scB A_\mu
  + \lb 2\frac{\dow f}{\dow \vec u^2} u^\mu u^\nu
    + f h^{\mu\nu} \rb \half \delta_\scB h_{\mu\nu}~.
\end{align}
There are a few lessons to be learnt from the equations of motion
\eqref{eq:delB-EOM}. Firstly, note that the equations of motion can be written
entirely in terms of $\delta_\scB n_\mu$, $\delta_\scB h_{\mu\nu}$, and
$\delta_\scB A_\mu$. Let us say that the background fields admit a timelike
Killing vector $K^\mu$, i.e.
$\lie_K n_\mu = \lie_K h_{\mu\nu} = \lie_K A_\mu = 0$, where $\lie_K$ denotes a
Lie derivative along $K^\mu$. Coupled to such a background, the equations of
motion admit a trivial ``equilibrium solution'' given by $u^\mu/T = K^\mu$ and
$\mu/T = K^\mu A_\mu$. Secondly, we can always eliminate any $(d+2)$ number of
linear combinations of $\delta_\scB n_\mu$, $\delta_\scB h_{\mu\nu}$, and
$\delta_\scB A_\mu$ from the higher derivative corrections to the constitutive
relations using equations of motion. This shall be useful later while writing
down the set of independent one-derivative corrections to the hydrodynamic
constitutive relations.

\subsubsection{Hydrodynamic frame transformations}
\label{sec:frame}

Recall that we had defined the hydrodynamic variables $u^i$, $T$, and $\mu$
using the thermodynamic relations \eqref{eq:thermodynamics-micro}. However,
these definitions are only well posed in equilibrium. Out of equilibrium, there
is no unique notion of fluid velocity, temperature, or chemical potential. This
is important because we can always redefine these quantities with terms
involving spacetime derivatives which will vanish in equilibrium, such as
\begin{equation}
  T \to T + \delta T~, \qquad
  \mu \to \mu + \delta\mu~, \qquad
  u^i \to u^i + \delta u^i~,
  \label{eq:frame-trans-fields}
\end{equation}
where $\delta T$, $\delta\mu$, $\delta u^i$ contain terms with at least one
derivative. We can also define $\delta u^\mu$ as the change in the covariant
fluid velocity, with $n_\mu \delta u^\mu = 0$. More details on the explicit
action of these redefinitions on the hydrodynamic constitutive relations can be
found in \cref{app:frame}.

Often, it is convenient to work in a ``hydrodynamic frame'' where one imposes
extra constraints on the derivative corrections that can enter the ideal order
constitutive relations \eqref{eq:consti-ideal}, so that this freedom is exactly
fixed. A natural choice is to ensure that the conserved densities: energy,
momentum, and particle number, do not obtain any corrections, by requiring
\begin{equation}
  \epsilon^\mu n_\mu = \epsilon~, \qquad
  \pi^\mu = \rho\, \vec u^\mu~, \qquad
  j^\mu n_\mu = n~,
  \label{eq:density-frame}
\end{equation}
which we call the \emph{density frame}. This frame ties the fluid velocity with
the flow of momentum. In the Galilean case, this is same as the ``mass frame''
with the fluid velocity aligned with the flow of mass. For relativistic
hydrodynamics, there are other more popular hydrodynamic frames used in the
literature, such as the Landau and Eckart frame, where the fluid velocity is
aligned with internal energy and charge flow, respectively. However these frames
are known to exhibit unphysical pathologies such as superluminal propagation and
unstable modes in the linear spectrum in a finite velocity
state~\cite{Hiscock:1985zz, Bemfica:2017wps, Kovtun:2019hdm, Bemfica:2019knx,
  Bemfica:2020zjp}.  By contrast, the density frame defined above is always
well-defined. We will return to these issues in \cref{sec:linearised}. More
details about hydrodynamic frame transformations in boost-agnostic hydrodynamics
can be found in \cref{app:frame}. As it turns out, the most useful choice for us
is to leave the hydrodynamic field redefinition freedom to be unfixed for
now. We shall return to it in the next subsection.

\subsection{Adiabaticity equation, thermodynamic frame, and discrete symmetries}
\label{sec:adiabaticity}

We can write down a covariant version of the second law of thermodynamics
given in \cref{eq:second-law-flat} as
\begin{equation}
  \lb \nabla_\mu + F^n_{\mu\lambda} v^\lambda \rb s^\mu
  = \frac{1}{\sqrt{\gamma}} \dow_\mu \lb \sqrt{\gamma}\, s^\mu \rb
  = \kB\Delta \geq 0~.
  \label{eq:covariantsecondlaw}
\end{equation}
Here $\Delta$ has to be a positive semi-definite quadratic form and $\kB$ is the
Boltzmann constant. The second law is imposed onshell, i.e. it is only required
to be satisfied by configurations satisfying the conservation equations
\eqref{eq:NC.Conservation}. Nonetheless, we can convert it into a offshell
statement by adding arbitrary combinations of conservation
equations. Introducing an arbitrary vector multiplier $\beta^\mu$ and a scalar
one $\Lambda_\beta$, we can write~\cite{Loganayagam:2011mu, Jain:2018jxj}
\begin{align}
\label{eq:towrite}
  \lb \nabla_\mu + F^n_{\mu\lambda} v^\lambda \rb s^\mu
  &- \kB \beta^\rho n_\rho \lB \lb \nabla_\mu + F^n_{\mu\lambda} v^\lambda \rb
  \epsilon^\mu + \ldots \rB
  + \kB \beta^\rho h_{\rho\nu} \lB
  \lb \nabla_\mu + F^n_{\mu\lambda} v^\lambda \rb
  \lb v^\mu \pi^\nu + \tau^{\mu\nu} \rb + \ldots \rB \nn\\
  &+ \kB \lb \Lambda_\beta + \beta^\rho A_\rho \rb
    \lb \nabla_\mu + F^n_{\mu\lambda} v^\lambda \rb j^\mu
    = \kB\Delta \geq 0~,
\end{align}
which will be satisfied offshell for some $\beta^\mu$ and $\Lambda_\beta$. It can
be checked that the ideal fluid constitutive relations \eqref{eq:consti-ideal}
satisfy this relation for $\Delta = \mathcal{O}(\dow^2)$, provided that we
choose $\kB\beta^\mu = u^\mu/T + \mathcal{O}(\dow)$ and
$\kB\Lambda_\beta = (\mu - u^\mu A_\mu)/T + \mathcal{O}(\dow)$.

Recall that we had an immense amount of redefinition freedom on our hands in the
choice of hydrodynamic variables $u^\mu$, $T$, and $\mu$ that we left unfixed at
the end of \cref{sec:frame}. We can fix this freedom by requiring the
multipliers $\beta^\mu$, $\Lambda_\beta$ to be exactly equal to their ideal
order values with no derivative corrections
\begin{equation}
  \beta^\mu = \frac{u^\mu}{\kB T}~, \qquad
  \Lambda_\beta = \frac{\mu - u^\mu A_\mu}{\kB T}~.
  \label{eq:beta-def}
\end{equation}
This is known as a \emph{thermodynamic frame}. This, however, is not a complete
fixing. We can imagine performing certain redefinitions of $u^\mu$, $T$, $\mu$,
and by extension of $\beta^\mu$, $\Lambda_\beta$, that only change the
constitutive relations satisfying the adiabaticity equation
\eqref{eq:adiabaticity} up to combinations of conservation equations. Such
redefinitions still need to be accounted for as they leave the dynamics
invariant. This can be unambiguously done following our discussion around
\cref{eq:delB-EOM} and eliminating any $(d+2)$ combinations among
$\delta_\scB A_\mu$, $\delta_{\scB} n_\mu$, and $\delta_\scB h_{\mu\nu}$ from
the hydrodynamic data, leaving us with $d(d+5)/2$ independent components. Note
that $v^\mu v^\nu \delta_\scB h_{\mu\nu}$ is trivially zero. Different choices
lead to different thermodynamic frames. Of particular interest to us is the
thermodynamic density frame, where we choose the independent data to be
\begin{equation}
  h^{\mu\nu} \delta_{\scB} A_\nu, \qquad
  h^{\mu\nu} \delta_{\scB} n_\nu~, \qquad
  h^{\mu\rho}h^{\nu\sigma} \delta_{\scB} h_{\rho\sigma}~.
  \label{eq:spatial-nhs-data}
\end{equation}
This matches up with the density frame defined in \cref{eq:density-frame} in the
non-hydrostatic sector (i.e. part of the constitutive relations that vanish in a
hydrostatic/equilibrium configuration), but differ substantially in the
hydrostatic sector. The discussion for thermodynamic Landau frame is presented
in \cref{app:frame}. In the core of this paper we will be working in the
thermodynamic density frame. The main reason for this choice is that this frame,
unlike the Landau or Eckart frames, does not exhibit unphysical instabilities in
the linearised mode spectrum~\cite{Hiscock:1985zz, Bemfica:2017wps,
  Kovtun:2019hdm, Bemfica:2019knx, Bemfica:2020zjp}.\footnote{Stability of
  various hydrodynamic frames in relativistic, Galilean, and Carrollian fluids
  was studied in~\cite{Poovuttikul:2019ckt}.}  This fact will be clear when
studying linearised fluctuations in section \ref{sec:linearised}.

Eq.~\eqref{eq:towrite} can be transformed into a more useful form by defining the free energy current
\begin{equation}
  N^\mu = \frac{1}{\kB} s^\mu - \beta^\nu n_\nu\, \epsilon^\mu
  + v^\mu \beta^\nu \pi_\nu
  + \beta^{\lambda} h_{\lambda\nu} \tau^{\mu\nu}
  + \lb \Lambda_\beta + \beta^\rho A_\rho \rb j^\mu~,
  \label{eq:freeenergy}
\end{equation}
which leads to the \emph{adiabaticity equation}
\begin{equation}
  \lb \nabla_\mu + F^n_{\mu\lambda} v^\lambda \rb N^\mu
  = - \epsilon^\mu \delta_\scB n_\mu
  + \lb v^\mu \pi^\nu + \half \tau^{\mu\nu} \rb \delta_\scB h_{\mu\nu}
  + j^\mu \delta_\scB A_\mu + \Delta~, \qquad
  \Delta \geq 0~.
  \label{eq:adiabaticity}
\end{equation}
The operator $\delta_\scB$ combines a Lie derivative $\lie_\beta$ along
$\beta^\mu$ and a gauge shift along $\Lambda_\beta$, i.e.
\begin{gather}
  \delta_\scB  n_\mu = \lie_\beta n_\mu~, \qquad
  \delta_\scB h_{\mu\nu} = \lie_\beta h_{\mu\nu}~, \qquad
  \delta_\scB  A_\mu = \lie_\beta A_\mu + \dow_\mu \Lambda_\beta~.
\end{gather}
This form of the second law of thermodynamics is more useful to implement on a
curved background. It can be checked that the constitutive relations
\eqref{eq:consti-ideal} are the most general solution of the adiabaticity
equation \eqref{eq:adiabaticity} at the leading derivative order with
$\Delta=0$. This also justifies their explicit form in the presence of
background sources. Note that $\Delta$ being zero at this derivative order means
that ideal fluids are non-dissipative, as we would physically expect.

\begin{table}[t]
  \centering
  \begin{tabular}{c|ccc|c|c}
    \toprule
    & C & P & T & PT & CPT \\
    \midrule
    $X^0$, $t$, $\tau$ & $+$ & $+$ & $-$ & $-$ & $-$ \\
    $X^i$, $x^i$, $\sigma^i$ & $+$ & $-$ & $+$ & $-$  & $-$ \\
    $\varphi$ & $-$ & $+$ & $-$ & $-$ & $+$ \\
    \midrule

    $u^i$, $\beta^i$, $\bbbeta^i$ & $+$ & $-$ & $-$ & $+$ & $+$ \\
    $T$, $\beta^t$, $\bbbeta^\tau$ & $+$ & $+$ & $+$ & $+$ & $+$ \\
    $\mu$, $\Lambda_\beta$, $\Lambda_\bbbeta$ & $-$ & $+$ & $+$ & $+$ & $-$ \\
    
    \midrule
    $\epsilon^t$, $n_t$ & $+$ & $+$ & $+$ & $+$ & $+$ \\
    $\epsilon^i$, $n_i$ & $+$ & $-$ & $-$ & $+$ & $+$ \\
    $\pi_i$, $v^i$ & $+$ & $-$ & $-$ & $+$ & $+$ \\
    $\tau^{ij}$, $h_{ij}$ & $+$ & $+$ & $+$ & $+$ & $+$ \\
    $j^t$, $b_t$ & $-$ & $+$ & $+$ & $+$ & $-$ \\
    $j^i$, $b_i$ & $-$ & $-$ & $-$ & $+$ & $-$ \\
    \bottomrule
  \end{tabular}
  \caption{Action of parity, time-reversal, and charge conjugation
    on various quantities in classical hydrodynamics and effective field
    theory. In the next section we introduce \SK double copies of various quantities in the effective theory,
    with labels ``$1/2$'' or ``$r/a$'', which have the same transformation properties as their unlabelled
    counterparts.\label{tab:CPT}}
\end{table}


Additional phenomenological requirements beyond Aristotelian symmetries, and the
second law of thermodynamics, are also usually imposed on the hydrodynamic
constitutive relations. Such is the case of discrete time-reversal (T), parity
(P), and charge conjugation (C) symmetries. In particular, underlying
microscopic theories are often taken to respect some kind of time-reversal
symmetry, like T, PT, or CPT. Denoting the action of these symmetries by
$\Theta$, in \cref{tab:CPT} we provide the transformation properties of various
quantities of interest under these symmetries.  Such symmetries are responsible
for imposing Onsager's conditions on the hydrodynamic correlation functions
(see~\cite{Kovtun:2012rj}) or for requiring the constitutive relations to be
$\Theta$-invariant in equilibrium
\cite{Banerjee:2015uta,Banerjee:2012iz,Jensen:2012jh}. These discrete symmetries
will be crucial when formulating the Schwinger-Keldysh EFT in the next
section. This completes our brief review of Aristotelian hydrodynamics -- the
explicit one-derivative order corrections will be considered in \cref{sec:1der}.

\section{Effective field theory for boost-agnostic hydrodynamics}
\label{sec:EFT}

In this section, we discuss the \SK effective field theory for boost-agnostic
hydrodynamics. Unlike the EFT for relativistic hydrodynamics developed over the
last decade~\cite{Dubovsky:2011sj, Grozdanov:2013dba, Crossley:2015evo,
  Glorioso:2017fpd, Glorioso:2016gsa, Gao:2017bqf, Glorioso:2017lcn,
  Gao:2018bxz, Jensen:2017kzi, Jensen:2018hse, Haehl:2018lcu}, the EFT
description of boost-agnostic hydrodynamics needs to treat time and space
directions on independent footing. A similar discussion for Galilean
hydrodynamics appeared recently in~\cite{Jain:2020vgc}, where the time and space
directions were indeed treated independently, but nonetheless had to be tied
down to respect the underlying Milne boost symmetry. As noted there, Milne
boosts actually make things quite hard for an effective field theorist; to make
this symmetry manifest, one needs to pass to a higher-dimensional
``null-background'' representation followed by a null reduction to obtain the
final results. Since boost-agnostic hydrodynamics does not worry about boosts
altogether, the ensuing EFT is formally simpler than its Galilean cousin. In
fact, the following discussion is mostly a reproduction of section~5
of~\cite{Jain:2020vgc}, but with the Milne boost symmetry revoked. The lack of a
symmetry does mean that many more terms can now enter the effective action at a
given derivative order that were previously not allowed, making the
boost-agnostic case structurally more richer; we will see an example of this for
one-derivative fluids in \cref{sec:1der}.

\subsection{Schwinger-Keldysh sigma model on the fluid worldvolume}

In this section we introduce a worldvolume formulation of the EFT. The fluid
worldvolume is a $(d+1)$-dimensional manifold endowed with coordinates
$\sigma^\alpha$.  These coordinates can be interpreted as labels associated with
each fluid element in the physical spacetime.  The dynamical fields living on
the worldvolume are $X^\mu_s(\sigma)$ and $\varphi_s(\sigma)$ with $s=1,2$, and
are \SK double copies of spacetime coordinates and U(1) phases of a given fluid
element. We can decompose these fields in average combinations according to
$X^\mu_{1,2} = X^\mu_r \pm \hbar/2\, X^\mu_a$ and
$\varphi_{1,2} = \varphi_r \pm \hbar/2\,\varphi_a$. The combination
$X^\mu_r(\sigma)$ denotes the physical spacetime coordinates and is akin to an
embedding map, while $\varphi_r(\sigma)$ denotes the physical U(1) phase of the
fluid elements. In turn, the average combinations $X^\mu_a(\sigma)$ and
$\varphi_a(\sigma)$ encode the stochastic degrees of freedom. The worldvolume
also contains two fixed reference fields, namely, a thermal vector field
$\bbbeta^\alpha(\sigma)$ and a chemical shift field
$\Lambda_\bbbeta(\sigma)$. These additional fields define the global rest frame
and global chemical potential associated with states in global thermal
equilibrium.

The EFT is required to be invariant under translations and rotations of the
coordinates $X^\mu_s(\sigma)$ as well as under global U(1) shifts of the phases
$\varphi_s(\sigma)$ acting independently on the two \SK spacetimes. In order to
understand their action within the effective field theory, one must introduce
double copies of Aristotelian sources as in \cref{sec:sources}.  In particular,
associated with each \SK spacetime we have the clock forms $n_{s\mu}(X_s)$,
degenerate spatial metrics $h_{s\mu\nu}(X_s)$ and gauge fields
$A_{s\mu}(X_s)$. Thus, under local \SK spacetime diffeomorphisms and gauge
transformations
\begin{gather}
  X_s^\mu(\sigma) \to X'^\mu_s(X_s(\sigma))~, \qquad \varphi_s(\sigma) \to
  \varphi_s(\sigma) - \Lambda_s(X_s(\sigma))~,
  \label{eq:SK-phys-diffeo}
\end{gather}
the action on the background sources is given by \cref{eq:NC.diffeo}. It is
useful to make the symmetries \eqref{eq:SK-phys-diffeo} manifest on the fluid
worldvolume by defining pullbacks of the background sources (with an additional
gauge transformation) such that
\begin{gather}
  \bbn_{s\alpha}(\sigma)
  = n_{s\mu}(X_s(\sigma))\,\dow_\alpha X^\mu_s(\sigma)~, \qquad
  \bbh_{s\alpha\beta}(\sigma) = h_{s\mu\nu}(X_s(\sigma))\,
  \dow_\alpha X^\mu_s(\sigma) \dow_\beta X^\nu_s(\sigma)~, \nn\\
  \bbA_{s\alpha}(\sigma) = A_{s\mu}(X_s(\sigma))\,\dow_\alpha X^\mu_s(\sigma) +
  \dow_\alpha\varphi_s(\sigma)~~.
  \label{eq:SKinvariants-FS}
\end{gather}
All the dependence on the dynamical and background fields in the effective
theory must enter via these invariants to respect \SK spacetime symmetries. This
fixes the structure of coupling between background and dynamical fields in the
effective theory.

The EFT on the fluid worldvolume is required to be locally reparametrisation invariant and invariant
under local shifts of the U(1) phases $\varphi_s(\sigma)$. In particular, under
\begin{subequations}
  \begin{gather}
    \sigma^\alpha \to \sigma'^\alpha(\sigma)~, \qquad \varphi_s(\sigma) \to
    \varphi_s(\sigma) + \lambda(\sigma)~,
  \end{gather}
  in which the two phases shift simultaneously. The pullback of background sources,
  \begin{gather}
    \bbn_{s\alpha}(\sigma) \to \bbn'_{s\alpha}(\sigma') =
    \frac{\dow\sigma^\beta}{\dow \sigma'^\alpha}\bbn_{s\beta}(\sigma)~, \qquad
    \bbh_{s\alpha\beta}(\sigma) \to \bbh'_{s\alpha\beta}(\sigma') =
    \frac{\dow\sigma^\gamma}{\dow \sigma'^\alpha} \frac{\dow \sigma^\delta}{\dow
      \sigma'^\beta}
    \bbh_{s\gamma\delta}(\sigma)~, \nn\\
    \bbA_{s\alpha}(\sigma) \to \bbA'_{s\alpha}(\sigma') =
    \frac{\dow\sigma^\beta}{\dow \sigma'^\alpha} \lb \bbA_{s\beta}(\sigma)
    + \dow_\beta \lambda(\sigma) \rb~, 
  \end{gather}
  transform as tensors under such reparametrisations and phase shifts while the worldvolume fields $\bbbeta^\alpha(\sigma)$ and $\Lambda_\bbbeta(\sigma)$ transform in the expected manner, namely
  \begin{gather}
   \bbbeta^\alpha(\sigma) \to \bbbeta'^\alpha(\sigma') =
    \frac{\dow\sigma'^\alpha(\sigma)}{\dow\sigma^\beta} \bbbeta^\beta(\sigma)~,
    \qquad \Lambda_\bbbeta(\sigma) \to \Lambda'_\bbbeta(\sigma') =
    \Lambda_\bbbeta(\sigma) - \bbbeta^\alpha(\sigma) \dow_\alpha
    \lambda(\sigma)~.
    \end{gather}
  \label{eq:fluid-spacetime-symm}%
\end{subequations}%
Given the transformation properties \eqref{eq:fluid-spacetime-symm}, it is possible to build
a gauge-invariant combination using the pullback of the gauge fields
$\bbA_{a\alpha} = (\bbA_{1\alpha}-\bbA_{2\alpha})/\hbar$. On the other hand, the combination
$\bbA_{r\alpha} = (\bbA_{1\alpha}+\bbA_{2\alpha})/2$ is not gauge-invariant. As such, when considering
an effective action, $\bbA_{r\alpha}$ can only enter via the gauge-invariant combinations 
$\bbbeta^\alpha \bbA_{r\alpha} + \Lambda_\bbbeta$ and $2\dow_{[\alpha}\bbA_{r\beta]}$.

It is possible to partially fix the reparametrisation freedom by choosing a set of 
worldvolume coordinates $\sigma^\alpha = (\tau,\sigma^i)$ and
setting $\bbbeta^\alpha = \beta_0\delta^\alpha_\tau$ as well as
$\Lambda_\bbbeta = \beta_0\mu_0$. Here, $\beta_0 = (\kB T_0)^{-1}$ is the
(constant) inverse temperature and $\mu_0$ the (constant) chemical potential of the global
thermal state. Given these choices, we are left with residual spatial reparametrisation freedom 
$\tau \to \tau + f(\vec\sigma)$ and $\sigma^i\to\sigma'^i(\vec\sigma)$ as well as with U(1)
phase shifts $\varphi_s \to \varphi_s + \lambda(\vec\sigma)$ (see \cite{Jain:2020vgc}).

The effective action $S$ for boost-agnostic hydrodynamics is the most generic
functional made out of the background and dynamical fields, respecting the \SK
spacetime symmetries \eqref{eq:SK-phys-diffeo} and the fluid worldvolume
symmetries \eqref{eq:fluid-spacetime-symm}. Using the invariants
$\mathbb{\Phi}_{s} = (\bbn_{s\alpha},\bbh_{s\alpha\beta},\bbA_{s\alpha})$ and
the reference thermal data $\bbB = (\bbbeta^\alpha,\Lambda_\bbbeta)$, the
effective action can be written in terms of a Lagrangian density
\begin{equation}
  S[\mathbb{\Phi}_1,\mathbb{\Phi}_2;\bbB]
  = \int \df^{d+1}\sigma\sqrt{\bbgamma_r}\,
  \mathcal{L}[\mathbb{\Phi}_1,\mathbb{\Phi}_2;\bbB]~~,
  \label{eq:SK-action-NC}
\end{equation}
where we have defined $\bbgamma_r = \det(\bbn_{r\alpha}\bbn_{r\beta} + \bbh_{r\alpha\beta})$ with
$\bbn_{r\alpha} = (\bbn_{1\alpha}+\bbn_{1\alpha})/2$ and
$\bbh_{r\alpha\beta} = (\bbh_{1\alpha\beta}+\bbh_{1\alpha\beta})/2$. The
Lagrangian $\mathcal{L}$ is a gauge-invariant scalar on the worldvolume. 
This form of the
action makes all the spacetime and worldvolume symmetries of the effective
theory manifest. However, the action is also required to obey a set of \SK
requirements on the account of describing generic thermal field
theories \cite{Crossley:2015evo,Jain:2020vgc}. These are
\begin{subequations}
  \begin{gather}
    S^*[\bbPhi_1,\bbPhi_2;\bbB] = -S[\bbPhi_2,\bbPhi_1;\bbB]~, \\
    S[\bbPhi,\bbPhi;\bbB] = 0~,  \\
    \mathrm{Im}\, S[\bbPhi_1,\bbPhi_2;\bbB] \geq 0~, \\
    S[\bbPhi_1,\bbPhi_2;\bbB]
    = S[\tilde\bbPhi_1,\tilde\bbPhi_2;\tilde\bbB]~,
    \label{eq:SK-KMS-FS}
  \end{gather}
  \label{eq:SK-cons-FS}%
\end{subequations}%
where the ``tilde'' KMS-conjugation in \cref{eq:SK-KMS-FS} is defined as
\begin{gather}
  \tilde\bbn_{1\alpha}(\sigma) = \Theta\bbh_{1\alpha}(\sigma)~, \quad
  \tilde\bbn_{2\alpha}(\sigma) = \Theta\bbn_{2\alpha}(\sigma) -
  i\hbar\,\Theta\lie_\bbbeta\bbn_{2\alpha}(\sigma)
  + \mathcal{O}(\hbar)~, \nn\\
  \tilde\bbh_{1\alpha\beta}(\sigma) = \Theta\bbh_{1\alpha\beta}(\sigma)~, \quad
  \tilde\bbh_{2\alpha\beta}(\sigma) = \Theta\bbh_{2\alpha\beta}(\sigma) -
  i\hbar\,\Theta\lie_\bbbeta \bbh_{2\alpha\beta}(\sigma)
  + \mathcal{O}(\hbar)~, \nn\\
  \tilde\bbA_{1\alpha}(\sigma) = \Theta\bbA_{1\alpha}(\sigma)~, \quad
  \tilde\bbA_{2\alpha}(\sigma) = \Theta\bbA_{2\alpha}(\sigma) -
  i\hbar\,\Theta\lie_\bbbeta \bbA_{2\alpha}(\sigma) -
  i\hbar\,\Theta\dow_\alpha\Lambda_\bbbeta(\sigma)
  + \mathcal{O}(\hbar), \nn\\
  \tilde\bbbeta^\alpha(\sigma) = \Theta\bbbeta^\alpha(\sigma)~, \qquad
  \tilde\Lambda_\bbbeta(\sigma) = \Theta\Lambda_\bbbeta(\sigma)~.
\end{gather}
Here $\Theta$ represents a discrete symmetry transformation involving a
time-flip, e.g. T, PT, or CPT. Its action on various quantities is given in
\cref{tab:CPT}. The operator $\lie_\bbbeta$ denotes a Lie derivative
along $\bbbeta^\alpha$. We can compactly denote these transformations as
\begin{equation}
  \tilde\bbPhi_1 = \Theta\bbPhi_1~, \qquad
  \tilde\bbPhi_2 = \Theta\bbPhi_2 - i\hbar\,\Theta\delta_\bbB\bbPhi_2
  + \mathcal{O}(\hbar)~, \qquad
  \tilde\bbB = \Theta\bbB~.
\end{equation}
The operator $\delta_\bbB$ combines the Lie derivative $\lie_\bbbeta$ along
$\bbbeta^\alpha$ and a gauge shift along $\Lambda_\bbbeta$.  Here we have
focused on the KMS transformations in the statistical limit ($\hbar\to0$); the
finite $\hbar$ quantum versions are the same as those in the Galilean case \cite{Jain:2020vgc}.


\subsection{Physical spacetime formulation}

The effective theory on the fluid worldvolume can be rewritten on the physical
spacetime.  The average coordinates $X^\mu_r(\sigma)$ are interpreted as an
embedding map, such that the location of the worldvolume in the physical
spacetime is given by $x^\mu = X^\mu_r(\sigma)$. Inverting this map implies that
the worldvolume coordinates $\sigma^{\alpha}= \sigma^\alpha(x)$ are seen as
dynamical fields from the physical spacetime point of view. Analogously, we can
express all other dynamical fields living on the worldvolume as functions of the
physical spacetime coordinates, in particular, the U(1) phase
$\varphi_r(x) = \varphi_r(\sigma(x))$ and the stochastic noise fields
$X_a^\mu(x) = X^\mu_a(\sigma(x))$ and $\varphi_a(x) = \varphi_a(\sigma(x))$.  It
is useful to split the worldvolume sources \eqref{eq:SKinvariants-FS} into
average and difference combinations
$\bbn_{1,2\,\alpha} = \bbn_{r\alpha} \pm \hbar/2\,\bbn_{a\alpha}$,
$\bbh_{1,2\,\alpha\beta} = \bbh_{r\alpha\beta} \pm
\hbar/2\,\bbh_{a\alpha\beta}$, and
$\bbA_{1,2\,\alpha} = \bbA_{r\alpha} \pm \hbar/2\,\bbA_{a\alpha}$. Using these,
one can define worldvolume gauge-invariant pushforwards onto the physical
spacetime using the inverse map $\sigma^\alpha(x)$. In particular the average
physical sources are given by\footnote{Here $N_{r,a\,\mu}$ should not be
  confused with the free energy current $N^\mu$ in eq.~\eqref{eq:freeenergy}.}
\begin{subequations}
  \begin{gather}
    N_{r\mu}(x)
  = \bbn_{r\alpha}(\sigma(x))\, \dow_\mu\sigma^\alpha(x)
    = n_{r\mu}(x) + \mathcal{O}(\hbar)~~, \nn\\
  H_{r\mu\nu}(x)
  = \bbh_{r\alpha\beta}(\sigma(x)) \, 
    \dow_\mu\sigma^\alpha(x) \dow_\nu\sigma^\beta(x)
    = h_{r\mu\nu}(x) + \mathcal{O}(\hbar)~~, \nn\\
  B_{r\mu}(x)
  = \bbA_{r\alpha}(\sigma(x)) \, \dow_{\mu}\sigma^\alpha(x)
    - \dow_{\mu}\varphi_r(x)
    = A_{r\mu}(x) + \mathcal{O}(\hbar)~~,
    \end{gather}
    while the stochastic sources are defined as
    \begin{gather}
    N_{a\mu}(x)
  = \bbn_{a\alpha}(\sigma(x))\, \dow_\mu\sigma^\alpha(x)
    = n_{a\mu}(x) + \lie_{X_a}n_{r\mu}(x) + \mathcal{O}(\hbar)~~, \nn\\
  H_{a\mu\nu}(x)
  = \bbh_{a\alpha\beta}(\sigma(x)) \, 
    \dow_\mu\sigma^\alpha(x) \dow_\nu\sigma^\beta(x)
    = h_{a\mu\nu}(x) + \lie_{X_a} h_{r\mu\nu} + \mathcal{O}(\hbar)~~, \nn\\
  B_{a\mu}(x)
  = \bbA_{a\alpha}(\sigma(x)) \, \dow_{\mu}\sigma^\alpha(x)
    = A_{a\mu}(x) + \dow_\mu \varphi_a(x)
    + \lie_{X_a}A_{r\mu}(x) + \mathcal{O}(\hbar)~~.
    \end{gather}
    \label{eq:newsources}%
\end{subequations}
In \cref{eq:newsources}, we have defined the Lie derivative along $X^\mu_a(x)$
as $\lie_{X_a}$ and decomposed the background fields as well into average and
difference combinations such that $n_{1,2\mu} = n_{r\mu} \pm \hbar/2 n_{a\mu}$,
$h_{1,2\mu\nu} = h_{r\mu\nu} \pm \hbar/2 h_{a\mu\nu}$, and
$A_{1,2\mu} = A_{r\mu} \pm \hbar/2 A_{a\mu}$ up to leading order in
$\hbar$. These average background fields can be identified with the classical
Aristotelian background fields of \cref{sec:sources}. We can also identify a
frame velocity $v^\mu_r$ and inverse spatial metric $h_r^{\mu\nu}$ on the
physical spacetime as the averages $v_r^\mu = (v_1^\mu + v_2^\mu)/2$ and
$h_r^{\mu\nu} = (h_1^{\mu\nu}+h_2^{\mu\nu})/2$. These satisfy the conditions
\eqref{eq:NC-normalisations} at leading order in $\hbar$.

Similarly, the hydrodynamic fields $\beta^\mu(x)$ and $\Lambda_\beta(x)$ are
obtained by pushforward of $\bbbeta^\alpha(\sigma)$ and
$\Lambda_\bbbeta(\sigma)$ such that\footnote{If we pick the frame
  $\bbbeta^\alpha(\sigma) = \beta_0\delta^\alpha_\tau$ and
  $\Lambda_\bbbeta = \beta_0\mu_0$, one obtains the physical spacetime
  counterparts $\beta^\mu = \beta_0 \dow_\tau X^\mu_r$ and
  $\Lambda_\beta = \beta_0(\mu_0 + \dow_\tau \varphi_r)$.}
\begin{equation}
  \beta^\mu(x) = \bbbeta^\alpha(\sigma(x)) \dow_\alpha X_r^\mu(\sigma(x))~, \qquad
  \Lambda_\beta(x) = \Lambda_\bbbeta(\sigma(x))
  + \bbbeta^\alpha(\sigma(x)) \dow_\alpha \varphi_r(\sigma(x))~.
 \label{eq:SK-beta-fields-NC}
\end{equation}
Additionally, the classical hydrodynamic fields introduced in \cref{sec:essentials}, namely,
the normalised fluid velocity $u^\mu(x)$ obeying $u^\mu N_{r\mu} = 1$, the local temperature $T(x)$, and the
chemical potential $\mu(x)$ are defined as\footnote{These definitions of the
  hydrodynamic fields are not boost invariant, since we are dealing with
  boost-agnostic hydrodynamics. The relation to Galilean and relativistic
  hydrodynamic fields is presented in \cref{sec:gal-rel}.}
\begin{gather}
  \kB T(x) = \frac{1}{\beta^\mu(x) N_{r\mu}(x)}~, \qquad
  u^\mu(x) = \frac{\beta^\mu(x)}{\beta^\lambda(x) N_{r\lambda}(x)}~, \qquad
  \mu(x)
  = \frac{\beta^\mu(x) B_{r\mu}(x) + \Lambda_\beta(x)}
  {\beta^\lambda(x) N_{r\lambda}(x)}~,
  \label{eq:SK-hydro-fields-NC}
\end{gather}
which are gauge-invariant and do not depend on the stochastic fields.

It is necessary to make sure that the \SK spacetime symmetries
\eqref{eq:SK-phys-diffeo} are correctly implemented in the physical
spacetime. This can be done by requiring the resulting EFT to be invariant under
``average'' coordinate and gauge transformations using $\sigma(x)$ and
$\varphi_r(x)$, i.e.
\begin{subequations}
  \begin{equation}
    x^\mu \to x'^\mu(x)~, \qquad
    \varphi_r(x) \to \varphi_r(x) - \Lambda(x)~.
  \end{equation}
  Under such transformations the ``average'' background structures and
  hydrodynamic fields transform according to
  \begin{gather}
    N_{r\mu}(x) \to N'_{r\mu}(x') =
    \frac{\dow x^\nu}{\dow x'^\mu} N_{r\nu}(x)~, \qquad
    H_{r\mu\nu}(x) \to H'_{r\mu\nu}(x') =
    \frac{\dow x^\rho}{\dow x'^\mu} \frac{\dow x^\sigma}{\dow x'^\nu}
    H_{r\rho\sigma}(x)~, \nn\\   
    B_{r\mu}(x) \to B'_{r\mu}(x') =
    \frac{\dow x^\nu}{\dow x'^\mu}
    \lb B_{r\nu}(x) + \dow_\nu \Lambda(x) \rb~, \nn\\   
    \beta^\mu(x) \to \beta'^\mu(x') =
    \frac{\dow x'^\mu(\sigma)}{\dow x^\nu} \bbbeta^\nu(x), \qquad
    \Lambda_\beta(x) \to \Lambda'_\beta(x') =
    \Lambda_\beta(x) - \beta^\mu(x) \dow_ \mu \Lambda(x)~,
  \end{gather}%
  while the ``difference'' stochastic parts transform as
  \begin{gather}
  N_{a\mu\nu}(x) \to N'_{a\mu\nu}(x') =
    \frac{\dow x^\nu}{\dow x'^\mu} N_{a\nu}(x)~, \qquad
  H_{a\mu\nu}(x) \to H'_{a\mu\nu}(x') =
    \frac{\dow x^\rho}{\dow x'^\mu} \frac{\dow x^\sigma}{\dow x'^\nu}
    H_{a\rho\sigma}(x)~, \nn\\
   B_{a\mu}(x) \to B'_{a\mu}(x') =
    \frac{\dow x^\nu}{\dow x'^\mu} B_{a\nu}(x)~.
  \end{gather}
  \label{eq:PS-symm-NC}%
\end{subequations}%
Let us introduce the compact notation
$\Phi_{r,a} = (N_{r,a\mu},H_{r,a\mu\nu},B_{r,a\mu})$ and
$\scB = (\beta^\mu,\Lambda_\beta)$, which are essentially the physical spacetime
versions of $\bbPhi_{1,2}$ and $\bbB$. In terms of these, the hydrodynamic
effective action \eqref{eq:SK-action-NC} can be rewritten in physical spacetime
language leading to
\begin{equation}
  S[\Phi_r,\Phi_a;\scB]
  = \int \df^{d+1}x\sqrt{\gamma_r}\,
  \mathcal{L}[\Phi_r,\Phi_a;\scB]~~,
  \label{eq:SK-action-PS-NC}
\end{equation}
with $\gamma_r = \det(n_{r\mu}n_{r\nu} + h_{r\mu\nu})$. The action is manifestly
invariant under worldvolume and physical spacetime symmetries. However,
it does need to satisfy the \SK constraints
\begin{subequations}
  \begin{gather}
    S^*[\Phi_r,\Phi_a;\scB] = -S[\Phi_r,-\Phi_a;\scB]~,
    \label{eq:SK-conj-null} \\
    S[\Phi_r,\Phi_a=0;\scB] = 0~, \label{eq:SK-lin-null} \\
    \mathrm{Im}\, S[\Phi_r,\Phi_a;\scB] \geq 0~, \label{eq:SK-pos-null} \\
    S[\Phi_r,\Phi_a;\scB] = S[\tilde\Phi_r,\tilde\Phi_a;\tilde\scB]~,
    \label{eq:SK-KMS-null}
  \end{gather}
  \label{eq:SK-constraints-null}%
\end{subequations}
where the KMS conjugation follows from
\begin{equation}
  \tilde\Phi_r = \Theta\Phi_r + \mathcal{O}(\hbar)~, \qquad
  \tilde\Phi_a = \Theta\Phi_a + i\Theta\delta_\scB\Phi_r
  + \mathcal{O}(\hbar)~, \qquad
  \tilde\scB = \Theta\scB + \mathcal{O}(\hbar)~.
\end{equation}
The operator $\delta_\scB$ denotes a Lie derivative along $\beta^\mu$ combined
with a gauge shift along $\Lambda_\beta$, i.e.
$\delta_{\scB}N_{r\mu} = \lie_\beta N_{r\mu}$,
$\delta_{\scB}H_{r\mu\nu} = \lie_\beta H_{r\mu\nu}$, and
$\delta_\scB B_{r\mu} = \lie_\beta B_{r\mu} + \dow_\mu \Lambda_\beta$. As a
reminder, $\Theta$ is a discrete symmetry transformation that the theory might
enjoy such as T, PT, or CPT (see \cref{tab:CPT}). We can define the ``$r/a$''
variants of the hydrodynamic operators by varying the action with respect to ``$a/r$''
background fields according to
\begin{align}
  \delta S
  &= \int \df^{d+1} x \sqrt{\gamma_r} \bigg[
  \rho^\mu_r\delta A_{a\mu}
  - \epsilon^\mu_r \delta n_{a\mu}
  + \lb v^\mu_r \pi^\nu_r + \half \tau^{\mu\nu}_r\rb \delta h_{a\mu\nu} \nn\\
  &\qquad\qquad\qquad\qquad
    + \rho^\mu_a\delta A_{r\mu}
  - \epsilon^\mu_a \delta n_{r\mu}
    + \lb v^\mu_r \pi^\nu_a
    + \half \tau^{\mu\nu}_a\rb \delta h_{r\mu\nu}
  \bigg]~.
\end{align}
The ``$r$'' operators are understood as the physical hydrodynamic observables,
while the ``$a$'' ones as the associated stochastic noise
counterparts. Out-of-equilibrium thermal correlations functions of these
operators can be computed by varying the \SK generating functional which takes the form
\begin{equation}
  \exp W[\phi_{r},\phi_{a}]
  = \int \mathcal{D}X_r \mathcal{D} X_a
  \mathcal{D}\varphi_r\mathcal{D}\varphi_a\,
  \exp\lb iS[\Phi_{r},\Phi_{a};\scB]\rb~~,
  \label{eq:SK-Z-null}
\end{equation}
where $\phi_{r,a} = (-n_{r,a\mu},h_{r,a\mu\nu},A_{r,a\mu})$.

\subsection{Schwinger-Keldysh effective action}

Based upon the considerations of the previous subsection, it is possible to find
the explicit structure of the effective action entering in
eq.~\eqref{eq:SK-Z-null}. The procedure is directly analogous to that of
Galilean fluids \cite{Jain:2020vgc}. KMS conjugation acts on the building blocks
of the effective action according to $\scB \to\Theta\scB$,
$\Phi_r \to \Theta\Phi_r$, $\Phi_a \to \Theta\Phi_a + i\Theta\delta_\scB \Phi_r$
for the hydrodynamic fields $\scB = (\beta^\mu,\Lambda_\beta)$ and the
invariants $\Phi_{r,a} = (N_{r,a\mu}, 1/2\,H_{r,a\mu\nu}, B_{r,a\mu})$. Thus the
most general effective action for hydrodynamics without boosts is given by a set
of totally-symmetric multi-linear operators $\mathcal{D}_m(\circ,\ldots)$ made
out of $\Phi_r$ and $\scB$, allowing $m$ number of arguments from the vector
space spanned by $i\delta_\scB\Phi_r$ and $\Phi_a$. In particular, the minimal
Lagrangian for classical hydrodynamics is given by
\begin{align}
  \mathcal{L}
  &= \cD_1(\Phi_a)
    + i \cD_{2}(\Phi_a,\Phi_a {+} i\delta_\scB\Phi_r)
    + \cD_{3}(\Phi_a{+}{\textstyle\frac{i}{2}}\delta_\scB \Phi_r,
    \Phi_a,\Phi_a {+} i \delta_\scB \Phi_r)
    + \mathcal{O}(\hbar)~~,
    \label{eq:L-G-NC}%
\end{align}
where the $\mathcal{D}_m(\circ,\ldots)$ operators satisfy the following constraints (see \cite{Jain:2020vgc} for more details)
\begin{subequations}
  \begin{gather}
    \mathcal{D}_1(\delta_\scB \Phi_r) = \frac{1}{\sqrt{\gamma_r}} \dow_\mu \lb
    \sqrt{\gamma_r}\, \mathcal{N}_0^\mu\rb
    \quad \text{for some vector~} \mathcal{N}_0^\mu~~, \\
    \mathcal{D}_{1}(\Phi_a), \quad
    \mathcal{D}_{2}(\Phi_a,\Phi_a), \quad
    \mathcal{D}_{3}(\Phi_a,\Phi_a,\Phi_a)
    \quad \text{are $\Theta$-even}~, \\
    \mathcal{D}_2(\Phi,\Phi)\big|_{\text{leading order}} \geq 0
    \quad \text{for arbitrary~} \Phi = (N_\mu,1/2\, H_{\mu\nu}, B_\mu)~~.
  \end{gather}
  \label{eq:operators}%
\end{subequations}
These constraints are consistent with the second law of thermodynamics
\eqref{eq:covariantsecondlaw}; see~\cite{Jain:2020vgc} for a derivation. In the
next section, we will provide the explicit form of the operators
$\mathcal{D}_{1,2}$. As we are focusing in first order corrections, we do not
provide the form of $\mathcal{D}_{3}$ as this operator contributes with second
order and higher correction terms in the gradient expansion. Since in this work
we are not interested in the most generic stochastic contributions, we have
skipped the operators $\mathcal{D}_{n>3}$ in our discussion, as these do not
contribute to the classical equations of hydrodynamics without boosts;
see~\cite{Jain:2020fsm} for a detailed discussion.

\section{One-derivative boost-agnostic hydrodynamics}
\label{sec:1der}

In this section, we discuss charged boost-agnostic hydrodynamics up to first
order in the derivative expansion. We start by writing down the most generic
classical constitutive relations for the system allowed by the adiabaticity
equation (second law of thermodynamics) discussed in \cref{sec:adiabaticity}. We
then proceed to write down the explicit effective action for one-derivative
hydrodynamics utilising the machinery from \cref{sec:EFT}.  We will briefly
discuss how these are related to the more familiar constitutive relations of
Galilean and relativistic hydrodynamics in \cref{sec:gal-rel}. As a simple
application of these results, in section \ref{sec:linearised} we study
fluctuations around an equilibrium state in boost agnostic hydrodynamics.

\subsection{Classical constitutive relations}

We want to work out the most generic constitutive relations that satisfy the
adiabaticity equation \eqref{eq:adiabaticity}, truncated at first order in the
derivative expansion. Focusing on the parity-even sector, the solutions can be
classified into three classes: (1) hydrostatic (hs; Class H$_\rmS$) that survive
in a hydrostatic configuration, i.e. when we set
$\delta_\scB n_\mu = \delta_\scB h_{\mu\nu} = \delta_\scB A_\mu = 0$; (2)
non-hydrostatic non-dissipative (nhsnd; Class $\overline\rmD$) that vanish in an
hydrostatic configuration, but do not contribute to entropy production quadratic
form $\Delta$; and finally (3) dissipative (diss; Class D) that also vanish in a
hydrostatic configuration, but contribute non-trivially to entropy
production.\footnote{In the parity-odd sector, there can be additional
  contributions coming from global anomalies (Class A) and transcendental
  anomalies (Class H$_\rmV$). These contributions are entirely fixed up to a few
  constants. See e.g.~\cite{Jain:2018jxj} for a discussion.}

\subsubsection{Hydrostatic transport}
\label{sec:hydrostatic}

The hydrostatic sector is completely characterised by the free-energy density
\begin{equation}
  \mathcal{N} 
  = p
  - F_0 v^\mu \dow_\mu \mu
  - F_1 v^\mu \dow_\mu T
  - F_2 v^\mu \dow_\mu \vec u^2~.
  \label{eq:hydrostatic-N}
\end{equation}
This is the most generic hydrostatic scalar that can be made out of the
constituent fields at one-derivative order. Here $p(T,\mu,\vec u^2)$ and
$F_{0,1,2}(T,\mu,\vec u^2)$ are arbitrary functions of zeroth order scalars. It
is easy to check that
\begin{align}
  \frac{1}{\sqrt{\gamma}}\dow_\mu (\sqrt{\gamma}\,
  \mathcal{N}\beta^\mu)
  &= \frac{1}{\sqrt{\gamma}}\delta_\scB (\sqrt{\gamma}\,\mathcal{N}) \nn\\
  &= \mathcal{N} \lb v^\mu \delta_\scB n_\mu
    + \half h^{\mu\nu} \delta_\scB h_{\mu\nu} \rb
    + \frac{\delta \mathcal{N}}{\delta A_\mu} \delta_\scB A_\mu
    + \frac{\delta \mathcal{N}}{\delta n_\mu} \delta_\scB n_\mu
    + \frac{\delta \mathcal{N}}{\delta h_{\mu\nu}} \delta_\scB h_{\mu\nu} \nn\\
  &\qquad
    - \frac{1}{\sqrt{\gamma}} \dow_\mu (\sqrt{\gamma}\, \Theta_{\mathcal{N}}^\mu)~.
\end{align}
The variational derivatives have been performed at constant
$\beta^\mu = u^\mu/T$ and $\Lambda_\beta = (\mu-u^\mu A_\mu)/T$. Here
$\Theta_{\mathcal{N}}^\mu$ denotes a total derivative term generated
during the Euler-Lagrange procedure. Comparing this with \cref{eq:adiabaticity},
we can read out
\begin{gather}
  N^\mu_{\text{hs}} = \mathcal{N} \beta^\mu + \Theta_{\mathcal{N}}^\mu~, \qquad
  \Delta_{\text{hs}} = 0~, \nn\\
  j^\mu_{\text{hs}}
  = \frac{\delta \mathcal{N}}{\delta A_\mu}, \qquad
  \epsilon^\mu_{\text{hs}}
  = - \frac{\delta \mathcal{N}}{\delta n_\mu} - \mathcal{N} v^\mu~, \qquad
  \pi^{\mu}_{\text{hs}}
  = h^\mu_{~\rho} n_{\sigma} \frac{\delta \mathcal{N}}{\delta
    h_{\rho\sigma}}~, \qquad
  \tau^{\mu\nu}_{\text{hs}}
  = 2 h^\mu_{~\rho} h^\nu_{~\sigma} \frac{\delta \mathcal{N}}{\delta h_{\rho\sigma}}
  + \mathcal{N} h^{\mu\nu}~.
\end{gather}
Explicitly, we find
\begin{align}
  j^{\mu}_{\text{hs}}
  &= n\, u^\mu
    - \frac{\dow F_0}{\dow\mu} u^\mu v^\lambda \dow_\lambda \mu
    + \frac{1}{\sqrt{\gamma}}\dow_\lambda \lb \sqrt{\gamma}\, F_0 v^\lambda \rb u^\mu
    - \frac{\dow F_1}{\dow\mu} u^\mu v^\lambda \dow_\lambda T
    - \frac{\dow F_2}{\dow\mu} u^\mu v^\lambda \dow_\lambda \vec u^2
    + \mathcal{O}(\dow^2)~, \nn\\[1ex]
  \epsilon^{\mu}_{\text{hs}}
  &= \epsilon\,u^\mu + p\,\vec u^\mu
    - w_{F_0} u^\mu v^\lambda \dow_\lambda \mu
    + \frac{\mu}{\sqrt{\gamma}}\dow_\lambda \lb \sqrt{\gamma}\, F_0 v^\lambda \rb
    u^\mu \nn\\
  &\qquad
    - w_{F_1} u^\mu v^\lambda \dow_\lambda T
    + \frac{T}{\sqrt{\gamma}}\dow_\lambda \lb \sqrt{\gamma}\,
    F_1 v^\lambda \rb u^\mu
    - w_{F_2} u^\mu v^\lambda \dow_\lambda \vec u^2
    + \frac{2\vec u^2}{\sqrt{\gamma}}\dow_\lambda\lb \sqrt{\gamma}\,
    F_2 v^\lambda \rb u^\mu 
    + \mathcal{O}(\dow^2)~, \nn\\[1ex]
  \pi^\mu_{\text{hs}}
  &= \rho\, \vec u^\mu
    - \lb 2 \frac{\dow F_0}{\dow\vec u^2} \vec u^\mu v^\lambda 
    - F_0 h^{\mu\lambda} \rb \dow_\lambda \mu
    - \lb 2 \frac{\dow F_1}{\dow\vec u^2} \vec u^\mu v^\lambda 
    - F_1 h^{\mu\lambda} \rb \dow_\lambda T \nn\\
  &\qquad
    - \lb 2 \frac{\dow F_2}{\dow\vec u^2} \vec u^\mu v^\lambda 
    - F_2 h^{\mu\lambda} \rb \dow_\lambda \vec u^2
    + \frac{2}{\sqrt{\gamma}}\dow_\lambda \lb \sqrt{\gamma}\, F_2 v^\lambda \rb
    \vec u^\mu
    + \mathcal{O}(\dow^2)~, \nn\\[1ex]
  \tau^{\mu\nu}_{\text{hs}}
  &= \rho\,\vec u^\mu \vec u^\nu + p\, h^{\mu\nu}
    - \lb 2 \frac{\dow F_0}{\dow\vec u^2} \vec u^\mu \vec u^\nu
    + F_0 h^{\mu\nu}\rb v^\lambda \dow_\lambda \mu
    - \lb 2 \frac{\dow F_1}{\dow\vec u^2} \vec u^\mu \vec u^\nu
    + F_1 h^{\mu\nu}\rb v^\lambda \dow_\lambda T  \nn\\
  &\qquad
    - \lb 2 \frac{\dow F_2}{\dow\vec u^2} \vec u^\mu \vec u^\nu
    + F_2 h^{\mu\nu}\rb v^\lambda \dow_\lambda \vec u^2
    + \frac{2}{\sqrt{\gamma}}\dow_\lambda \lb \sqrt{\gamma}\, F_2 v^\lambda \rb
    \vec u^\mu \vec u^\nu
    + \mathcal{O}(\dow^2)~,
\end{align}
and
\begin{equation}
  \Theta_{\mathcal{N}}^\mu
  = \frac{1}{\kB T} v^\mu \lb F_0 u^\lambda \dow_\lambda \mu
  + F_1 u^\lambda \dow_\lambda T
  + F_2 u^\lambda \dow_\lambda \vec u^2 \rb,
\end{equation}
where we have used the identity
$\delta_\scB v^\mu = - v^\mu v^\nu \delta_\scB n_\mu - h^{\mu\lambda} v^\nu
\delta_\scB h_{\lambda\nu}$ and defined 
\begin{equation}
  w_{F_i}
  = T \frac{\dow F_i}{\dow T}
  + \mu \frac{\dow F_i}{\dow \mu}
  + 2 \vec u^2 \frac{\dow F_i}{\dow \vec u^2}~.
  \label{eq:hydrostatic_c}
\end{equation}
The readers can convince themselves that these are the most generic
parity-preserving hydrostatic constitutive relations allowed by the adiabaticity
equation. In the uncharged limit, that is, when $F_0=0$ and
$F_{1,2}\equiv F_{1,2}(T,\vec u^2)$, the hydrostatic contributions
\eqref{eq:hydrostatic_c} agree with those in \cite{deBoer:2020xlc}.\footnote{The
  comparison requires using the identification \eqref{eq:covariantST} and
  flipping $v^\mu\to-v^\mu$ to match the conventions of \cite{deBoer:2020xlc}.}

If the underlying microscopic theory respects a discrete $\Theta$ symmetry, such
as T, PT, or CPT, this will need to be imposed on the free-energy density
$\mathcal{N}$. For instance, for $\Theta=\rmT$ or $\Theta=\text{PT}$, all three
one-derivative coefficients must vanish, i.e. $F_{0,1,2} = 0$. On the other hand
for $\Theta = \text{CPT}$, we can only state that $F_{0,1,2}$ must be odd
functions of the chemical potential $\mu$. Note that if $j^\mu$ corresponds to
the particle number current, it does not make sense to discuss CPT.

\subsubsection{Non-hydrostatic non-dissipative transport}

Next we have the non-hydrostatic non-dissipative transport made out of linear
combinations of $\delta_{\scB} A_\mu$, $\delta_\scB n_\mu$, and
$\delta_\scB h_{\mu\nu}$ that satisfy \cref{eq:adiabaticity} with
$\Delta_{\text{nhsnd}} = 0$. Recall that due to our thermodynamic frame-fixing
condition, all the dependence on these can only appear via their spatial
components in \cref{eq:spatial-nhs-data}. Inspecting \cref{eq:adiabaticity}, it
immediately follows that
$n_\mu j^\mu_{\text{nhsnd}} = n_\mu \epsilon^\mu_{\text{nhsnd}} =
\pi^\mu_{\text{nhsnd}} = 0$. For the remaining contributions, we find
\begin{equation}
  \def\arraystretch{1.2}
  \begin{pmatrix}
    j^\mu_{\text{nhsnd}} \\
    \epsilon^\mu_{\text{nhsnd}} \\
    \tau^{\mu\nu}_{\text{nhsnd}}
  \end{pmatrix}
  = - \kB T
  \begin{pmatrix}
    0 & \bar D_{j\epsilon}^{\mu\rho} & \bar D_{j\tau}^{\mu(\rho\sigma)} \\
    - \bar D^{\rho\mu}_{j\epsilon} & 0 & \bar D_{\epsilon\tau}^{\mu(\rho\sigma)} \\
    - \bar D_{j\tau}^{\rho(\mu\nu)} & - \bar D_{\epsilon\tau}^{\rho(\mu\nu)}
    & \bar D^{(\mu\nu)(\rho\sigma)}_{\tau\tau}
  \end{pmatrix}
  \begin{pmatrix}
    \delta_\scB A_\rho \\ - \delta_\scB n_\rho \\ \half\delta_\scB h_{\rho\sigma}
  \end{pmatrix},
  \label{eq:matrix-nhsnd}
\end{equation}
where
\begin{align}
  \bar D^{\mu\rho}_{j\epsilon}
  &= \bar\fv_{01} P^{\mu\rho}
    + \bar\fs_{01} \hat u^\mu \hat u^\rho~, \nn\\
  \bar D_{j\tau}^{\mu(\rho\sigma)}
  &= 2\bar\fv_{02} P^{\mu(\rho} \hat u^{\sigma)}
    + \bar\fs_{02} \hat u^\mu \hat u^\rho \hat u^\sigma
    + \bar\fs_{03} \hat u^\mu P^{\rho\sigma}~, \nn\\
  \bar D_{\epsilon\tau}^{\mu(\rho\sigma)}
  &= 2\bar\fv_{12} P^{\mu(\rho} \hat u^{\sigma)}
    + \bar\fs_{12} \hat u^\mu \hat u^\rho \hat u^\sigma
    + \bar\fs_{13} \hat u^\mu P^{\rho\sigma}~, \nn\\
  \bar D^{(\mu\nu)(\rho\sigma)}_{\tau\tau}
  &= \bar\fs_{23} \lb  \hat u^\mu \hat u^\nu P^{\rho\sigma}
    - P^{\mu\nu} \hat u^\rho \hat u^\sigma\rb.
\end{align}
Here $\hat u^\mu = \vec u^\mu/|\vec u|$ and
$P^{\mu\nu} = h^{\mu\nu} - \hat u^\mu \hat u^\nu$, with
$|\vec u| = \sqrt{\vec u^2}$.  The transport coefficients appearing here are
arbitrary functions of $T$, $\mu$, and $\vec u^2$. The coefficient matrix in
\cref{eq:matrix-nhsnd} is antisymmetric, which ensures that there is no
contribution to $\Delta_{\text{nhsnd}}$.

As for the discrete symmetries $\Theta$, Onsager's relations require that the
constitutive relations in \cref{eq:matrix-nhsnd} are even under $\Theta$. With
$\Theta = \rmT$ or $\Theta = \text{PT}$, the entire non-hydrostatic
non-dissipative sector is set to zero. On the other hand for
$\Theta = \text{CPT}$, all the transport coefficients appearing here must be odd
functions of $\mu$.

\subsubsection{Dissipative transport}

The dissipative sector is quite similar to the non-hydrostatic non-dissipative
sector, but with the coefficient matrix being symmetric, leading to entropy
production. We again find that
$n_\mu j^\mu_{\text{diss}} = n_\mu \epsilon^\mu_{\text{diss}} =
\pi^\mu_{\text{diss}} = 0$, along with
\begin{align}
  \def\arraystretch{1.2}
  \begin{pmatrix}
    j^\mu_{\text{diss}} \\ \epsilon^\mu_{\text{diss}} \\ \tau^{\mu\nu}_{\text{diss}}
  \end{pmatrix}
  = -
  \kB T\begin{pmatrix}
    D_{jj}^{\mu\rho} & D_{j\epsilon}^{\mu\rho} & D_{j\tau}^{\mu(\rho\sigma)} \\
    D^{\rho\mu}_{j\epsilon} & D_{\epsilon\epsilon}^{\mu\rho}
    & D_{\epsilon\tau}^{\mu(\rho\sigma)} \\
    D_{j\tau}^{\rho(\mu\nu)} & D_{\epsilon\tau}^{\rho(\mu\nu)}
    & D^{(\mu\nu)(\rho\sigma)}_{\tau\tau}
  \end{pmatrix}
  \begin{pmatrix}
    \delta_\scB A_\rho \\ - \delta_\scB n_\rho \\ \half\delta_\scB h_{\rho\sigma}
  \end{pmatrix}~,
\end{align}
with
\begin{align}
  D^{\mu\rho}_{jj}
  &= \fv_{00} P^{\mu\rho}
    + \fs_{00} \hat u^\mu \hat u^\rho~, \nn\\
  D^{\mu\rho}_{j\epsilon}
  &= \fv_{01} P^{\mu\rho}
    + \fs_{01} \hat u^\mu \hat u^\rho~, \nn\\
  D^{\mu\rho}_{\epsilon\epsilon}
  &= \fv_{11} P^{\mu\rho}
    + \fs_{11} \hat u^\mu \hat u^\rho~, \nn\\
  D_{j\tau}^{\mu(\rho\sigma)}
  &= 2\fv_{02} P^{\mu(\rho} \hat u^{\sigma)}
    + \fs_{02} \hat u^\mu \hat u^\rho \hat u^\sigma
    + \fs_{03} \hat u^\mu P^{\rho\sigma}~, \nn\\
  D_{\epsilon\tau}^{\mu(\rho\sigma)}
  &= 2\fv_{12} P^{\mu(\rho} \hat u^{\sigma)}
    + \fs_{12} \hat u^\mu \hat u^\rho \hat u^\sigma
    + \fs_{13} \hat u^\mu P^{\rho\sigma}~, \nn\\
  D^{(\mu\nu)(\rho\sigma)}_{\tau\tau}
  &= 2\ft \lb P^{\rho(\mu} P^{\nu)\sigma}
    - {\textstyle\frac{1}{d-1}} P^{\mu\nu}P^{\rho\sigma} \rb
    + 4\fv_{22} \hat u^{(\mu} P^{\nu)(\rho} \hat u^{\sigma)} \nn\\
  &\qquad
    + \fs_{22} \hat u^\mu \hat u^\nu \hat u^\rho \hat u^\sigma
    + \fs_{23} \lb P^{\mu\nu} \hat u^\rho \hat u^\sigma
    + \hat u^\mu \hat u^\nu P^{\rho\sigma}\rb
    + \fs_{33} P^{\mu\nu} P^{\rho\sigma}~.
\end{align}
The associated entropy production quadratic form is given by
\begin{equation}
  \Delta_{\text{diss}}
  = \kB T
  \begin{pmatrix}
    \delta_\scB A_\rho \\ - \delta_\scB n_\rho \\ \half\delta_\scB h_{\rho\sigma}
  \end{pmatrix}^\rmT
  \begin{pmatrix}
    D_{nn}^{\mu\rho} & D_{n\epsilon}^{\mu\rho} & D_{n\pi}^{\mu(\rho\sigma)} \\
    D^{\rho\mu}_{n\epsilon} & D_{\epsilon\epsilon}^{\mu\rho}
    & D_{\epsilon\pi}^{\mu(\rho\sigma)} \\
    D_{n\pi}^{\rho(\mu\nu)} & D_{\epsilon\pi}^{\rho(\mu\nu)}
    & D^{(\mu\nu)(\rho\sigma)}_{\pi\pi}
  \end{pmatrix}
  \begin{pmatrix}
    \delta_\scB A_\rho \\ - \delta_\scB n_\rho \\ \half\delta_\scB h_{\rho\sigma}
  \end{pmatrix}
  \geq 0~.
\end{equation}
Imposing the positivity constraint merely requires that the dissipative
coefficient matrix is positive semi-definite. Explicitly, this leads to
\begin{equation}
  \begin{pmatrix}
    \fs_{00} & \fs_{01} & \fs_{02} & \fs_{03} \\
    \fs_{01} & \fs_{11} & \fs_{12} & \fs_{13} \\
    \fs_{02} & \fs_{12} & \fs_{22} & \fs_{23} \\
    \fs_{03} & \fs_{13} & \fs_{23} & \fs_{33}
  \end{pmatrix}
  \geq 0, \qquad
  \begin{pmatrix}
    \fv_{00} & \fv_{01} & \fv_{02} \\
    \fv_{01} & \fv_{11} & \fv_{12} \\
    \fv_{02} & \fv_{12} & \fv_{22}
  \end{pmatrix}
  \geq 0~, \qquad
  \ft \geq 0~,
  \label{eq:const2nd}
\end{equation}
where the positive semi-definiteness of a matrix means that all its eigenvalues
are positive semi-definite.

The discrete symmetry $\Theta$ requirements here work opposite to the
non-hydrostatic non-dissipative sector. With $\Theta =\rmT$ or
$\Theta = \text{PT}$ symmetry, the dissipative sector transport coefficients are
left invariant, while with $\Theta = \text{CPT}$ symmetry, all the dissipative
transport coefficients must be even functions of $\mu$.

Therefore, we have a total of 30 coefficients at one-derivative order: 4
hydrostatic (including the ideal order pressure $p$), 9 non-hydrostatic
non-dissipative, and 17 dissipative transport coefficients. We note that this
counting differs from that of \cite{Novak:2019wqg}.\footnote{The authors in
  \cite{Novak:2019wqg} counted 9 non-dissipative transport coefficients and 20
  dissipative transport coefficients. Further discussion on this can be found in
  \cref{app:Landau}.} In the (uncharged) limit in which the U(1) current is
removed, we have that
\begin{equation}
F_0=\bar\fv_{01}=\bar\fs_{01}=\bar\fv_{02}=\bar\fs_{02}=\bar\fs_{03}=\fv_{00}=\fs_{00}=\fv_{01}=\fs_{01}=\fs_{02}=\fv_{02}=\fv_{03}=0~~.
\label{eq:unchargedlimit}
\end{equation}
This amounts to a total of 17 transport coefficients: 3 hydrostatic (including
ideal order pressure), 4 non-hydrostatic non-dissipative, and 10 dissipative
transport coefficients, agreeing with the counting of \cite{deBoer:2020xlc} when 
focusing in special case in which the additional $U(1)$ is not present. A
precise comparison is given in \cref{app:Landau}.

\subsubsection{Entropy current}

The free-energy current associated with the one-derivative order constitutive
relations above is simply given as
\begin{equation}
  N^\mu = \mathcal{N} \beta^\mu + \Theta_{\mathcal{N}}^\mu.
\end{equation}
Note that there is no contribution to $N^\mu$ due to dissipative and
non-dissipative non-hydrostatic constitutive relations. Correspondingly, the
entropy current is given as
\begin{align}
  s^\mu
  &= \mathcal{N} \frac{u^\mu}{T} + \kB \Theta_{\mathcal{N}}^\mu
    + \frac{u^\nu}{T} n_\nu \epsilon^\mu
    - v^\mu \frac{u^\nu}{T} \pi_\nu
    - \frac{u^\lambda}{T} h_{\lambda\nu}\tau^{\mu\nu}
    - \frac{\mu}{T} j^\mu \nn\\
  &= s^\mu_{\text{can}} + s^\mu_{\text{non-can}}~~,
\end{align}
where
\begin{equation}
  s^\mu_{\text{can}}
  = \frac{p}{T}u^\mu
  + \frac{u^\nu}{T} n_\nu \epsilon^\mu
  - v^\mu \frac{u^\nu}{T} \pi_\nu
  - \frac{u^\lambda}{T} h_{\lambda\nu}\tau^{\mu\nu}
  - \frac{\mu}{T} j^\mu~~,
\end{equation}
is known as the canonical entropy current, while
\begin{align}
  s^\mu_{\text{non-can}}
  &= (\mathcal{N}-p) \frac{u^\mu}{T} + \kB \Theta_{\mathcal{N}}^\mu \nn\\
  &= \frac{v^\mu}{T} \lb F_0 \vec u^\lambda \dow_\lambda \mu
    + F_1 \vec u^\lambda \dow_\lambda T
    + F_2 \vec u^\lambda \dow_\lambda \vec u^2 \rb \nn\\
  &\qquad
    - \frac{\vec u^\mu}{T} \lb F_0 v^\mu \dow_\mu \mu
    + F_1 v^\mu \dow_\mu T
    + F_2 v^\mu \dow_\mu \vec u^2 \rb~,
    \label{eq:EC-non-can}
\end{align}
is known as the non-canonical part of the entropy current.

\subsection{Explicit effective action}
\label{sec:SKeffective}

We implement the following derivative counting in the EFT. The hydrodynamic
fields $\beta^\mu$, $\Lambda_\beta$ and the noise fields $X^\mu_a$, $\varphi_a$
are taken to be $\mathcal{O}(\dow^0)$. On the other hand, the average background
fields $n_{r\mu}$, $h_{r\mu\nu}$, $A_{r\mu}$ are treated at
$\mathcal{O}(\dow^0)$, while their difference partners $n_{a\mu}$,
$h_{a\mu\nu}$, $A_{a\mu}$ at $\mathcal{O}(\dow^1)$. In compact notation, this
means that $\Phi_r$, $\scB$ are $\mathcal{O}(\dow^0)$, while $\Phi_a$,
$\delta_\scB \Phi_r$ are $\mathcal{O}(\dow^1)$. Note that the classical
constitutive relations are given by a variational derivative of the effective
action with respect to the ``$a$'' type background fields. Therefore, the
derivative order of the effective action must be one more than that of the
constitutive relations. It follows that one-derivative hydrodynamics is entirely
characterised by the operators $\mathcal{D}_1(\circ)$ truncated at
one-derivative order and $\mathcal{D}_2(\circ,\circ)$ truncated at zeroth order;
$\mathcal{D}_3(\circ,\circ,\circ)$ only contributes to two-derivative
constitutive relations and higher.

Due to eq.~\eqref{eq:operators} we know that $\mathcal{D}_1(\circ)$ is the most
generic operator such that $\mathcal{D}_1(\delta_\scB \Phi_r)$ is a total
derivative. Recalling that in the statistical limit
$\Phi_r = (-n_{r\mu},1/2\,h_{r\mu\nu},A_{r\mu}) + \mathcal{O}(\hbar)$, we see
that this requirement is precisely the adiabaticity equation. The most generic
solution is therefore characterised by the adiabatic (hydrostatic $+$
non-hydrostatic non-dissipative) constitutive relations. Explicitly
\begin{equation}
  \mathcal{D}_1(\Phi_a)
  = j^\mu_{\text{hs}+\text{nhsnd}} B_{a\mu}
  - \epsilon^\mu_{\text{hs}+\text{nhsnd}} N_{a\mu}
  + \lb v^\mu \pi^\nu_{\text{hs}}
  + \half \tau^{\mu\nu}_{\text{hs}+\text{nhsnd}} \rb
  H_{a\mu\nu}~. 
\end{equation}
On the other hand, $\mathcal{D}_2(\circ,\circ)$ needs to be most generic
symmetric positive semi-definite bilinear operator. The contribution from the
same to the effective action is given as
\begin{equation}
  i\mathcal{D}_2(\Phi_a,\Phi_a{+}i\delta_\scB \Phi_r)
  = i\kB T
  \begin{pmatrix}
    B_{a\mu} \\ - N_{a\mu} \\ \half H_{a\mu\nu}
  \end{pmatrix}^\rmT
  \begin{pmatrix}
    D_{jj}^{\mu\rho} & D_{j\epsilon}^{\mu\rho} & D_{j\tau}^{\mu(\rho\sigma)} \\
    D^{\rho\mu}_{j\epsilon} & D_{\epsilon\epsilon}^{\mu\rho}
    & D_{\epsilon\tau}^{\mu(\rho\sigma)} \\
    D_{j\tau}^{\rho(\mu\nu)} & D_{\epsilon\tau}^{\rho(\mu\nu)}
    & D^{(\mu\nu)(\rho\sigma)}_{\tau\tau}
  \end{pmatrix}
  \begin{pmatrix}
    B_{a\rho} + i\delta_\scB A_{r\rho} \\ - N_{a\rho} - i\delta_\scB n_{r\rho}
    \\ \half H_{a\rho\sigma} + \frac{i}{2}\delta_\scB h_{r\rho\sigma}
  \end{pmatrix}.
\end{equation}
The $\Theta$-requirements on various constitutive relations discussed in the
previous subsection follow from here by demanding that both $\mathcal{D}_1$ and
$\mathcal{D}_2$ operators are $\Theta$-even, in accordance with the \SK
requirements in \cref{eq:operators}.

In flat spacetime, with $\Theta=\rmT$ or $\Theta=\text{PT}$ discrete symmetry
(that leads to vanishing of hydrostatic and non-hydrostatic non-dissipative
sectors), the effective Lagrangian for one-derivative order boost-agnostic
hydrodynamics takes the form
\begin{align}
  \mathcal{L}
  &= n\, \dow_t \varphi_a
    + n\, u^i \dow_i \varphi_a
    - \epsilon\, \dow_t X^t_a
    - \lb \epsilon + p \rb u^i \dow_i X^t_a
    + \rho\, u^i \dow_t X_{ai}
    + \lb \rho\,u^j u^i
    + p\, \delta^{ji} \rb
    \dow_j X_{ai} \nn\\
  &\qquad
    + i\kB T
  \begin{pmatrix}
    \dow_i\varphi_a \\ - \dow_i X_a^t \\ \dow_i X_{aj}
  \end{pmatrix}^\rmT
  \begin{pmatrix}
    D_{jj}^{ik} & D_{j\epsilon}^{ik} & D_{j\tau}^{i(kl)} \\
    D^{ki}_{j\epsilon} & D_{\epsilon\epsilon}^{ik}
    & D_{\epsilon\tau}^{i(kl)} \\
    D_{j\tau}^{k(ij)} & D_{\epsilon\tau}^{k(ij)}
    & D^{(ij)(kl)}_{\tau\tau}
  \end{pmatrix}
  \begin{pmatrix}
    \dow_k \varphi_a + \frac{i}{\kB}\dow_k\frac{\mu}{T} \\
    - \dow_kX_a^t - \frac{i}{\kB} \dow_k \frac{1}{T}
    \\ \dow_k X_{al} + \frac{i}{\kB} \dow_k \frac{u_l}{T}
  \end{pmatrix}.
  \label{eq:flat-action}
\end{align}
We will use these considerations in \cref{sec:linearised} to study linearised
fluctuations.

\section{Special limits}
\label{sec:gal-rel}

The spectrum of transport coefficients for a boost-agnostic fluid is quite
rich. We have a total of 30 coefficients at one-derivative order.  For a clearer
picture of these coefficients, it is helpful to make contact with the special
cases of fluids respecting Galilean or Lorentz boost symmetries.  In both these
instances, the spectrum only contains the thermodynamic pressure $p$ in the
hydrostatic sector and 3 dissipative transport coefficients: shear viscosity
$\eta$, bulk viscosity $\zeta$, and thermal/electric conductivity
$\kappa$/$\sigma$. In particular, there are no allowed one-derivative terms in
the hydrostatic or non-hydrostatic non-dissipative sectors. We will also discuss
fluids respecting anisotropic Lifshitz scaling symmetry, in which case the
transport coefficients reduce to: 3 hydrostatic (including pressure $p$), 6
non-hydrostatic non-dissipative, and 13 dissipative. The temperature dependence
of all these 22 coefficients is fixed by the scaling symmetry.

\subsection{Galilean fluids}

Galilean (Milne) boost symmetry acts on the background fields according to (see,
e.g.~\cite{Jain:2020vgc})
\begin{gather}
  n_\mu \to n_\mu, \qquad h_{\mu\nu} \to h_{\mu\nu} - 2 n_{(\mu} \psi_{\nu)} +
  n_\mu n_\nu \psi^2, \qquad
  A_\mu \to A_\mu + m \lb \psi_{\mu} - \frac{1}{2} n_\mu \psi^2 \rb, \nn\\
  v^\mu \to v^\mu + \psi^\mu, \qquad h^{\mu\nu} \to h^{\mu\nu},
  \label{eq:MilneBoost}
\end{gather}
for arbitrary parameters $\psi^\mu$ satisfying $\psi^\mu n_\mu = 0$, with the
definitions $\psi_\mu = h_{\mu\nu}\psi^\nu$, and
$\psi^2 = h_{\mu\nu}\psi^\mu \psi^\nu$. Here $m$ is an arbitrary constant
signifying mass per unit charge/particle. These are essentially the rules
governing how the background sources must change when we move to a different
frame of reference given by $v^\mu \to v^\mu + \psi^\mu$.  Using the
(equilibrium) effective action variation in \cref{eq:basic-delS}, this leads to
the Milne boost Ward identity
\begin{equation}
  \pi_\mu = m\,h_{\mu\nu} j^\nu.
  \label{eq:Milne-Ward}
\end{equation}
The hydrodynamic fields in the representation $\beta^\mu$, $\Lambda_\beta$
correspond to the local thermodynamic frame and hence do not transform under
boosts. However $u^\mu$, $T$, $\mu$ fields can potentially transform, which can
be derived using their definitions in \cref{eq:beta-def}. We find
\begin{equation}
  u^\mu \to u^\mu, \qquad
  T \to T, \qquad
  \mu \to \mu + m\, \vec u^\mu \psi_\mu - \frac{m}{2} \psi^2.
\end{equation}
It is convenient to define the boost-invariant Galilean chemical potential
$\mu_\gal = \mu + m/2\,\vec u^2$. All the transport coefficients in a Galilean
fluid are functions of $T$ and $\mu_\gal$. The equation of state is given as
$p(T,\mu,\vec u^2) = p(T,\mu_\gal)$. Using the thermodynamic relations
\eqref{eq:thermodynamics}, one then derives
\begin{equation}
  \df p = s\, \df T + n\, \df\mu_\gal~, \qquad
  \epsilon_\gal = Ts + \mu_\gal n - p~, \qquad
  \rho = m\, n~,
\end{equation}
where we have identified the Galilean internal energy density
$\epsilon_\gal = \epsilon - 1/2\, \rho\, \vec u^2$. The relation $\rho = m\, n$
can be understood as the ``Galilean equation of state''.


Constitutive relations for Galilean hydrodynamics in curved space-time are
already known; see for instance~\cite{Jain:2020vgc} and references
therein. Their translation to the boost-agnostic representation discussed in
this paper is quite straightforward because the language and hydrodynamic
frame that we have employed for boost-agnostic fluids is quite similar to the
one used for Galilean hydrodynamics. The complete set of one-derivative order
constitutive relations are given as
\begin{align}
  j^\mu
  &= \frac{\rho}{m} u^\mu, \nn\\
  \epsilon^\mu
  &= \lb\epsilon_\gal + {\textstyle\half}\rho \vec u^2 \rb u^\mu
    + p\, \vec u^\mu
    + T^2 \kappa\, h^{\mu\nu} \delta_\scB n_\nu \nn\\
  &\qquad
    - \lb 2\eta h^{\rho(\mu} h^{\nu)\sigma}
    + \lb \zeta - {\textstyle\frac{2}{d}}\eta\rb
    h^{\mu\nu} h^{\rho\sigma} \rb \vec u_\nu
    \half \lb \delta_\scB h_{\rho\sigma}
    - 2 \vec u_{(\rho}\delta_\scB n_{\sigma)} \rb
    , \nn\\
  \pi^\mu
  &= \rho\, \vec u^\mu, \nn\\
  \tau^{\mu\nu}
  &= \rho\, \vec u^\mu \vec u^\nu
    - \lb 2\eta h^{\rho(\mu} h^{\nu)\sigma}
    + \lb \zeta - {\textstyle\frac{2}{d}}\eta\rb
    h^{\mu\nu} h^{\rho\sigma} \rb \vec u_\nu
    \half \lb \delta_\scB h_{\rho\sigma}
    - 2 \vec u_{(\rho}\delta_\scB n_{\sigma)} \rb.
\end{align}
Note that the constitutive relations are already in the density frame
\eqref{eq:density-frame}. The mapping of ideal order thermodynamic coefficients
is already given above. At one-derivative order, we find
\begin{gather}
  F_0 = F_1 = F_2 = 0~, \nn\\
  \bar\fs_{01} = \bar\fs_{02} = \bar\fs_{03} = \bar\fs_{12} = \bar\fs_{13} =
  \bar\fs_{23} = \bar\fv_{01} = \bar\fv_{02} = \bar\fv_{12} = 0~, \nn\\
  \fs_{00} = \fs_{01} = \fs_{02} = \fs_{03} = \fv_{00} = \fv_{01} = \fv_{02} =
  0~, \nn\\
  \fs_{22} = \frac{\fs_{12}}{|\vec u|} 
  = \zeta + 2{\textstyle\frac{d-1}{d}}\eta, \qquad
  \fs_{23} = \frac{\fs_{13}}{|\vec u|} 
  = \zeta - {\textstyle\frac{2}{d}}\eta~, \qquad
  \fs_{33} = \zeta + {\textstyle\frac{2}{d(d-1)}}\eta~, \nn\\
  \fs_{11} = T\kappa + u^2 \lb \zeta + 2{\textstyle\frac{d-1}{d}}\eta \rb~, \qquad
  \fv_{11} = T\kappa + u^2\eta~, \qquad
  \fv_{22} = \frac{\fv_{12}}{|\vec u|} = \ft = \eta~.
  \label{eq:Galilean-constraints}
\end{gather}
We see that the hydrostatic and non-hydrostatic non-dissipative sectors are
entirely absent. The coefficients appearing in the charge/mass current in the
dissipative sector are also absent due to the Milne boost Ward identity
\eqref{eq:Milne-Ward}. The remaining 10 non-zero dissipative coefficients are
determined in terms of $\eta$, $\zeta$, and $\kappa$.

\subsection{Relativistic fluids}
\label{sec:rel}

Our generic discussion of boost-agnostic fluids is also capable of handling
relativistic hydrodynamics. However, the discussion is considerably more
involved than the Galilean case owing to the inherent non-linearity of
relativistic hydrodynamics. Lorentz boost symmetry can be imposed by requiring
the theory to be invariant under the transformation of background
sources\footnote{These can be derived using the reverse logic and noting that
  the relativistic metric $g_{\mu\nu}$ and inverse metric $g^{\mu\nu}$ are
  invariant and related to the Aristotelian sources as in
  \cref{eq:rel-sources}. Milne boosts \eqref{eq:MilneBoost} follow from the
  Lorentz boosts \eqref{eq:LorentzBoost} by identifying
  $A_\mu^\rel = m c^2 n_\mu + A^\gal_\mu$ and taking $c\to\infty$.}
\begin{gather}
  n_\mu \to n_\mu - \frac{1}{c^2}\lb \psi_{\mu} - \half n_\mu \psi^2\rb, \qquad
  h_{\mu\nu} \to h_{\mu\nu} - 2 n_{(\mu}\psi_{\nu)} + n_\mu n_\nu \psi^2, \qquad
  A_\mu \to A_\mu, \nn\\
  v^\mu \to v^\mu + \psi^\mu, \qquad
  h^{\mu\nu} \to
  h^{\mu\nu} + \frac{1}{c^2} \lb 2 v^{(\mu}\psi^{\nu)} + \psi^\mu \psi^\nu \rb.
  \label{eq:LorentzBoost}
\end{gather}
When implemented on the action variation \eqref{eq:basic-delS} at the linear
level, this imposes the center-of-energy conservation by setting
\begin{equation}
  \pi_\mu = \frac{1}{c^2} h_{\mu\nu}\epsilon^\nu~.
  \label{eq:Lorentz-Ward}
\end{equation}
The relativistic metric sources can be defined as
\begin{equation}
  g_{\mu\nu} = - c^2 n_\mu n_\nu + h_{\mu\nu}~, \qquad
  g^{\mu\nu} = - v^\mu v^\nu/c^2 + h^{\mu\nu}~,
  \label{eq:rel-sources}
\end{equation}
which are invariant under the above transformations. In particular
$\sqrt{-g} = c\sqrt{\gamma}$. In relativistic theories, one typically works with
an energy-momentum tensor $T^{\mu}_{~\nu}$ and charge current $J^\mu$. These are
related to our Aristotelian quantities as
\begin{equation}
  \epsilon^\mu = - T^{\mu}_{~\nu} v^\nu~, \qquad
  \pi_\nu = n_\mu T^{\mu}_{~\rho} h^\rho_{~\nu}~, \qquad
  \tau^{\mu\nu} = h^\mu_{~\rho} T^{\rho}_{~\sigma} h^{\sigma\nu}~, \qquad
  j^\mu = J^\mu~.
  \label{eq:current-decomposition}
\end{equation}
It can be explicitly checked that the relativistic conservation equations
$\nabla^\rel_\mu T^{\mu}_{~\nu} = F_{\nu\rho} J^\rho$,
$\nabla^\rel_\mu J^{\mu} = 0$, where $\nabla_\mu^{\rel}$ is the covariant
derivative associated with $g_{\mu\nu}$, reduces to the respective
boost-agnostic versions stated in \cref{eq:NC.Conservation}. Similar to the
Galilean discussion, requiring $\beta^\mu$, $\Lambda_\beta$ to be invariant, we
can derive the transformation of the hydrodynamic fields as
\begin{equation}
  T \to \frac{T}{1
  - \frac{1}{c^2}\lb \vec u^\mu \psi_{\mu} - \half \psi^2\rb}~ , \qquad
  \mu \to \frac{\mu}{1
    - \frac{1}{c^2}\lb \vec u^\mu \psi_{\mu} - \half \psi^2\rb}~, \qquad
  u^\mu \to \frac{u^\mu}{1
    - \frac{1}{c^2}\lb \vec u^\mu \psi_{\mu} - \half \psi^2\rb}~.
  \label{eq:rel-hydro-variables}
\end{equation}
We can define the relativistic versions of hydrodynamic fields according to
\begin{equation}
\label{eq:urel}
  u^\mu_\rel = \gamma_u u^\mu~, \qquad
  T_\rel = \gamma_u T~, \qquad
  \mu_\rel = \gamma_u \mu~,
\end{equation}
where $\gamma_u = 1/\sqrt{1-\vec u^2/c^2}$ is the Lorentz factor. Note that
$g_{\mu\nu} u^\mu_\rel u^\nu_\rel = - c^2$. The equation of state of a
relativistic fluid is expressed as $p(T,\mu,\vec u^2) = p(T_\rel,\mu_\rel)$. We
find the thermodynamic relations
\begin{equation}
\label{eq:thermorel}
  \df p = s_\rel \df T_\rel + n_\rel \df\mu_\rel~, \qquad
  \epsilon_\rel
  = T_\rel s_\rel + \mu_\rel n_\rel - p~, \qquad
  \rho = \frac{\epsilon + p}{c^2}
  = \frac{\gamma_u^2}{c^2}(\epsilon_\rel+p)~,
\end{equation}
where relativistic proper densities are defined as $n_\rel = n/\gamma_u$,
$s_\rel = s/\gamma_u$, and $\epsilon_\rel = \epsilon - \rho\, \vec u^2$. The
``relativistic equation of state'' is given as $\epsilon + p = \rho\, c^2$.

The constitutive relations for a relativistic fluid in the Landau frame 
$(T^{\mu}{}_\nu)_{\text{diss}} u^\nu_\rel = J^\mu_{\text{diss}} u^\nu_\rel
g_{\mu\nu} = 0$, are given as (see e.g. \cite{Bhattacharyya:2012nq})
\begin{align}
  T^{\mu}_{~\rho} g^{\rho\nu}
  &= \frac{1}{c^2}(\epsilon_\rel+p_\rel) u^\mu_\rel u^\nu_\rel
    + p\, g^{\mu\nu}
    - T_\rel \lb
    2\eta \Delta^{\rho(\mu}\Delta^{\nu)\sigma}
    + (\zeta-{\textstyle\frac{2}{d}}\eta) \Delta^{\mu\nu} \Delta^{\rho\sigma} \rb
    \half \delta_\scB g_{\rho\sigma}~,
    \nn\\
  J^\mu
  &= n_\rel u^\mu_\rel
    - T_\rel \sigma\, \Delta^{\mu\nu} \delta_\scB A_\nu~.
\end{align}
Expressing these according to the definitions \eqref{eq:urel},
\eqref{eq:thermorel}, and \eqref{eq:current-decomposition}, we find
\begin{align}
  j^\mu
  &= n u^\mu
    - T\,\gamma_u \sigma\, \lb \bar\Delta^{\mu\nu}
    + \gamma_u^2\frac{\vec u^2}{c^2} v^\mu v^\nu
    + 2\frac{\gamma_u^2}{c^2} v^{(\mu} \vec u^{\nu)} \rb
    \delta_\scB A_\nu~, \nn\\
  \epsilon^\mu
  &= \lb \epsilon_\rel + \rho\,\vec u^2\rb u^\mu
    + p\, \vec u^\mu \nn\\
  &\qquad
    - T \lb \frac{1}{c^2}v^\mu \vec u_\alpha + h^\mu_\alpha \rb
    \vec u_\nu \eta^{\alpha\nu\rho\sigma}
    \half \lB
    \delta_\scB h_{\rho\sigma}
    - 2\vec u_{(\sigma} \lb \delta_{\scB} n_{\rho)}
    - \frac{1}{c^2} v^\lambda \delta_\scB h_{\rho)\lambda} \rb
    - \frac{2}{c^2} \vec u_{\rho} \vec u_\sigma
    v^\lambda \delta_{\scB} n_{\lambda} \rB~, \nn\\
  \pi^\mu
  &= \rho\, \vec u^\mu
    - \frac{T}{c^2} \vec u_\nu \eta^{\mu\nu\rho\sigma}
    \half \lB 
    \delta_\scB h_{\rho\sigma}
    - 2\vec u_{(\sigma} \lb \delta_{\scB} n_{\rho)}
    - \frac{1}{c^2} v^\lambda \delta_\scB h_{\rho)\lambda} \rb
    - \frac{2}{c^2} \vec u_{\rho} \vec u_\sigma
    v^\lambda \delta_{\scB} n_{\lambda} \rB~, \nn\\
  \tau^{\mu\nu}
  &= \rho\, \vec u^\mu \vec u^\nu + p\, h^{\mu\nu}
    - T \eta^{\mu\nu\rho\sigma} \half \lB
    \delta_\scB h_{\rho\sigma}
    - 2\vec u_{(\sigma} \lb \delta_{\scB} n_{\rho)}
    - \frac{1}{c^2} v^\lambda \delta_\scB h_{\rho)\lambda} \rb
    - \frac{2}{c^2} \vec u_{\rho} \vec u_\sigma
    v^\lambda \delta_{\scB} n_{\lambda} \rB~.
    \label{eq:constrel}
\end{align}
Here, we have defined
\begin{gather}
  \eta^{\mu\nu\rho\sigma}
  = 2\gamma_u \eta \bar\Delta^{\rho(\mu} \bar\Delta^{\nu)\sigma}
  + \gamma_u(\zeta-{\textstyle\frac{2}{d}}\eta) \bar\Delta^{\mu\nu}
  \bar\Delta^{\rho\sigma}, \qquad
  \bar\Delta^{\mu\nu} = h^{\mu\nu} + \frac{\gamma_u^2}{c^2} \vec u^\mu \vec
  u^\nu~.
\end{gather}
In \cref{sec:1der}, we obtained the generic constitutive relations in the
boost-agnostic representation. However, these results were presented in the
thermodynamic density frame and not the Landau frame. To make contact between
\eqref{eq:constrel} and the explicit transport coefficients discussed in
\cref{sec:1der}, we need to perform a frame transformation to the density
frame. Before doing this explicitly, one can already infer that
\begin{subequations}
  \begin{gather}
    F_0 = F_1 = F_2 = 0~, \nn\\
    \bar\fs_{01} = \bar\fs_{02} = \bar\fs_{03} = \bar\fs_{12} = \bar\fs_{13} =
    \bar\fs_{23} = \bar\fv_{01} = \bar\fv_{02} = \bar\fv_{12} = 0~, \nn\\
    \fs_{01} = \fs_{11} = \fs_{12} = \fs_{13}
    = \fv_{01} = \fv_{11} = \fv_{12} = 0~.
  \end{gather}
\end{subequations}
These follow from the fact that relativistic fluids, like their Galilean
counterparts, do not have any transport coefficients in the hydrostatic and
non-hydrostatic non-dissipative sectors at one-derivative order. Also, certain
coefficients in the dissipative sector are zero due to the Lorentz boost Ward
identity \eqref{eq:Lorentz-Ward}. The remaining 10 coefficients are
non-trivially related to $\eta$, $\zeta$, $\sigma$ according to
\begin{align}
  \fs_{00}
  &= \frac{1}{\gamma_u} \alpha_3^2(1-\alpha_1)^2
    \lb \zeta + {\textstyle2\frac{d-1}{d}}\eta \rb
    + \frac{1}{\gamma_u} (1+\alpha_2)^2
    \sigma~, \nn\\
  \fs_{02}
  &= 
    -\frac{(1-\alpha_1)^2 \alpha_3}{\gamma_u^2}
    \lb \zeta + {\textstyle2\frac{d-1}{d}}\eta \rb
    - \frac{(1+\alpha_2)\alpha_2}{\gamma_u^2\alpha_3}
    \sigma~, \nn\\
  \fs_{03}
  &=  - (1-\alpha_1)\alpha_3
    \lb \zeta - {\textstyle\frac{2}{d}}\eta \rb
    + (1-\alpha_1) \alpha_1\alpha_3
    \lb \zeta + {\textstyle2\frac{d-1}{d}}\eta \rb
    - \frac{(1+\alpha_2)\alpha_2}{\alpha_3}\sigma~,
    \nn\\
  \fs_{22}
  &= \frac{(1-\alpha_1)^2}{\gamma_u^3}
    \lb \zeta + {\textstyle2\frac{d-1}{d}}\eta \rb
    + \frac{\alpha_2^2}{\gamma_u^3\alpha_3^2} \sigma~, \nn\\
  \fs_{23}
  &= \frac{(1-\alpha_1)}{\gamma_u}
    \lb \zeta - {\textstyle\frac{2}{d}}\eta \rb
    - \frac{(1-\alpha_1) \alpha_1}{\gamma_u}
    \lb \zeta + {\textstyle2\frac{d-1}{d}}\eta \rb
    + \frac{\alpha_2^2}{\gamma_u\alpha_3^2}\sigma~, \nn\\
  \fs_{33}
  &= \gamma_u\lb \zeta + {\textstyle\frac{2}{d(d-1)}}\eta  \rb
    - 2 \gamma_u \alpha_1 \lb \zeta - {\textstyle\frac{2}{d}}\eta \rb
    + \gamma_u\alpha_1^2 \lb \zeta + {\textstyle2\frac{d-1}{d}}\eta \rb
    + \gamma_u\frac{\alpha_2^2}{\alpha_3^2} \sigma~, \nn\\
  \fv_{00}
  &=
    \gamma_u \sigma
    + \gamma_u \alpha_3^2 \eta~ ,
    \qquad
    \fv_{02} = - \alpha_3\eta~, \qquad
    \fv_{22} = \frac{\eta}{\gamma_u}~, \qquad
    \ft = \gamma_u\eta~,
    \label{eq:Rel-constraints}
\end{align}
where we have defined the thermodynamic coefficients 
\begin{align}
  \alpha_1
  &= \gamma_u^2\frac{\vec u^2}{c^2}
  \lb \frac{\dow p}{\dow\epsilon}
    + \frac{1}{|\vec u|}\frac{\dow p}{\dow|\pi|} \rb
    = 
  \frac{- \frac{\vec u^2}{c^2}\lb \frac{\dow p}{\dow\epsilon_\rel} 
    + \frac{\gamma_u n}{\epsilon+p}
    \frac{\dow p}{\dow n_\rel}\rb }{1 - \frac{\vec u^2}{c^2}
    \lb \frac{\dow p}{\dow\epsilon_\rel} 
    + \frac{\gamma_u n}{\epsilon+p}
    \frac{\dow p}{\dow n_\rel} \rb }
    ~, \nn\\
  \alpha_2
  &= \gamma_u^2\frac{\vec u^2}{c^2}
    \frac{n}{\epsilon+p} \frac{\dow p}{\dow n}
    = \frac{\frac{\vec u^2}{c^2}
    \frac{\gamma_u n}{\epsilon+p}\frac{\dow p}{\dow n_\rel}}
    {1 - \frac{\vec u^2}{c^2}
    \lb \frac{\dow p}{\dow\epsilon_\rel} 
    + \frac{\gamma_u n}{\epsilon+p}
    \frac{\dow p}{\dow n_\rel} \rb }~, \qquad
  \alpha_3 = \gamma_u \frac{n|\vec u|}{\epsilon+p}~.
\end{align}
Further details about this derivation can be found in \cref{app:Landau}.

\subsection{Lifshitz fluids}

The final example we want to consider is that of a Lifshitz fluid, which is invariant under
anisotropic scaling of spacetime coordinates $t\to \lambda^z t$,
$x^i\to \lambda x^i$. Covariantly, we can define the Lifshitz symmetry as its
action on the background sources
\begin{gather}
  n_\mu \to \lambda^{-z}\, n_\mu~, \qquad
  h_{\mu\nu} \to \lambda^{-2}\, h_{\mu\nu}~, \qquad
  A_\mu \to A_\mu,  \nn\\
  v^\mu \to \lambda^{z} v^\mu~, \qquad
  h^{\mu\nu} \to \lambda^{2} h^{\mu\nu}~.
\end{gather}
Plugging these into the variational expression \cref{eq:basic-delS}, we can
derive the Lifshitz Ward identity
\begin{equation}
  \tau^{\mu\nu} h_{\mu\nu} = z\, \epsilon^\mu n_\mu~,
  \label{eq:Lifshitz-Ward}
\end{equation}
which is the covariant version of the respective identities introduced
in~\cite{deBoer:2020xlc,Novak:2019wqg}. Demanding the partition function or
effective action to be invariant under this scaling leads to the scaling
behaviour of the conserved currents
\begin{equation}
  \epsilon^\mu \to \lambda^{d+2z}\epsilon^\mu~, \qquad
  \pi_\mu \to \lambda^{d}\pi_\mu~, \qquad
  \tau^{\mu\nu} \to \lambda^{d+z+2}\tau^{\mu\nu}~, \qquad
  j^\mu \to \lambda^{d+z}j^\mu~.
\end{equation}
Requiring that $\beta^\mu$, $\Lambda_\beta$ remain invariant, results in the
scaling properties of the hydrodynamic fields
\begin{equation}
  u^\mu \to \lambda^{z} u^\mu~, \qquad
  T \to \lambda^{z} T~, \qquad
  \mu \to \lambda^{z} \mu~.
\end{equation}
Note that $\vec u^2 \to \lambda^{2z-2} \vec u^2$ and
$\hat u^\mu \to \lambda\hat u^\mu$. This implies that the scalar ratios $\mu/T$ and
$\vec u^2/T^{2-2/z}$ are scale invariant. Implementing this for ideal fluids,
one can infer that the equation of state of a Lifshitz fluid takes the form
\begin{equation}
  p(T,\mu,\vec u^2) = T^{d/z+1}\, \hat p(\mu/T,\vec u^2/T^{2-2z})~.
\end{equation}
This leads to the thermodynamic identity
\begin{equation}
  \rho\,\vec u^2 + p\,d = z\, \epsilon~,
\end{equation}
which can also be derived directly using \cref{eq:Lifshitz-Ward}.

As for the one-derivative order transport coefficients, the scaling behaviour
goes as
\begin{equation}
  \upsilon(T,\mu,\vec u^2) =  T^{w_\upsilon/z}\, \hat \fc(\mu/T,\vec u^2/T^{2-2z})~,
\end{equation}
where the weight factor $w_\lambda$ for the various coefficients is given by
\begin{equation}
  w_\upsilon =
  \begin{cases}
    d+z & \text{for} \quad \upsilon = p~, \\
    d-z & \text{for} \quad \upsilon = F_0, F_1~, \\
    d-2z+2 & \text{for} \quad \upsilon = F_2~, \\
    d-2 & \text{for} \quad \upsilon = \fv_{00}, \fs_{00}~, \\
    d+z-2 & \text{for} \quad \upsilon = \fv_{01}, \fs_{01}~,
    \bar\fv_{01}, \bar\fs_{01}, \\
    d+2z-2 & \text{for} \quad \upsilon = \fv_{11}, \fs_{11}~, \\
    d-1 &\text{for} \quad \upsilon = \fv_{02}, \fs_{02}, \fs_{03}~,
    \bar\fv_{02}, \bar\fs_{02}, \bar\fs_{03}~,\\
    d+z-1 &\text{for} \quad \upsilon = \fv_{12}, \fs_{12}, \fs_{13}~,
    \bar\fv_{12}, \bar\fs_{12}, \bar\fs_{13}, \\
    d   &\text{for} \quad \upsilon = \ft, \fv_{22}, \fs_{22}, \fs_{23},
    \fs_{33}, \bar\fs_{23}~.
  \end{cases}
\end{equation}
We have included the ideal order pressure for completeness. However, not all of
these coefficients are independent. Firstly, the derivative corrections must
satisfy the Lifshitz Ward identities \eqref{eq:Lifshitz-Ward}. In particular, in
the density frame, this implies that the non-hydrostatic corrections (including
both dissipative and non-hydrostatic non-dissipative) must satisfy that
$\tau^{\mu\nu}_{\text{nhs}} h_{\mu\nu} = 0$.\footnote{The condition is slightly
  more non-trivial in the Landau frame employed
  in~\cite{deBoer:2020xlc,Novak:2019wqg}:
  $\tau^{\mu\nu}_{\text{nhs}} h_{\mu\nu} = z\,\epsilon^\mu_{\text{nhs}} n_\mu$.}
Furthermore, one-derivative scalars in the hydrostatic free energy density
\eqref{eq:hydrostatic-N} can only come via the scale-covariant combinations
\begin{equation}
  v^\mu\dow_\mu \bfrac{\mu}{T}~, \qquad
  v^\mu \dow_\mu \bfrac{\vec u^{2}}{T^{2-2/z}}~.
\end{equation}
Similarly, all the non-hydrostatic data must appear in combinations
\begin{equation}
  \delta_\scB A_\mu~, \qquad
  \delta_\scB n_\mu - \frac{z}{2d} n_\mu h^{\rho\sigma}\delta_\scB h_{\rho\sigma}~, \qquad
  \delta_\scB h_{\mu\nu}
  - \frac{1}{d} h_{\mu\nu} h^{\rho\sigma} \delta_\scB h_{\rho\sigma}~.
\end{equation}
This leads to 8 constraints among the transport coefficients: 1 in the
hydrostatic sector, 3 in the non-hydrostatic non-dissipative sector, and 4 in
the dissipative sector, namely
\begin{gather}
  F_1 = - 2\frac{(z-1)\vec u^2}{z\, T} F_2 - \frac{\mu}{T} F_0~, \nn\\
  \bar\fs_{02} = - (d-1) \bar\fs_{03}~, \qquad
  \bar\fs_{12} = - (d-1) \bar\fs_{13}~, \qquad
  \bar\fs_{23} = 0~, \nn\\
  \fs_{02} = - (d-1)\fs_{03}~, \qquad
  \fs_{12} = - (d-1)\fs_{13}~, \qquad
  \fs_{22} = - (d-1)\fs_{23}  = (d-1)^2\fs_{33}~.
  \label{eq:Lifhitz-constraints}
\end{gather}
In the uncharged limit \eqref{eq:unchargedlimit}, the total number of transport
coefficients after Lifshitz scaling agrees with that of \cite{deBoer:2020xlc}.
Scale invariant Galilean fluids are compatible with $z=2$ Lifshitz symmetry. An
easy way to see this is that requiring $\mu_\gal = \mu + m/2\,\vec u^2$ to scale
homogeneously forces us to set $z=2$. The thermodynamic equation of state, in
this case, leads to $\epsilon_\gal = d/2\, p$. Comparing the Galilean
coefficients \eqref{eq:Galilean-constraints} to \cref{eq:Lifhitz-constraints} we
read out that $\zeta = 0$. Similarly, scale invariant relativistic fluids are
compatible with $z=1$ isotropic scaling symmetry, since the relativistic
hydrodynamic fields in \eqref{eq:rel-hydro-variables} should scale
homogeneously. The equation of state becomes $\epsilon_\rel = d\, p$, while at
the one-derivative order we again find $\zeta = 0$ by imposing the Lifshitz
constraints \eqref{eq:Lifhitz-constraints} on \cref{eq:Rel-constraints}.

\section{Linearised fluctuations}
\label{sec:linearised}

In \cref{sec:1der}, we determined the explicit effective action and all
transport coefficients appearing at first order in the derivative expansion for
boost-agnostic fluids. In this section we study fluctuations around equilibrium
anisotropic states and determine the mode structure. We ignore hydrostatic and
non-hydrostatic non-dissipative contributions; these can also be systematically
switched off by imposing T or PT symmetry. We also turn off the background
fields and the resultant effective action is given in
\cref{eq:flat-action}. Additionally, for simplicity, we do not turn on all
dissipative transport coefficients but instead consider small deviations away
from Galilean constitutive relations.  We begin by obtaining the linearised
equations and later study the possible modes in a 3+1 dimensional fluid living
in flat spacetime.

\subsection{Linearised equations}

We want to perturb the Lagrangian \eqref{eq:flat-action} around an equilibrium
anisotropic state with non-zero fluid velocity
$u^i$.\footnote{Ref.~\cite{deBoer:2020xlc} considered the isotropic case with
  $u^i=0$. } The equilibrium configuration is characterised by
\begin{equation}
\tau=t~~,~~\sigma^i=x^i~~,~~\varphi_r=0~~,~~X^t_a=X^i_a=\varphi_a=0~~,
\end{equation}
with constant temperature $T=T_0$, constant chemical potential $\mu=\mu_0$, and
constant non-zero spatial velocity $u^i=u^{i}_0$, as well as constant energy
density, pressure, and charge density. In order to understand fluctuations
around this state, we consider small perturbations of the particle/charge
density $\delta n$, energy density $\delta \epsilon$, momentum density
$\delta \pi_i$ and the stochastic variables $\delta X^t_a$, $\delta X^i_a$, and
$\delta\varphi_a$. The variation of the Lagrangian \eqref{eq:flat-action} under
these small perturbations and underlying assumptions becomes
\begin{align}
  \mathcal{\delta L}
  = \varphi^I_a K_I{}^J \mathcal{O}_J
  + \frac{i}{2} \varphi_a^I G_{IJ} \varphi_a^J~~,
\end{align}
where we have introduced the operators
\begin{align}
  \mathcal{O}_I
  &=
    \begin{pmatrix}
      \delta n \\ \delta\epsilon \\ \hat u^i\delta\pi_i \\ P^{j}_i\delta\pi_j
    \end{pmatrix}, \qquad
  \varphi_a^I
  =
  \begin{pmatrix}
    \delta \varphi_a \\ - \delta X_a^t \\ \hat u_i \delta X_{a}^i \\ P^i_j \delta X_{a}^j
  \end{pmatrix}, \nn\\
  M_I{}^J
  &=
  \begin{pmatrix}
    |\vec\pi|\frac{\dow\hat n}{\dow n} \hat\dow_u
    & |\vec\pi|\frac{\dow\hat n}{\dow \epsilon} \hat\dow_u
    & \lb 2\vec\pi^2 \frac{\dow\hat n}{\dow \vec\pi^2} + \hat n \rb \hat\dow_u
    & \hat n\, \hat\dow_j \\
    |\vec\pi|\frac{\dow\hat w}{\dow n}  \hat\dow_u
    & |\vec\pi|\frac{\dow\hat w}{\dow \epsilon}  \hat\dow_u
    & \lb 2\vec\pi^2 \frac{\dow\hat w}{\dow \vec\pi^2} + \hat w \rb \hat\dow_u
    & \hat w\, \hat\dow_j \\
    \frac{\dow(\rho\vec u^2 + p)}{\dow n} \hat\dow_u
    & \frac{\dow(\rho\vec u^2 + p)}{\dow \epsilon} \hat\dow_u
    & 2|\vec\pi|
    \frac{\dow(\rho\vec u^2+p)}{\dow\vec\pi^2} \hat\dow_u
    & |\vec u| \hat\dow_j \\
    \frac{\dow p}{\dow n} \hat\dow_i
    & \frac{\dow p}{\dow\epsilon} \hat\dow_i
    & 2|\vec\pi|\frac{\dow p}{\dow\vec\pi^2} \hat\dow_i
    & P_{ij} |\vec u|\hat\dow_u
  \end{pmatrix}, \nn\\
  \frac{G_{IJ}}{\kB T}
  &= -2
  \begin{pmatrix}
    \fv_{00} \hat\dow^2 + \fs_{00} \hat\dow_u^2
    & \fv_{01} \hat\dow^2 + \fs_{01} \hat\dow_u^2
    & \fv_{02} \hat\dow^2 + \fs_{02} \hat\dow_u^2
    & (\fv_{02}{+}\fs_{03}) \hat\dow_u\hat\dow_j \\
    \fv_{01} \hat\dow^2 + \fs_{01} \hat\dow_u^2
    & \fv_{11} \hat\dow^2 + \fs_{11} \hat\dow_u^2
    & \fv_{12} \hat\dow^2 + \fs_{12} \hat\dow_u^2
    & (\fv_{12}{+}\fs_{13}) \hat\dow_u\hat\dow_j \\
    \fv_{02} \hat\dow^2 + \fs_{02} \hat\dow_u^2
    & \fv_{12} \hat\dow^2 + \fs_{12} \hat\dow_u^2
    & \fv_{22} \hat\dow^2 + \fs_{22} \hat\dow_u^2
    & (\fv_{22}{+}\fs_{23}) \hat\dow_u\hat\dow_j \\
    (\fv_{02}{+}\fs_{03}) \hat\dow_u\hat\dow_i
    & (\fv_{12}{+}\fs_{13}) \hat\dow_u\hat\dow_i
    & (\fv_{22}{+}\fs_{23}) \hat\dow_u\hat\dow_i
    & \lb \fs_{33} {+} {\textstyle\frac{d-3}{d-1}}\ft \rb \hat\dow_i \hat\dow_j
    + P_{ij} (\ft\, \hat\dow^2 + \fv_{22}\dow_u^2)
  \end{pmatrix}, \nn\\
  \chi_{IJ}
  &=
    \begin{pmatrix}
    T \dow(\mu/T)/\dow n
    & T \dow(\mu/T)/\dow \epsilon
    & 2T|\vec\pi| \dow(\mu/T)/\dow \vec\pi^2
    & 0 \\
    1/T\, \dow T/\dow n
    & 1/T\, \dow T/\dow \epsilon
    & 2/T\, |\vec\pi| \dow T/\dow \vec\pi^2
    & 0 \\
    T|\vec\pi| \dow(T^{-1}\rho^{-1})/\dow n
    & T|\vec\pi| \dow(T^{-1}\rho^{-1})/\dow \epsilon
    & 
    2T\vec\pi^2 \dow(T^{-1}\rho^{-1})/\dow\vec\pi^2 + \frac{1}{\rho}
    & 0 \\
    0 & 0 & 0 & P^{ij}/\rho
  \end{pmatrix}^{-1}, \nn\\
  K_I{}^J
  &= - \delta^J_I \dow_t - M_I{}^J - \frac{1}{2T} G_{IK} (\chi^{-1})^{KJ}~~.
\end{align}
Here we have defined $\hat n = n/\rho$ and $\hat w = (\epsilon+p)/\rho$, and
introduced the differential operators $\hat\dow_u = \hat u^k \dow_k$,
$\hat\dow^i = P^{ij}\dow_j$, and $\hat\dow^2 = P^{ij}\dow_i\dow_j$. It can be
explicitly checked that the susceptibility matrix $\chi_{IJ}$ is symmetric. All
quantities appearing in $M_I{}^J$, $G_{IJ}$, and $ \chi_{IJ}$ should be
understood as being evaluated in the equilibrium configuration; we have dropped
the subscript ``$0$'' for clarity. Varying the perturbed Lagrangian with respect
to the stochastic variables $\varphi_a^I$, one obtains the linearised equations
of motion. We will now use these to find the mode structure.

\subsection{Mode structure}

Since the linearised equations are given by $K_I{}^J \mathcal{O}_J=0$, it is
possible to find the dispersion relations by looking at the zeros of
$\text{det}(K_I{}^J)$. In 3+1 dimensions, we find a pair of sound modes (with a
different velocity along or opposite the fluid flow), one number-density
diffusion mode, one shear mode along the fluid velocity, and another shear mode
transverse to the fluid velocity. Assuming plane-wave perturbations, in general,
the modes have the following structure
\begin{gather}
  \omega - u^i k_i = v^{\pm}_s(\theta)k - \frac{i}{2}\Gamma_s(\theta)k^2, \nn\\
  \omega - u^i k_i = -iD_0(\theta)k^2, \qquad
  \omega - u^i k_i = -iD_1(\theta)k^2, \qquad
  \omega - u^i k_i = -iD_2(\theta)k^2.
\end{gather}
Here $\theta$ is the angle between $u^i$ and $k^i$, $\Gamma_s(\theta)$ is the
attenuation constant, and $D_{0,1,2}(\theta)$ are diffusion
constants. Explicitly, we find that the velocity of sound $v_s^\pm(\theta)$ are
the solutions of the quadratic equation
\begin{equation}
  v_s^\pm(\theta)^2 + X |\vec u| \cos\theta\, v_s^{\pm}(\theta)
  + Y\vec u^2\cos^2\theta + Z = 0~~,
\end{equation}
for which the functions X, Y, Z are not particularly illuminating. The general solution is given by
\begin{equation}
\begin{split}
v_s^\pm(\theta)=&\pm\sqrt{v_{s,0}^2\left(1-|\vec u|^2\cos^2\theta\left(\frac{\partial \rho}{\partial \epsilon} +2\rho \frac{\partial \rho}{\partial\vec \pi^2}\right)\right)+|\vec u|^2\cos^2\theta\left(\frac{\partial p}{\partial \epsilon}+2\frac{\partial p}{\partial \vec \pi^2}+\frac{v_{s,1}}{2}\right)^2}\\
&-\frac{|\vec u|}{2}\cos\theta~ v_{s,1}~~,
\end{split}
\label{eq:soundspeed}
\end{equation}
where we have defined
\begin{equation}
v_{s,0}=\sqrt{\frac{\partial p}{\partial n}\frac{n}{\rho}+\frac{\partial p}{\partial \epsilon}\frac{w}{\rho}+2\rho |\vec u|^2\frac{\partial p}{\partial \vec \pi^2}}~~,~~v_{s,1}=\frac{\partial \rho}{\partial n}\frac{n}{\rho}+\frac{\partial \rho}{\partial \epsilon}\frac{w}{\rho}-\left(1+\frac{\partial p}{\partial \epsilon}+2\rho\frac{\partial p}{\partial \vec \pi^2}-2\rho|\vec u|^2\frac{\partial \rho}{\partial \vec \pi^2}\right)~~.
\end{equation}
The pair of sound modes in eq.~\eqref{eq:soundspeed} propagate with different
speeds due to the presence of a non-zero equilibrium fluid speed $|\vec u|$. The
two speeds are equal if we impose the Galilean or relativistic equations of state
discussed in \cref{sec:gal-rel}. Hence, the distinction between the parallel and
anti-parallel sound speeds, $v_s^+\neq v_s^-$ is an imprint of broken boost
symmetry. In the isotropic case $|\vec u|\to0$, the two sound speeds are again
equal to each other and reduce to the results of \cite{deBoer:2017abi}.

In turn, the diffusion constants $D_{0,1}(\theta)$ are solutions of the quadratic equation
\begin{equation}
  D_{0,1}(\theta)^2 + A(\theta) D_{0,1}(\theta) + B(\theta) = 0~~, 
\end{equation}
for which the functions $A(\theta)$ and $B(\theta)$ are some cumbersome
functions of the thermodynamic variables, while the transverse shear diffusion
constant $D_2(\theta)$ is simply given by
\begin{equation}
  D_2(\theta) = \frac{\ft}{\rho} + \frac{\fv_{22}-\ft}{\rho} \cos^2\theta~~.
  \label{eq:D2}
\end{equation}
In order to provide an analytically tractable example of $D_{0,1}(\theta)$ and
the attenuation constant $\Gamma_s(\theta)$, we consider slight departures away
from the transport properties of a Galilean fluid, characterised only by three
transport coefficients at first order, namely, $\kappa, \eta, \zeta$ (see
\cref{sec:gal-rel}). This is still a non-trivial example because we are taking
into account the modified thermodynamics due to the absence of a boost
symmetry. In this special case, we can split the attenuation and diffusion
constants into Galilean contributions and corrections due to the absence of
boosts in a small velocity expansion such that
\begin{equation}
  \Gamma_s(\theta)
  = \Gamma_\gal(\theta)+|\vec u|^2 \Gamma_{\text{u}}(\theta)~~,~~
  D_{0,1}(\theta)
  = D_{0,1\gal}(\theta)+|\vec u|^2 D_{0,1\text{u}}(\theta)~~,
\end{equation}
where, focusing on the case of $\theta=\pi/2$, the Galilean contributions are given by
\begin{align}
  &\Gamma_{\gal}\left(\frac{\pi}{2}\right)=\frac{\kappa}{\rho
    v_{s,\gal}^2}\frac{\partial p}{\partial
    \epsilon}\left(\frac{\partial T}{\partial n}n+\frac{\partial
    T}{\partial\epsilon}w\right)+\frac{3\zeta+4\eta}{3\rho}~~,~~
    v_{s,\gal}^2
    = \left(\frac{\partial p}{\partial n} \frac{n}{\rho} 
    +\frac{\partial p}{\partial\epsilon}\frac{w}{\rho} \right)~~, \nn\\
  &D_{0,\gal}\left(\frac{\pi}{2}\right)=\frac{n
    \kappa}{v_{s,\gal}^2\rho}\left(\frac{\partial p}{\partial
    n}\frac{\partial T}{\partial \epsilon}-\frac{\partial p}{\partial
    \epsilon}\frac{\partial T}{\partial
    n}\right)~~,~~
    D_{1,\gal}\left(\frac{\pi}{2}\right)=\frac{\eta}{\rho}~~.
\end{align}
If one imposes Galilean thermodynamics as in \cref{sec:gal-rel}, one obtains the
diffusion and attenuation constants presented in \cite{Jain:2020vgc}. On the
other hand, the corrections due to the absence of a boost symmetry are given by
\begin{align}
  \Gamma_{\text{u}}\left(\frac{\pi}{2}\right)
  &= -\frac{1}{\rho^3v_{s,\gal}^4}\Bigg[\eta\frac{\partial p}{\partial
    \epsilon}\rho v_{s,\gal}^2\left(\frac{\partial \rho}{\partial
    n}n+\frac{\partial \rho}{\partial \epsilon}w\right) \nn\\
  &+\rho^2 v_{s,\gal}^2\left(2\eta\frac{\partial p}{\partial \vec \pi
    ^2}\left(\frac{\partial \rho}{\partial n}n+\frac{\partial
    \rho}{\partial \epsilon}w\right)+\frac{\partial
    p}{\partial \epsilon}\left(T\kappa
    \frac{\partial\rho}{\partial \epsilon}-\eta\right)\right)
    \nn\\
  &+\rho^3 v_{s,\gal}^2\left(\kappa\frac{\partial p}{\partial \epsilon}\frac{\partial T}{\partial \epsilon}-2\eta\frac{\partial p}{\partial \vec \pi^2}\right)+2\rho^3\kappa \frac{\partial p}{\partial \epsilon}\frac{\partial p}{\partial \vec \pi^2}\left(\frac{\partial T}{\partial n}n+\frac{\partial T}{\partial \epsilon}w\right)\Bigg]~~,
\end{align}
\begin{align}
D_{0,u}\left(\frac{\pi}{2}\right)=&-n\kappa\Bigg[\eta\rho^2v_{s,\gal}^2\left(\frac{\partial p}{\partial n}\frac{\partial \rho}{\partial \epsilon}-\frac{\partial p}{\partial \epsilon}\frac{\partial \rho}{\partial n}\right)\left(Tv_{s,\gal}^2\frac{\partial \rho}{\partial \epsilon}+\frac{\partial \rho}{\partial \epsilon}\left(\frac{\partial T}{\partial n}n+\frac{\partial T}{\partial \epsilon}w\right)\right) \nn\\
&+\rho^3v_{s,\gal}^2\Big(2\eta\frac{\partial p}{\partial \vec \pi^2}\left(\frac{\partial T}{\partial n}n+\frac{\partial T}{\partial \epsilon}w\right)\left(\frac{\partial p}{\partial n}\frac{\partial \rho}{\partial \epsilon}-\frac{\partial p}{\partial \epsilon}\frac{\partial \rho}{\partial n}\right)\nn\\
&+\frac{\partial p}{\partial \epsilon}\left(T\kappa \frac{\partial \rho}{\partial \epsilon}-\eta\right)\left(\frac{\partial p}{\partial \epsilon}\frac{\partial T}{\partial n}+\frac{\partial p}{\partial n}\frac{\partial T}{\partial \epsilon}\right)\Big)\nn\\
&+\kappa\rho^4\frac{\partial p}{\partial\epsilon}\left(\frac{\partial p}{\partial n}\frac{\partial T}{\partial \epsilon}-\frac{\partial p}{\partial \epsilon}\frac{\partial T}{\partial n}\right)\left(\frac{\partial T}{\partial \epsilon}v_{s,\gal}^2+2\frac{\partial p}{\partial \vec\pi^2}\left(\frac{\partial T}{\partial n}n+\frac{\partial T}{\partial \epsilon}w\right)\right)\Bigg]\nn\\
&\times~v_{s,\gal}^4\rho^3\left(\frac{\partial p}{\partial n}n\left(\eta-\kappa\rho\frac{\partial T}{\partial \epsilon}\right)+\frac{\partial p}{\partial \epsilon}\left(\eta w+\kappa\rho\frac{\partial T}{\partial n}n\right)\right)^{-1}~~,
\end{align}
\begin{align}
D_{1,u}\left(\frac{\pi}{2}\right)=&\eta\Bigg[2\frac{\partial p}{\partial \vec \pi^2} T\rho\left(\eta\frac{\partial \rho}{\partial \epsilon}w-\eta\rho+n\kappa \rho\frac{\partial T}{\partial n}\frac{\partial \rho}{\partial \epsilon}+\kappa\rho^2\frac{\partial T}{\partial \epsilon}+\frac{\partial \rho}{\partial n}n\left(\eta-\frac{\partial T}{\partial \epsilon}\kappa\rho\right)\right)\nn\\
&+\frac{\partial p}{\partial n}n\Big(\frac{\partial \rho}{\partial \epsilon}\left(\kappa T^2\frac{\partial \rho}{\partial \epsilon}-2T\left(\eta-\frac{\partial T}{\partial \epsilon}\kappa\rho\right)\right) \nn\\
&-\rho\left(\frac{\partial T}{\partial \epsilon}+2T\frac{\partial \rho}{\partial \vec \pi^2}+2\rho\frac{\partial T}{\partial \vec\pi^2}\right)\left(\eta-\frac{\partial T}{\partial \epsilon}\kappa\rho\right)\Big)\nn\\
&-\frac{\partial p}{\partial \epsilon}\Big(\frac{\partial \rho}{\partial n}n T\left(T\kappa \frac{\partial \rho}{\partial \epsilon}+\kappa\rho\frac{\partial T}{\partial \epsilon}-\eta\right)+T\frac{\partial \rho}{\partial \epsilon}\left(w\eta +\kappa \rho\left(\frac{\partial T}{\partial n}n-T\right)\right)\nn\\
&+\rho\left(\left(\frac{\partial T}{\partial \epsilon}+2\rho\frac{\partial T}{\partial \vec\pi^2}\right)\left(w\eta+\kappa\rho\frac{\partial T}{\partial n}n\right)+T\left(\eta+2\eta w\frac{\partial \rho}{\partial \vec\pi^2}-\kappa\rho \frac{\partial T}{\partial \epsilon}+2n\kappa\rho\frac{\partial \rho}{\partial \vec\pi^2}\frac{\partial T}{\partial n}\right)\right)\Bigg]\nn\\
&\times~\rho T\left(n\frac{\partial p}{\partial n}\left(\eta-\kappa\rho\frac{\partial T}{\partial \epsilon}\right)+\frac{\partial p}{\partial \epsilon}\left(w\eta +n\kappa\rho\frac{\partial T}{\partial n}\right)\right)^{-1}~~.
\end{align}
As one can explicitly observe, the longitudinal shear diffusion constant
$D_{1}(\theta)$ receives corrections due to a non-vanishing fluid velocity,
while the transverse shear diffusion constant $D_{2}(\theta)$ in \cref{eq:D2}
does not. This is again an imprint of the broken boost symmetry.

We would like to point out that the linear mode analysis presented above has
been done at finite equilibrium fluid velocity $u^i = u^i_0$, and yet we do not
encounter any additional gapped unphysical poles in the upper-half complex
$\omega$ plane. This is in contrast to the Landau and Eckart frames typically
employed in relativistic hydrodynamics, that are unstable in a finite
fluid-velocity state; see~\cite{Kovtun:2019hdm,Poovuttikul:2019ckt}. This
affirms that the density frame introduced in this paper is a stable hydrodynamic
frame, applicable to hydrodynamic theories with arbitrary boost symmetry
structure -- Galilean, Lorentzian, or absence thereof.

\section{Outlook}
\label{sec:discussion}

The main goal of this paper was to formulate a \SK effective field theory for
hydrodynamics without boosts. The formal construction, presented in
\cref{sec:EFT}, was based on the recently developed EFT for Galilean
hydrodynamics~\cite{Jain:2020vgc}. In the process of building this EFT, we
 provided a spacetime covariant framework for hydrodynamics without boosts
and a rigorous offshell analysis of the independent transport coefficients
together with the constraints that need to be satisfied for the second law of
thermodynamics to hold (see \cref{eq:const2nd}). An accurate counting of
transport coefficients reveals that there are 4 hydrostatic coefficients
(including the ideal order pressure), 9 non-hydrostatic non-dissipative
coefficients, and 17 independent dissipative transport coefficients. Thus,
boost-agnostic hydrodynamics is characterised by a total of 30 independent
transport coefficients up to first order in a gradient expansion.

Part of this work, specifically \cref{sec:1der}, can also be seen as an
extension of the covariant formulation of \cite{deBoer:2017ing, deBoer:2020xlc}
to include an additional particle/charge current. In the uncharged limit, our
results agree with those of \cite{deBoer:2020xlc}. As such, we provided a
general covariant framework for treating simultaneously Lorentzian, Galilean,
and Lifshitz fluids; the respective boost and scaling limits were performed in
\cref{sec:gal-rel}. In addition, we studied the general spectrum of linear modes
around an anisotropic finite-velocity state as an application of this theory in
\cref{sec:linearised}. This analysis revealed specific imprints of the absence
of boost invariance that were previously unknown. In particular, in 3+1
spacetime dimensions, we found a pair of sounds with different velocities
depending on whether the sound wave propagates along the equilibrium fluid
velocity $u^i_0$ or opposite to it. Furthermore, the shear modes, which usually
have multiplicity 2 in Galilean or relativistic fluids, now split into a shear
mode along the fluid velocity and another transverse to it. Such imprints are
clear smoking guns for potential experimental realisations of hydrodynamic
systems without boost invariance, and were not visible when fluctuating around
isotropic equilibrium states as in \cite{deBoer:2017abi}. In relation to this,
we note that we have provided our results in a new hydrodynamic frame that is
linearly stable, irrespective of the boost symmetry in place, making the system
of equations of motion and constitutive relations ideal for performing numerical
simulations without running into unphysical artefacts.

Besides a unified framework that can treat different physical systems on the
same footing, one of main goals of this work was to set the stage for future
non-trivial extensions. As revealed in the introduction, systems of interest
with broken boost symmetry exhibit intertwined patterns of symmetry breaking
that can include spontaneous/explicit translation symmetry breaking,
superfluidity, or liquid crystal phases. One of our main motivators are the
hydrodynamic theories of flocking, such as the Toner-Tu model
\cite{PhysRevE.58.4828}. In such settings, not only is the boost symmetry
explicitly broken, but there are also additional non-conserved driving forces
responsible for the activity that breaks the spacetime translation symmetry. As
far as we are aware, though widely used, such models lack a rigorous derivation
in terms of an effective field theory or even a complete classical understanding
and characterisation of the allowed transport. Another system of interest is
that of quantum matter exhibiting charge density wave
phases~\cite{Delacretaz:2017zxd}, in which, besides the absence of a boost
symmetry, spatial translations are also typically broken explicitly and
spontaneously. \SK EFT provides a route for understanding these systems, as it
offers a controlled framework for symmetry breaking, moving away from classical
hydrodynamics, and exploring its consequences (see e.g.~\cite{Landry:2019iel}
for an ideal order non-dissipative discussion). It will also be interesting to
explore the purely non-equilibrium non-classical stochastic effects arising from
the EFT analysis, such as those reported in~\cite{Jain:2020fsm}, in the context
of broken boost scenarios. We leave these explorations for future work.

A natural extension of this work is to consider the case of spontaneous breaking
of Lorentz boost symmetry, and more generally, of Poincar\'{e} symmetry. This is
directly relevant for condensed matter systems where a classification of phases
of matter has been partially provided in \cite{Nicolis:2015sra}. Such extension
would result in a finite temperature version of the same and, as the world is
Lorentz invariant, would be highly relevant to pursue for real-world
applications.

In the absence of any controlled experiment that could probe all 30 transport
coefficients, it would be interesting to understand better the phenomenology and
mode structure of fluids without a boost symmetry. To this aim, it would be
relevant to understand whether holographic models, in the spirit of
\cite{Bhattacharyya:2008jc}, exhibiting explicit Lorentz boost symmetry breaking
could be constructed. An analysis of the black hole spectrum of quasinormal
modes within such models would provide reasonable theoretical input for the
equation of state and transport coefficients, allowing to better probe the
physics of these systems.

\acknowledgements

We would like to thank Jan de Boer and Niels A. Obers for useful discussions. JA
is partly supported by the Netherlands Organization for Scientific Research
(NWO) and by the Dutch Institute for Emergent Phenomena (DIEP) cluster at the
University of Amsterdam. AJ is supported in part by the NSERC Discovery Grant
program of Canada.

\appendix

\section{Frame transformations and comparison with previous works}
\label{app:frame}

In this appendix we discuss frame transformations in boost-agnostic
hydrodynamics in detail and discuss the translation of our results to those
of~\cite{Novak:2019wqg,deBoer:2020xlc}. We give a general procedure to convert
constitutive relations in arbitrary frame to our density frame. The analysis can
equivalently be adapted to arrive at other hydrodynamic frames.

\subsection{Generalities of hydrodynamic frame transformation}

We know that the hydrostatic part of the constitutive relations, in particular
the leading derivative order ideal fluid, can be generated from a hydrostatic
generating functional. To wit, we can start from a free energy density
$\mathcal{N}$ and use the variational formulae
\begin{equation}
  \frac{1}{\sqrt{\gamma}} \delta \lb\sqrt{\gamma}\, p\rb
  = j^\mu_0 \delta A_\mu
  - \epsilon^\mu_0 \delta n_\mu
  + \half \lb 2 v^\mu \pi_0^\nu + \tau^{\mu\nu}_0 \rb \delta h_{\mu\nu}
  + \fh_\mu \delta \beta^\mu
  + \fn \lb \delta \Lambda_\beta+ A_\mu \delta\beta^\mu \rb,
  \label{eq:delN-general}
\end{equation}
where ``0'' denotes the ideal part of the constitutive relations. Here
$\fh_\mu$ and $\fn$ denote variations with respect to hydrodynamic fields, that
do not contribute if we replace $\delta \to \delta_\scB$, leading to the
adiabaticity equation. This form is particularly useful because it allows us to
directly read out the equations of motion associated with the hydrostatic part
of the constitutive relations. Employing gauge and diffeomorphism invariance of
$\mathcal{N}$, the equations of motion \eqref{eq:NC.Conservation} can be
re-expressed as
\begin{equation}
  \frac{1}{\sqrt{\gamma}} \delta_\scB \lb \sqrt{\gamma}\, \fn \rb
  = \mathcal{O}(\dow^2)~, \qquad
  \frac{1}{\sqrt{\gamma}} \delta_\scB \lb \sqrt{\gamma}\, \fh_\mu\rb
  + \fn\,\delta_\scB A_\mu = \mathcal{O}(\dow^2)~.
\end{equation}
This form of the equations of motion was already derived for ideal fluids in
\cref{eq:delB-EOM}. Note that this is only the hydrostatic part of the equations
of motion and will admit derivative corrections. But it will be useful for us in
our discussion of frame transformations. Using \cref{eq:delN-general}, the
equations of motion can also be expressed as
\begin{align}
\label{eq:EOMappendix}
  \frac{1}{\sqrt{\gamma}} \delta_\scB \lb \sqrt{\gamma}\, \fn \rb
  &= \frac{\delta j^\rho_0}{\delta\Lambda_\beta}\, \delta_\scB A_\rho
    - \frac{\delta \epsilon^\rho_0}{\delta\Lambda_\beta}\, \delta_\scB n_\rho
    + \half \frac{\delta(2 v^\rho \pi_0^\sigma +
    \tau^{\rho\sigma}_0)}{\delta\Lambda_\beta}\,
    \delta_\scB h_{\rho\sigma} =  \mathcal{O}(\dow^2)~, \nn\\
  \frac{1}{\sqrt{\gamma}} \delta_\scB \lb \sqrt{\gamma}\,
  (\fh_\mu + \fn\, A_\mu)\rb
  &= \frac{\delta j^\rho_0}{\delta\beta^\mu}\, \delta_\scB A_\rho
    - \frac{\delta \epsilon^\rho_0}{\delta\beta^\mu}\, \delta_\scB n_\rho
    + \half \frac{\delta(2 v^\rho \pi_0^\sigma +
    \tau^{\rho\sigma}_0)}{\delta\beta^\mu}\,
    \delta_\scB h_{\rho\sigma} =  \mathcal{O}(\dow^2)~.
\end{align}
For explicit computations, it is convenient to instead work with $u^\mu$, $T$,
$\mu$, in terms of which we can recast these as
\begin{equation}
  \def\arraystretch{1.2}
  \begin{pmatrix}
    \frac{1}{T} \frac{\delta j^\rho_0}{\delta(\mu/T)} 
    & \frac{1}{T} \frac{\delta \epsilon^\rho_0}{\delta(\mu/T)}
    & \frac{1}{T} \frac{\delta(2 v^\rho \pi_0^\sigma +
      \tau^{\rho\sigma}_0)}{\delta(\mu/T)} \\
    T \frac{\delta j^\rho_0}{\delta T} 
    & T \frac{\delta \epsilon^\rho_0}{\delta T}
    & T \frac{\delta(2 v^\rho \pi_0^\sigma +
      \tau^{\rho\sigma}_0)}{\delta T} \\
    \frac{1}{T} h^\tau_\mu \frac{\delta j^\rho_0}{\delta(u^\tau/T)} 
    & \frac{1}{T} h^\tau_\mu \frac{\delta \epsilon^\rho_0}{\delta(u^\tau/T)}
    & \frac{1}{T} h^\tau_\mu \frac{\delta(2 v^\rho \pi_0^\sigma +
      \tau^{\rho\sigma}_0)}{\delta(u^\tau/T)}
  \end{pmatrix}
  \begin{pmatrix}
    \delta_\scB A_\rho \\
    - \delta_\scB n_\rho \\
    \half\delta_\scB h_{\rho\sigma}
  \end{pmatrix}
  =  \mathcal{O}(\dow^2)~.
\end{equation}
We define the matrices
\begin{equation}
  \def\arraystretch{1.2}
  \chi =
  \begin{pmatrix}
    \frac{1}{T} \frac{n_\lambda \delta j^\lambda_0}{\delta(\mu/T)} 
    & \frac{1}{T} \frac{n_\lambda \delta \epsilon^\lambda_0}{\delta(\mu/T)}
    & \frac{1}{T} \frac{\delta\pi_0^\lambda}{\delta(\mu/T)} \\
    T \frac{n_\lambda\delta j^\lambda_0}{\delta T} 
    & T \frac{n_\lambda \delta \epsilon^\lambda_0}{\delta T}
    & T \frac{\delta\pi_0^\lambda}{\delta T} \\
    \frac{1}{T} h^\tau_\mu \frac{n_\lambda\delta j^\lambda_0}{\delta(u^\tau/T)} 
    & \frac{1}{T} h^\tau_\mu \frac{n_\lambda\delta
      \epsilon^\lambda_0}{\delta(u^\tau/T)}
    & \frac{1}{T} h^\tau_\mu \frac{\delta\pi_0^\lambda}{\delta(u^\tau/T)}
  \end{pmatrix}, \quad
  \chi_S
  = \begin{pmatrix}
    \frac{1}{T} \frac{h^\rho_\lambda \delta j^\lambda_0}{\delta(\mu/T)} 
    & \frac{1}{T} \frac{h^\rho_\lambda \delta \epsilon^\lambda_0}{\delta(\mu/T)}
    & \frac{1}{T} \frac{\delta\tau^{\rho\sigma}_0}{\delta(\mu/T)} \\
    T \frac{h^\rho_\lambda\delta j^\lambda_0}{\delta T} 
    & T \frac{h^\rho_\lambda \delta \epsilon^\lambda_0}{\delta T}
    & T \frac{\delta\tau^{\rho\sigma}_0}{\delta T} \\
    \frac{1}{T} h^\tau_\mu \frac{h^\rho_\lambda\delta j^\lambda_0}{\delta(u^\tau/T)} 
    & \frac{1}{T} h^\tau_\mu \frac{h^\rho_\lambda\delta
      \epsilon^\lambda_0}{\delta(u^\tau/T)}
    & \frac{1}{T} h^\tau_\mu \frac{\delta\tau^{\rho\sigma}_0}{\delta(u^\tau/T)}
  \end{pmatrix}.
\end{equation}
Here $\chi$ is the same susceptibility matrix defined in
\cref{sec:linearised}. Let us also define
\begin{equation}
  \def\arraystretch{1.2}
  M = \chi^{-1}\chi_S =
  \begin{pmatrix}
    \rho \frac{\dow\hat n}{\dow n} \vec u^\rho
    & \rho \frac{\dow\hat w}{\dow n} \vec u^\rho
    & \frac{\dow p}{\dow n} h^{\rho\sigma}
    -\frac{\dow\rho}{\dow n} \vec u^\rho \vec u^\sigma \\
    \rho \frac{\dow\hat n}{\dow\epsilon} \vec u^\rho
    & \rho \frac{\dow\hat w}{\dow\epsilon} \vec u^\rho
    & \frac{\dow p}{\dow \epsilon} h^{\rho\sigma}
    -\frac{\dow\rho}{\dow \epsilon} \vec u^\rho \vec u^\sigma \\
    \frac{\rho}{|\vec u|} \frac{\dow\hat n}{\dow |\pi|} \vec u_\mu \vec u^\rho
    + \hat n h_\mu^{\rho}
    & \frac{\rho}{|\vec u|} \frac{\dow\hat w}{\dow |\pi|} \vec u_\mu \vec u^\rho
    + \hat w h_\mu^{\rho}
    & \frac{\vec u_\mu}{|\vec u|}
    \lb \frac{\dow p}{\dow |\pi|} h^{\rho\sigma} 
    - \frac{\dow\rho}{\dow|\pi|} \vec u^\rho \vec u^\rho \rb
    + 2 \vec u^{(\rho} h_\mu^{\sigma)}
  \end{pmatrix}~.
\end{equation}
Here we have used $w = \epsilon+p$, $\hat n = n/\rho$, and
$\hat w = (\epsilon+p)/\rho$.  This allows us to re-express the equations of
motion \eqref{eq:EOMappendix} as
\begin{equation}
  \begin{pmatrix}
    v^\rho\delta_\scB A_\rho \\
    - v^\rho\delta_\scB n_\rho \\
    v^\rho\delta_\scB h_{\rho\sigma}
  \end{pmatrix}
  = - M
  \begin{pmatrix}
    \delta_\scB A_\rho \\
    - \delta_\scB n_\rho \\
    \half\delta_\scB h_{\rho\sigma}
  \end{pmatrix}
  +  \mathcal{O}(\dow^2)~.
\end{equation}

On the other hand, one-derivative order frame transformations of the
hydrodynamic fields $u^\mu$, $T$, and $\mu$ act as
\begin{equation}
  \def\arraystretch{1.2}
  \delta
  \begin{pmatrix}
    j^\mu \\
    \epsilon^\mu  \\
    2v^{(\mu} \pi^{\nu)} + \tau^{\mu\nu}
  \end{pmatrix}
  =
  \begin{pmatrix}
    \frac1T \frac{\delta j^\mu_0}{\delta(\mu/T)}
    & T \frac{\delta j^\mu_0}{\delta T}
    & \frac1T \frac{\delta j^\mu_0}{\delta(u^\lambda/T)} \\
    \frac1T h^\lambda_\rho \frac{\delta \epsilon^\mu_0}{\delta(\mu/T)}
    & T \frac{\delta \epsilon^\mu_0}{\delta T} 
    & \frac1T \frac{\delta \epsilon^\mu_0}{\delta(u^\lambda/T)} \\
    \frac1T h^\lambda_\rho
    \frac{\delta(2 v^\mu\pi_0^\nu + \tau^{\mu\nu}_0)}{\delta(\mu/T)}
    & T \frac{\delta(2 v^\mu\pi_0^\nu + \tau^{\mu\nu}_0)}{\delta T} 
    & \frac1T h^\lambda_\rho
    \frac{\delta(2 v^\mu\pi_0^\nu + \tau^{\mu\nu}_0)}{\delta(u^\lambda/T)}
  \end{pmatrix}
  \begin{pmatrix}
    T \delta(\mu/T) \\ \frac1T \delta T \\
    T \delta(u^\rho/T)
  \end{pmatrix}~.
\end{equation}
Using the decomposition of the constitutive relations into hydrostatic ``hs''
and non-hydrostatic ``nhs'' pieces (where ``nhs'' contains both ``diss'' and
``nhsnd''), the density frame is defined as
\begin{equation}
  n_\mu j^\mu_\nhs = n_\mu \epsilon^\mu_\nhs = \pi^\mu_\nhs = 0~.
\end{equation}
Denoting the respective corrections in the generic frame with tilde, we get
\begin{equation}
  \def\arraystretch{1.2}
  \begin{pmatrix}
    T \delta(\mu/T) \\ \frac1T \delta T \\
    T \delta(u^\rho/T)
  \end{pmatrix}
  = - \chi^{-\rmT}
  \begin{pmatrix}
    n_\lambda \tilde\jmath^\lambda_\nhs \\
    n_\lambda \tilde\epsilon^\lambda_\nhs  \\
    \tilde\pi^{\lambda}_\nhs
  \end{pmatrix}~.
\end{equation}
The superscript ``$\rmT$'' denotes a transpose and ``$-\rmT$'' an inverse
transpose.  The non-hydrostatic corrections in the thermodynamic density frame
can be written out explicitly as
\begin{equation}
  \def\arraystretch{1.2}
  \begin{pmatrix}
    j^\mu_\nhs \\
    \epsilon^\mu_\nhs  \\
    \tau^{\mu\nu}_\nhs
  \end{pmatrix}
  =
  \begin{pmatrix}
    h^\mu_{~\lambda} \tilde\jmath^\lambda_\nhs \\
    h^\mu_{~\lambda} \tilde\epsilon^\lambda_\nhs  \\
    \tilde\tau^{\mu\nu}_\nhs
  \end{pmatrix}
  - M^\rmT
  \begin{pmatrix}
    n_\lambda \tilde\jmath^\lambda_\nhs \\
    n_\lambda \tilde\epsilon^\lambda_\nhs  \\
    \tilde\pi^{\lambda}_\nhs
  \end{pmatrix}~.
  \label{eq:frametrans_half}
\end{equation}

To obtain a mapping between the respective transport coefficients, we need to do
a final manipulation requiring the usage of the equations of motion. We start by
decomposing the generic frame non-hydrostatic constitutive relations as
\begin{equation}
  \def\arraystretch{1.2}
  \begin{pmatrix}
    \tilde\jmath^\mu_\nhs \\ \tilde\epsilon^\mu_\nhs \\
    2v^{(\mu}\tilde\pi^{\nu)}_\nhs + \tilde\tau^{\mu\nu}_\nhs
  \end{pmatrix}
  = -
  \kB T\begin{pmatrix}
    \tilde C^{\mu\rho}_{nn} & \tilde C^{\mu\rho}_{n\epsilon}
    & \tilde C^{\mu(\rho\sigma)}_{n\pi} \\
    \tilde C^{\mu\rho}_{\epsilon n} & \tilde C^{\mu\rho}_{\epsilon\epsilon}
    & \tilde C^{\mu(\rho\sigma)}_{\epsilon\pi} \\
    \tilde C^{(\mu\nu)\rho}_{\pi n} & \tilde C^{(\mu\nu)\rho}_{\epsilon\pi}
    & \tilde C^{(\mu\nu)(\rho\sigma)}_{\pi\pi}
  \end{pmatrix}
  \begin{pmatrix}
    \delta_\scB A_\rho \\ - \delta_\scB n_\rho \\ \half\delta_\scB h_{\rho\sigma}
  \end{pmatrix}~.
\end{equation}
Note that the coefficient matrix is asymmetric because it contains both ``diss''
and ``nhsnd'' pieces. Let us decompose it further into space and time parts
\begin{align}
  \tilde\fC_{TT}
  &= 
    \begin{pmatrix}
      n_\mu n_\rho \tilde C^{\mu\rho}_{nn}
      & n_\mu n_\rho  \tilde C^{\mu\rho}_{n\epsilon}
      & n_\mu n_\rho h^\tau_\sigma  \tilde C^{\mu(\rho\sigma)}_{n\pi} \\
      n_\mu n_\rho h^\tau_\sigma \tilde C^{\mu\rho}_{\epsilon n}
      & n_\mu n_\rho \tilde C^{\mu\rho}_{\epsilon\epsilon}
      & n_\mu n_\rho  \tilde C^{\mu(\rho\sigma)}_{\epsilon\pi} \\
      n_\mu h^\beta_\nu n_\rho  \tilde C^{(\mu\nu)\rho}_{\pi n}
      & n_\mu h^\beta_\nu n_\rho  \tilde C^{(\mu\nu)\rho}_{\epsilon\pi}
      & n_\mu h^\beta_\nu n_\rho h^\tau_\sigma  \tilde C^{(\mu\nu)(\rho\sigma)}_{\pi\pi}
    \end{pmatrix}, \nn\\
  \tilde\fC_{TS}
  &= 
    \begin{pmatrix}
      n_\mu h^\lambda_\rho \tilde C^{\mu\rho}_{nn}
      & n_\mu h^\lambda_\rho  \tilde C^{\mu\rho}_{n\epsilon}
      & n_\mu h^\lambda_\rho h^\tau_\sigma  \tilde C^{\mu(\rho\sigma)}_{n\pi} \\
      n_\mu h^\lambda_\rho h^\tau_\sigma \tilde C^{\mu\rho}_{\epsilon n}
      & n_\mu h^\lambda_\rho \tilde C^{\mu\rho}_{\epsilon\epsilon}
      & n_\mu h^\lambda_\rho \tilde C^{\mu(\rho\sigma)}_{\epsilon\pi} \\
      n_\mu h^\beta_\nu h^\lambda_\rho  \tilde C^{(\mu\nu)\rho}_{\pi n}
      & n_\mu h^\beta_\nu h^\lambda_\rho  \tilde C^{(\mu\nu)\rho}_{\epsilon\pi}
      & n_\mu h^\beta_\nu h^\lambda_\rho h^\tau_\sigma  \tilde C^{(\mu\nu)(\rho\sigma)}_{\pi\pi}
    \end{pmatrix}, \nn\\
  \tilde\fC_{ST}
  &= 
    \begin{pmatrix}
      h^\alpha_\mu n_\rho \tilde C^{\mu\rho}_{nn}
      & h^\alpha_\mu n_\rho  \tilde C^{\mu\rho}_{n\epsilon}
      & h^\alpha_\mu n_\rho h^\tau_\sigma  \tilde C^{\mu(\rho\sigma)}_{n\pi} \\
      h^\alpha_\mu n_\rho h^\tau_\sigma \tilde C^{\mu\rho}_{\epsilon n}
      & h^\alpha_\mu n_\rho  \tilde C^{\mu\rho}_{\epsilon\epsilon}
      & h^\alpha_\mu n_\rho  \tilde C^{\mu(\rho\sigma)}_{\epsilon\pi} \\
      h^\alpha_\mu h^\beta_\nu n_\rho  \tilde C^{(\mu\nu)\rho}_{\pi n}
      & h^\alpha_\mu h^\beta_\nu n_\rho  \tilde C^{(\mu\nu)\rho}_{\epsilon\pi}
      & h^\alpha_\mu h^\beta_\nu n_\rho h^\tau_\sigma  \tilde C^{(\mu\nu)(\rho\sigma)}_{\pi\pi}
    \end{pmatrix}, \nn\\
  \tilde\fC_{SS}
  &= 
    \begin{pmatrix}
      h^\alpha_\mu h^\lambda_\rho \tilde C^{\mu\rho}_{nn}
      & h^\alpha_\mu h^\lambda_\rho \tilde C^{\mu\rho}_{n\epsilon}
      & h^\alpha_\mu h^\lambda_\rho h^\tau_\sigma  \tilde C^{\mu(\rho\sigma)}_{n\pi} \\
      h^\alpha_\mu h^\lambda_\rho h^\tau_\sigma \tilde C^{\mu\rho}_{\epsilon n}
      & h^\alpha_\mu h^\lambda_\rho \tilde C^{\mu\rho}_{\epsilon\epsilon}
      & h^\alpha_\mu h^\lambda_\rho \tilde C^{\mu(\rho\sigma)}_{\epsilon\pi} \\
      h^\alpha_\mu h^\beta_\nu h^\lambda_\rho  \tilde C^{(\mu\nu)\rho}_{\pi n}
      & h^\alpha_\mu h^\beta_\nu h^\lambda_\rho \tilde C^{(\mu\nu)\rho}_{\epsilon\pi}
      & h^\alpha_\mu h^\beta_\nu h^\lambda_\rho h^\tau_\sigma
      \tilde C^{(\mu\nu)(\rho\sigma)}_{\pi\pi}
    \end{pmatrix}.
\end{align}
This explicitly results in the compact expressions
\begin{align}
  \def\arraystretch{1.2}
  \begin{pmatrix}
    n_\mu \tilde\jmath^\mu_\nhs \\ n_\mu \tilde\epsilon^\mu_\nhs \\
    \tilde\pi^{\mu}_\nhs
  \end{pmatrix}
  &= - \kB T\, \tilde\fC_{TT}
  \begin{pmatrix}
    v^\rho \delta_\scB A_\rho \\
    - v^\rho \delta_\scB n_\rho \\
    v^\rho \delta_\scB h_{\rho\sigma}
  \end{pmatrix}
  - \kB T\, \tilde\fC_{TS}
  \begin{pmatrix}
    \delta_\scB A_\rho \\ - \delta_\scB n_\rho \\ \half\delta_\scB h_{\rho\sigma}
  \end{pmatrix}~, \nn\\
  \begin{pmatrix}
    h^\mu_\lambda \tilde\jmath^\lambda_\nhs \\
    h^\mu_\lambda\tilde\epsilon^\lambda_\nhs \\
    \tilde\tau^{\mu\nu}_\nhs
  \end{pmatrix}
  &= - \kB T\, \tilde\fC_{ST}
  \begin{pmatrix}
    v^\rho \delta_\scB A_\rho \\
    - v^\rho \delta_\scB n_\rho \\
    v^\rho \delta_\scB h_{\rho\sigma}
  \end{pmatrix}
  - \kB T\, \tilde\fC_{SS}
  \begin{pmatrix}
    \delta_\scB A_\rho \\ - \delta_\scB n_\rho \\ \half\delta_\scB h_{\rho\sigma}
  \end{pmatrix}~.
\end{align}
Plugging these into \cref{eq:frametrans_half} and using the equations of motion, we
get
\begin{align}
  \def\arraystretch{1.2}
  \begin{pmatrix}
    j^\mu_\hs \\
    \epsilon^\mu_\hs  \\
    \tau^{\mu\nu}_\hs
  \end{pmatrix}
  &= - \kB T
    \begin{pmatrix}
      - M \\ 1
    \end{pmatrix}^\rmT
  \begin{pmatrix}
    \tilde\fC_{TT} & \tilde\fC_{TS} \\
    \tilde\fC_{ST} & \tilde\fC_{SS}
  \end{pmatrix}
                \begin{pmatrix}
                  - M \\ 1
                \end{pmatrix}
  \begin{pmatrix}
    \delta_\scB A_\rho \\ - \delta_\scB n_\rho \\ \half\delta_\scB h_{\rho\sigma}
  \end{pmatrix}.
\end{align}
All in all, the transformation from a general frame to the density frame is
given by the transformation of non-hydrostatic transport coefficients
\begin{equation}
  \fC =
  \begin{pmatrix}
    - M \\ 1
  \end{pmatrix}^\rmT
  \begin{pmatrix}
    \tilde\fC_{TT} & \tilde\fC_{TS} \\
    \tilde\fC_{ST} & \tilde\fC_{SS}
  \end{pmatrix}
  \begin{pmatrix}
    - M \\ 1
  \end{pmatrix}~,
\end{equation}
where $\fC$ is related to combinations of dissipative and non-hydrostatic
non-dissipative coefficient matrices from \cref{sec:1der}, in particular
\begin{equation}
  \def\arraystretch{1.2}
  \fC = 
  \begin{pmatrix}
    D^{\mu\rho}_{nn} &
    D^{\mu\rho}_{n\epsilon} + \bar D_{n\epsilon}^{\mu\rho}
    & D_{n\pi}^{\mu(\rho\sigma)} + \bar D_{n\pi}^{\mu(\rho\sigma)} \\
    D^{\mu\rho}_{n\epsilon} - \bar D^{\rho\mu}_{n\epsilon}
    & D_{\epsilon\epsilon}^{\mu\rho}
    & D_{\epsilon\pi}^{\mu(\rho\sigma)} + \bar D_{\epsilon\pi}^{\mu(\rho\sigma)} \\
    D_{n\pi}^{\mu(\rho\sigma)} - \bar D_{n\pi}^{\rho(\mu\nu)}
    & D_{\epsilon\pi}^{\mu(\rho\sigma)} - \bar D_{\epsilon\pi}^{\rho(\mu\nu)}
    & D^{(\mu\nu)(\rho\sigma)}_{\pi\pi} + \bar D^{(\mu\nu)(\rho\sigma)}_{\pi\pi}
  \end{pmatrix}~.
\end{equation}

\subsection{Landau frame}
\label{app:Landau}

To discuss the constitutive relations in Landau frame, it is convenient to
define a covariant energy-momentum tensor~\cite{Novak:2019wqg,deBoer:2020xlc}
for boost-agnostic fluids
\begin{equation}
  T^\mu_{~\,\nu} \equiv -\epsilon^\mu n_\nu
  + v^\mu \pi_\nu + \tau^{\mu\lambda} h_{\lambda\nu}~,
  \label{eq:covariantST}
\end{equation}
where $\pi_\mu=h_{\mu\nu}\pi^\nu$. This energy-momentum tensor satisfies the conservation equations
\begin{gather}
  \lb \nabla_\mu + F^n_{\mu\lambda} v^\lambda \rb T^\mu_{~\,\nu}
  = F_{\nu\mu} j^\mu - F^n_{\nu\mu} \epsilon^\mu
    - \half \pi^\lambda \lie_v h_{\nu\lambda}~, \nn\\
  \implies \frac{1}{\sqrt{\gamma}} \dow_\mu \lb \sqrt{\gamma}\, T^\mu_{~\,\nu} \rb
  + \epsilon^\mu \dow_\nu n_\mu
  - \half \lb v^\mu \pi^\rho + \pi^\rho  v^\mu + \tau^{\mu\rho} \rb \dow_\nu h_{\mu\rho}
  = F_{\nu\mu} j^\mu~.
\end{gather}
In addition, we have the charge current $j^\mu$ that still satisfies the same
conservation equation as given in \cref{eq:NC.Conservation}.  The thermodynamic
density frame that we have employed in this work can be expressed as
$n_\mu (T^\mu_{~\,\nu})_{\text{nhs}} = 0$, $n_\mu j^\mu_{\text{nhs}} = 0$, where
``nhs'' collectively denotes dissipative and non-dissipative non-hydrostatic
contributions. By contrast, we can define the thermodynamic Landau frame as
$(T^\mu_{~\,\nu})_{\text{nhs}} u^\nu = 0$,
$n_\mu j^\mu_{\text{nhs}} = j^\mu \vec u_\mu/c^2$, leading to\footnote{The
  thermodynamic Landau frame agrees with the true Landau frame used
  in~\cite{Novak:2019wqg,deBoer:2020xlc}, defined as
  $T^\mu_{~\nu} u^\nu = (\epsilon - \rho\, \vec u^2) u^\mu$,
  $n_\mu j^\mu = n + \frac{1}{c^2} j^\mu \vec u_\mu$, only in the
  non-hydrostatic sector. The Landau frame definition of~\cite{Novak:2019wqg}
  has an arbitrary function $B$ in the U(1) sector, which we have taken to be
  $(1-\vec u^2/c^2)^{-1}$. Since $c$ is an arbitrary parameter at this stage,
  not having specialised to relativistic fluids, we can recover the generality
  of~\cite{Novak:2019wqg} by choosing $c$ appropriately.}
\begin{equation}
  n_\mu\epsilon^\mu_{\text{nhs}} = \pi^\mu_{\text{nhs}} \vec u_\mu~, \qquad
  h^\mu_{~\nu} \epsilon^\nu_{\text{nhs}}
  = \tau^{\mu\nu}_{\text{nhs}} \vec u_\nu~, \qquad
  n_\mu j^\mu_{\text{nhs}} = \frac{1}{c^2} j^\mu_{\text{nhs}} \vec u_\mu~.
  \label{eq:Landau-frame}
\end{equation}
Equivalently, this amounts to working with
\begin{equation}
  h^{\mu\nu} \delta_{\scB} A_\nu
  + \frac{1}{c^2} \vec u^\mu v^\nu \delta_{\scB} A_\nu~, \qquad
  \delta_{\scB} h_{\mu\nu}
  - 2 \vec u_{(\mu} \delta_\scB n_{\nu)}~.
\end{equation}
as the set of independent non-hydrostatic data.

The hydrostatic constitutive relations are still the same as
\cref{sec:hydrostatic}, but the non-hydrostatic non-dissipative and dissipative
constitutive relations in the thermodynamic Landau frame are given as
follows. Firstly, we have the non-hydrostatic non-dissipative densities
\begin{align}
  \begin{pmatrix}
    n_\nu \tilde\jmath^\nu_{\text{nhsnd}} \\
    n_\nu \tilde\epsilon^\nu_{\text{nhsnd}} \\
    \tilde\pi^\mu_{\text{nhsnd}}
  \end{pmatrix}
  &= -\kB T
  \begin{pmatrix}
    0
    & \frac{1}{c^2} \tilde{\bar D}_{j\pi}^{\mu\rho} \vec u_\mu \vec u_\rho
    & \frac{1}{c^2} \tilde{\bar D}_{j\pi}^{\mu\rho} \vec u_\mu \\
    - \frac{1}{c^2}\tilde{\bar D}_{j\pi}^{\rho\mu} \vec u_\rho \vec u_\mu
    & 0
    & 0 \\
    - \frac{1}{c^2}\tilde{\bar D}_{j\pi}^{\rho\mu} \vec u_\rho
    & 0 & 0
  \end{pmatrix}
  \begin{pmatrix}
    v^\sigma \delta_\scB A_\sigma \\
    - v^\sigma \delta_\scB n_\sigma \\
    v^\sigma \delta_\scB h_{\rho\sigma}
  \end{pmatrix} \nn\\
  &\qquad
    -\kB T
    \begin{pmatrix}
      0
      & \frac{1}{c^2}\tilde{\bar D}_{j\tau}^{\mu\rho\sigma} \vec u_\mu \vec u_\sigma
      & \frac{1}{c^2}\tilde{\bar D}_{j\tau}^{\mu\rho\sigma} \vec u_\mu \\
      -\tilde{\bar D}_{j\pi}^{\rho\mu} \vec u_\mu
      & \tilde{\bar D}_{\pi\tau}^{\mu\rho\sigma} \vec u_\mu \vec u_\sigma
      & \tilde{\bar D}_{\pi\tau}^{\mu\rho\sigma} \vec u_\mu \\
      - \tilde{\bar D}_{j\pi}^{\rho\mu}
      & \tilde{\bar D}_{\pi\tau}^{\mu\rho\sigma} \vec u_\sigma
      & \tilde{\bar D}_{\pi\tau}^{\mu\rho\sigma}
    \end{pmatrix}
  \begin{pmatrix}
    \delta_\scB A_\rho \\ - \delta_\scB n_\rho \\ \half\delta_\scB h_{\rho\sigma}
  \end{pmatrix},
\end{align}
and non-hydrostatic non-dissipative fluxes
\begin{align}
  \begin{pmatrix}
    \tilde\jmath^\mu_{\text{nhsnd}} \\
    \tilde\epsilon^\mu_{\text{nhsnd}} \\
    \tilde\tau^{\mu\nu}_{\text{nhsnd}}
  \end{pmatrix}
  &= -\kB T
  \begin{pmatrix}
    0
    & \tilde{\bar D}_{j\pi}^{\mu\rho} \vec u_\rho
    & \tilde{\bar D}_{j\pi}^{\mu\rho} \\
    - \frac{1}{c^2} \tilde{\bar D}_{j\tau}^{\rho\mu\nu} \vec u_\nu \vec u_\rho
    & - \tilde{\bar D}_{\pi\tau}^{\rho\mu\nu} \vec u_\nu \vec u_\rho
    & - \tilde{\bar D}_{\pi\tau}^{\rho\mu\nu} \vec u_\nu \\
    - \frac{1}{c^2} \tilde{\bar D}_{j\tau}^{\rho\mu\nu} \vec u_\rho
    & - \tilde{\bar D}_{\pi\tau}^{\rho\mu\nu} \vec u_\rho
    & - \tilde{\bar D}_{\pi\tau}^{\rho\mu\nu}
  \end{pmatrix}
  \begin{pmatrix}
    v^\sigma \delta_\scB A_\sigma \\
    - v^\sigma \delta_\scB n_\sigma \\
    v^\sigma \delta_\scB h_{\rho\sigma}
  \end{pmatrix} \nn\\
  &\qquad
    -\kB T
    \begin{pmatrix}
      0
    & \tilde{\bar D}_{j\tau}^{\mu\rho\sigma} \vec u_\sigma
    & \tilde{\bar D}_{j\tau}^{\mu\rho\sigma} \\
    -\tilde{\bar D}_{j\tau}^{\rho\mu\nu} \vec u_\nu
    & \tilde{\bar D}_{\tau\tau}^{\mu\nu\rho\sigma} \vec u_\nu \vec u_\sigma
    & \tilde{\bar D}_{\tau\tau}^{\mu\nu\rho\sigma} \vec u_\nu \\
    - \tilde{\bar D}_{j\tau}^{\rho\mu\nu}
    & \tilde{\bar D}_{\tau\tau}^{\mu\nu\rho\sigma} \vec u_\sigma
    & \tilde{\bar D}_{\tau\tau}^{\mu\nu\rho\sigma}
  \end{pmatrix}
  \begin{pmatrix}
    \delta_\scB A_\rho \\ - \delta_\scB n_\rho \\ \half\delta_\scB h_{\rho\sigma}
  \end{pmatrix},
\end{align}
where we have defined the transport coefficient matrices
\begin{align}
  \tilde{\bar D}_{j\pi}^{\mu\rho}
  &= \tilde{\bar\fv}_{01} P^{\mu\rho}
    + \tilde{\bar\fs}_{01} \hat u^\mu \hat u^\rho~, \nn\\
  \tilde{\bar D}_{j\tau}^{\mu\rho\sigma}
  &= 2 \tilde{\bar\fv}_{02} P^{\mu(\rho} \hat u^{\sigma)}
    + \tilde{\bar\fs}_{02} \hat u^\mu \hat u^\rho \hat u^\sigma
    + \tilde{\bar\fs}_{03} \hat u^\mu P^{\rho\sigma}, \nn\\
  \tilde{\bar D}_{\pi\tau}^{\mu\rho\sigma}
  &= 2 \tilde{\bar\fv}_{12} P^{\mu(\rho} \hat u^{\sigma)}
    + \tilde{\bar\fs}_{12} \hat u^\mu \hat u^\rho \hat u^\sigma
    + \tilde{\bar\fs}_{13} \hat u^\mu P^{\rho\sigma}, \nn\\
  \tilde{\bar D}^{\mu\nu\rho\sigma}_{\tau\tau}
  &= \tilde{\bar\fs}_{23}\lb
    \hat u^\mu \hat u^\nu P^{\rho\sigma}
    - P^{\mu\nu} \hat u^\rho \hat u^\sigma \rb.
\end{align}
On the other hand, in the dissipative sector we have the densities
\begin{align}
  \begin{pmatrix}
    n_\nu \tilde\jmath^\nu_{\text{diss}} \\
    n_\nu \tilde\epsilon^\nu_{\text{diss}} \\
    \tilde\pi^\mu_{\text{nhsnd}}
  \end{pmatrix}
  &= -\kB T
  \begin{pmatrix}
    \frac{1}{c^4}\tilde D_{jj}^{\mu\rho} \vec u_\mu \vec u_\rho
    & \frac{1}{c^2} \tilde D_{j\pi}^{\mu\rho} \vec u_\mu \vec u_\rho
    & \frac{1}{c^2} \tilde D_{j\pi}^{\mu\rho} \vec u_\mu \\
    \frac{1}{c^2}\tilde D_{j\pi}^{\rho\mu} \vec u_\mu \vec u_\rho
    & \tilde D_{\pi\pi}^{\mu\rho} \vec u_\mu \vec u_\rho
    & \tilde D_{\pi\pi}^{\mu\rho} \vec u_\mu \\
    \frac{1}{c^2}\tilde D_{j\pi}^{\rho\mu} \vec u_\rho
    & \tilde D_{\pi\pi}^{\mu\rho} \vec u_\rho
    & \tilde D_{\pi\pi}^{\mu\rho}
  \end{pmatrix}
  \begin{pmatrix}
    v^\sigma \delta_\scB A_\sigma \\
    - v^\sigma \delta_\scB n_\sigma \\
    v^\sigma \delta_\scB h_{\rho\sigma}
  \end{pmatrix} \nn\\
  &\qquad
    -\kB T
    \begin{pmatrix}
    \frac{1}{c^2}\tilde D_{jj}^{\mu\rho} \vec u_\mu
    & \frac{1}{c^2}\tilde D_{j\tau}^{\mu\rho\sigma} \vec u_\mu \vec u_\sigma
    & \frac{1}{c^2}\tilde D_{j\tau}^{\mu\rho\sigma} \vec u_\mu \\
    \tilde D_{j\pi}^{\rho\mu} \vec u_\mu
    & \tilde D_{\pi\tau}^{\mu\rho\sigma} \vec u_\mu \vec u_\sigma
    & \tilde D_{\pi\tau}^{\mu\rho\sigma} \vec u_\mu \\
    \tilde D_{j\pi}^{\rho\mu}
    & \tilde D_{\pi\tau}^{\mu\rho\sigma} \vec u_\sigma
    & \tilde D_{\pi\tau}^{\mu\rho\sigma}
  \end{pmatrix}
  \begin{pmatrix}
    \delta_\scB A_\rho \\ - \delta_\scB n_\rho \\ \half\delta_\scB h_{\rho\sigma}
  \end{pmatrix},
\end{align}
and fluxes
\begin{subequations}
  \begin{align}
    \begin{pmatrix}
      \tilde\jmath^\mu_{\text{diss}} \\
      \tilde\epsilon^\mu_{\text{diss}} \\
      \tilde\tau^{\mu\nu}_{\text{diss}}
    \end{pmatrix}
    &= -\kB T
      \begin{pmatrix}
        \frac{1}{c^2} \tilde D_{jj}^{\mu\rho} \vec u_\rho
        & \tilde D_{j\pi}^{\mu\rho} \vec u_\rho
        & \tilde D_{j\pi}^{\mu\rho} \\
        \frac{1}{c^2} \tilde D_{j\tau}^{\rho\mu\nu} \vec u_\nu \vec u_\rho
        & \tilde D_{\pi\tau}^{\rho\mu\nu} \vec u_\nu \vec u_\rho
        & \tilde D_{\pi\tau}^{\rho\mu\nu} \vec u_\nu \\
        \frac{1}{c^2} \tilde D_{j\tau}^{\rho\mu\nu} \vec u_\rho
        & \tilde D_{\pi\tau}^{\rho\mu\nu} \vec u_\rho
        & \tilde D_{\pi\tau}^{\rho\mu\nu}
      \end{pmatrix}
          \begin{pmatrix}
            v^\sigma \delta_\scB A_\sigma \\
            - v^\sigma \delta_\scB n_\sigma \\
            v^\sigma \delta_\scB h_{\rho\sigma}
          \end{pmatrix} \nn\\
    &\qquad -\kB T
      \begin{pmatrix}
        \tilde D_{jj}^{\mu\rho}
        & \tilde D_{j\tau}^{\mu\rho\sigma} \vec u_\sigma
        & \tilde D_{j\tau}^{\mu\rho\sigma} \\
        \tilde D_{j\tau}^{\rho\mu\nu} \vec u_\nu
        & \tilde D_{\tau\tau}^{\mu\nu\rho\sigma} \vec u_\nu \vec u_\sigma
        & \tilde D_{\tau\tau}^{\mu\nu\rho\sigma} \vec u_\nu \\
        \tilde D_{j\tau}^{\rho\mu\nu}
        & \tilde D_{\tau\tau}^{\mu\nu\rho\sigma} \vec u_\sigma
        & \tilde D_{\tau\tau}^{\mu\nu\rho\sigma}
      \end{pmatrix}
          \begin{pmatrix}
            \delta_\scB A_\rho \\ - \delta_\scB n_\rho \\
            \half\delta_\scB h_{\rho\sigma}
          \end{pmatrix}~.
  \end{align}
  \label{eq:Landau-diss}%
\end{subequations}
The coefficient matrices in the dissipative sector are given as
\begin{align}
  \tilde D_{jj}^{\mu\rho}
  &= \tilde\fv_{00} P^{\mu\rho}
    + \tilde\fs_{00} \hat u^\mu \hat u^\rho~, \nn\\
  \tilde D_{j\pi}^{\mu\rho}
  &= \tilde\fv_{01} P^{\mu\rho}
    + \tilde\fs_{01} \hat u^\mu \hat u^\rho~, \nn\\
  \tilde D_{\pi\pi}^{\mu\rho}
  &= \tilde\fv_{11} P^{\mu\rho}
    + \tilde\fs_{11} \hat u^\mu \hat u^\rho~, \nn\\
  \tilde D_{j\tau}^{\mu\rho\sigma}
  &= 2\tilde\fv_{02} P^{\mu(\rho} \hat u^{\sigma)}
    + \tilde\fs_{02} \hat u^\mu \hat u^\rho \hat u^\sigma
    + \tilde\fs_{03} \hat u^\mu P^{\rho\sigma}, \nn\\
  \tilde D_{\pi\tau}^{\mu\rho\sigma}
  &= 2\tilde\fv_{12} P^{\mu(\rho} \hat u^{\sigma)}
    + \tilde\fs_{12} \hat u^\mu \hat u^\rho \hat u^\sigma
    + \tilde\fs_{13} \hat u^\mu P^{\rho\sigma}, \nn\\
  \tilde C^{\mu\nu\rho\sigma}_{\tau\tau}
  &= 2\tilde\ft \lb P^{\rho(\mu} P^{\nu)\sigma}
    - {\textstyle\frac{1}{d-1}} P^{\mu\nu}P^{\rho\sigma} \rb
    + 4\tilde\fv_{22} \hat u^{(\mu} P^{\nu)(\rho} \hat u^{\sigma)} \nn\\
  &\qquad
    + \tilde\fs_{22} \hat u^\mu \hat u^\nu \hat u^\rho \hat u^\sigma
    + \tilde\fs_{23} \lb P^{\mu\nu} \hat u^\rho \hat u^\sigma
    + \hat u^\mu \hat u^\nu P^{\rho\sigma} \rb
    + \tilde\fs_{33} P^{\mu\nu} P^{\rho\sigma}~.
\end{align}

We can use the generic procedure chalked out in the previous subsection to map
the thermodynamic Landau frame coefficients to the thermodynamic density frame
used in the bulk of the paper. The coefficient coupling to the traceless tensor
remains unchanged during the map
\begin{equation}
  \ft = \tilde\ft~.
\end{equation}
However, in the vector sector, we find
\begin{align}
  \fv_{00}
  &= \tilde\fv_{00} + \hat n^2 \tilde\fv_{11}
    - 2\hat n \tilde\fv_{01}, \nn\\
  \fv_{01}
  &= |\vec u|\tilde\fv_{02} + \hat n \hat w \tilde\fv_{11}
    - \hat n \tilde\fv_{12}|\vec u| - \hat w \tilde\fv_{01}, \nn\\
  \fv_{02}
  &= \tilde\fv_{02} + \hat n |\vec u| \tilde\fv_{11} - \hat n \tilde\fv_{12}
    - |\vec u| \tilde\fv_{01}~, \nn\\
  \fv_{11}
  &= \vec u^2 \tilde\fv_{22}
    + \hat w^2 \tilde\fv_{11} - 2\hat w \tilde\fv_{12}|\vec u|~, \nn\\
  \fv_{12}
  &= |\vec u| \tilde\fv_{22}
    + \hat w |\vec u| \tilde\fv_{11}
    - \lb \hat w + \vec u^2\rb \tilde\fv_{12}~, \nn\\
  \fv_{22}
  &= \tilde\fv_{22}
    + \vec u^2   \tilde\fv_{11} - 2|\vec u|\tilde\fv_{12}~, \nn\\
  \bar\fv_{01}
  &= |\vec u|\tilde{\bar\fv}_{02}
    - \hat n \tilde{\bar\fv}_{12}|\vec u|
    - \hat w \tilde{\bar\fv}_{01}~, \nn\\
  \bar\fv_{02}
  &= \tilde{\bar\fv}_{02}
    - \hat n \tilde{\bar\fv}_{12}
    - |\vec u| \tilde{\bar\fv}_{01}~, \nn\\
  \bar\fv_{12}
  &= |\vec u| \tilde{\bar\fv}_{22}
    - \lb \hat w + \vec u^2 \rb \tilde{\bar\fv}_{12}~.
\end{align}
The mapping in the scalar sector is much messier to write down explicitly. We
instead use the matrix representation for clarity; we first isolate the scalar
part of the $M$ matrix as
\begin{align}
  M^\fs =
  \begin{pmatrix}
    |\vec\pi| \frac{\dow\hat n}{\dow n} 
    & |\vec\pi| \frac{\dow\hat w}{\dow n}
    & \frac{\dow p}{\dow n} - \frac{\dow\rho}{\dow n} \vec u^2
    & \frac{\dow p}{\dow n} \\
    |\vec\pi| \frac{\dow\hat n}{\dow\epsilon}
    & |\vec\pi| \frac{\dow\hat w}{\dow\epsilon}
    & \frac{\dow p}{\dow \epsilon}
    -\frac{\dow\rho}{\dow \epsilon} \vec u^2
    & \frac{\dow p}{\dow \epsilon} \\
    |\vec\pi| \frac{\dow\hat n}{\dow |\pi|} + \hat n
    & |\vec\pi| \frac{\dow\hat w}{\dow |\pi|} + \hat w
    & \frac{\dow p}{\dow |\pi|}
    - \frac{\dow\rho}{\dow|\pi|} \vec u^2 + 2 |\vec u|
    & \frac{\dow p}{\dow |\pi|}
  \end{pmatrix}.
\end{align}
We define the coefficient matrices in the Landau frame
\begin{gather}
  \tilde\fD_{TT}^\fs =
  \begin{pmatrix}
    \frac{\vec u^2}{c^4}\tilde\fs_{00}
    & \frac{\vec u^2}{c^2} \tilde\fs_{01}
    & \frac{|\vec u|}{c^2} \tilde\fs_{01} \\
    \frac{\vec u^2}{c^2} \tilde\fs_{01}
    & \vec u^2 \tilde\fs_{11}
    & |\vec u|\tilde\fs_{11} \\
    \frac{|\vec u|}{c^2} \tilde\fs_{01}
    & |\vec u|\tilde\fs_{11}
    & \tilde\fs_{11}
  \end{pmatrix},  \qquad
  \tilde{\bar\fD}_{TT}^\fs =
  \begin{pmatrix}
    0
    & \frac{\vec u^2}{c^2} \tilde{\bar\fs}_{01}
    & \frac{|\vec u|}{c^2} \tilde{\bar\fs}_{01} \\
    - \frac{\vec u^2}{c^2} \tilde{\bar\fs}_{01}
    & 0 & 0 \\
    - \frac{|\vec u|}{c^2} \tilde{\bar\fs}_{01}
    & 0 & 0
  \end{pmatrix}, \nn\\
  \tilde\fD_{TS}^\fs
  = \begin{pmatrix}
    \frac{|\vec u|}{c^2} \tilde\fs_{00}
    & \frac{\vec u^2}{c^2} \tilde\fs_{02}
    & \frac{|\vec u|}{c^2} \tilde\fs_{02}
    & \frac{|\vec u|}{c^2} \tilde\fs_{03} \\
    |\vec u| \tilde\fs_{01}
    & \vec u^2 \tilde\fs_{12}
    & |\vec u| \tilde\fs_{12}
    & |\vec u| \tilde\fs_{13} \\
    \tilde\fs_{01}
    & |\vec u| \tilde\fs_{12}
    & \tilde\fs_{12}
    & \tilde\fs_{13}
  \end{pmatrix}, \qquad
  \tilde{\bar\fD}_{TS}^\fs
    =
  \begin{pmatrix}
    0
    & \frac{\vec u^2}{c^2} \tilde{\bar\fs}_{02}
    & \frac{|\vec u|}{c^2} \tilde{\bar\fs}_{02}
    & \frac{|\vec u|}{c^2} \tilde{\bar\fs}_{03} \\
    - |\vec u| \tilde{\bar\fs}_{01}
    & \vec u^2 \tilde{\bar\fs}_{12}
    & |\vec u| \tilde{\bar\fs}_{12}
    & |\vec u| \tilde{\bar\fs}_{13} \\
    - \tilde{\bar\fs}_{01}
    & |\vec u| \tilde{\bar\fs}_{12}
    & \tilde{\bar\fs}_{12}
    & \tilde{\bar\fs}_{13}
  \end{pmatrix}, \nn\\
  \tilde\fD_{SS}^\fs =
  \begin{pmatrix}
    \tilde\fs_{00}
    & |\vec u| \tilde\fs_{02}
    & \tilde\fs_{02}
    & \tilde\fs_{03} \\
    |\vec u| \tilde\fs_{02}
    & \vec u^2\tilde\fs_{22}
    & |\vec u|\tilde\fs_{22}
    & |\vec u| \tilde\fs_{23}  \\
    \tilde\fs_{02}
    & |\vec u|\tilde\fs_{22}
    & \tilde\fs_{22} & \tilde\fs_{23} \\
    \tilde\fs_{03} 
    & |\vec u| \tilde\fs_{23} 
    & \tilde\fs_{23}
    & \tilde\fs_{33}
  \end{pmatrix}, \qquad
  \tilde{\bar\fD}_{SS}^\fs =
  \begin{pmatrix}
    0
    & |\vec u| \tilde{\bar\fs}_{02}
    & \tilde{\bar\fs}_{02}
    & \tilde{\bar\fs}_{03} \\
    - |\vec u| \tilde{\bar\fs}_{02}
    & 0
    & 0
    & |\vec u| \tilde{\bar\fs}_{23} \\
    - \tilde{\bar\fs}_{02}
    & 0
    & 0 & \tilde{\bar\fs}_{23} \\
    - \tilde{\bar\fs}_{03}
    & - |\vec u| \tilde{\bar\fs}_{23}
    & - \tilde{\bar\fs}_{23}
    & 0
  \end{pmatrix}~,
\end{gather}
and in the density frame
\begin{gather}
  \fD_{SS}^\fs =
  \begin{pmatrix}
    \fs_{00} & \fs_{02} & \fs_{02} & \fs_{03} \\
    \fs_{02} & \fs_{11} & \fs_{12} & \fs_{13} \\
    \fs_{02} & \fs_{12} & \fs_{22} & \fs_{23} \\
    \fs_{03} & \fs_{13} & \fs_{23} & \fs_{33}
  \end{pmatrix}, \qquad
  \bar\fD_{SS}^\fs =
  \begin{pmatrix}
    0 & \bar\fs_{02} & \bar\fs_{02} & \bar\fs_{03} \\
    -\bar\fs_{02} & 0 & \bar\fs_{12} & \bar\fs_{13} \\
    -\bar\fs_{02} & -\bar\fs_{12} & 0 & \bar\fs_{23} \\
    -\bar\fs_{03} & -\bar\fs_{13} & -\bar\fs_{23} & 0
  \end{pmatrix}.
\end{gather}
The mapping is given in terms of these as
\begin{align}
  \fD_{SS}^\fs
  &= \tilde\fD_{SS}^\fs
  - (M^\fs)^\rmT\tilde\fD_{TS}^\fs
  - (\tilde\fD_{TS}^\fs)^\rmT M^\fs
    + (M^\fs)^\rmT\tilde\fD_{TT}^\fs M^\fs~, \nn\\
  \bar\fD_{SS}^\fs
  &= \tilde{\bar\fD}_{SS}^\fs
  - (M^\fs)^\rmT\tilde{\bar\fD}_{TS}^\fs
  + (\tilde{\bar\fD}_{TS}^\fs)^\rmT M^\fs
  + (M^\fs)^\rmT\tilde{\bar\fD}_{TT}^\fs M^\fs~.
\end{align}

In \cref{sec:rel}, we have used this procedure to obtain the mapping for a
relativistic fluid in the Landau frame to the density frame. The transport
coefficients for a relativistic fluid in the Landau frame are given as
\begin{gather}
  F_0 = F_1 = F_2 = 0, \nn\\
  \tilde{\bar\fs}_{01} = \tilde{\bar\fs}_{02} = \tilde{\bar\fs}_{03}
  = \tilde{\bar\fs}_{12} = \tilde{\bar\fs}_{13} =
  \tilde{\bar\fs}_{23} = \tilde{\bar\fv}_{01}
  = \tilde{\bar\fv}_{02} = \tilde{\bar\fv}_{12} = 0~, \nn\\
  \tilde\fs_{00} = \gamma_u^3\sigma, \qquad
  \tilde\fs_{01} = \tilde\fs_{02} = \tilde\fs_{03} = 0, \nn\\
  \tilde\fs_{22}
  = \frac{c^2}{|\vec u|} \tilde\fs_{12}
  = \frac{c^4}{\vec u^2} \tilde\fs_{11}
  = \gamma_u^5 \lb \zeta + 2{\textstyle\frac{d-1}{d}}\eta \rb, \qquad
  \tilde\fs_{23}
  = \frac{c^2}{|\vec u|} \tilde\fs_{13}
  = \gamma_u^3 \lb \zeta - {\textstyle\frac{2}{d}}\eta \rb, \qquad
  \tilde\fs_{33} = \gamma_u \lb \zeta
  + {\textstyle\frac{2}{d(d-1)}}\eta \rb, \nn\\
  \fv_{00} = \gamma_u\sigma, \qquad
  \tilde\fv_{01} = \tilde\fv_{02} = 0, \qquad
  \tilde\fv_{22}
  = \frac{c^2}{|\vec u|} \tilde\fv_{12}
  = \frac{c^4}{\vec u^2}\tilde\fv_{11} 
  = \gamma_u^3\eta, \qquad
  \tilde\ft = \gamma_u \eta~.
\end{gather}
We can use the formulas mentioned above to recover the respective transport
coefficients in the density frame reported in \cref{eq:Rel-constraints}.  While
performing the mapping, it is useful to note that the relativistic equation of
state implies the identities
\begin{gather}
  \frac{\dow p}{\dow\epsilon}
  = \frac{\frac{1}{\gamma_u^2} \frac{\dow p}{\dow\epsilon_\rel}
    + 2 \frac{\vec u^2}{c^2} \frac{\dow p}{\dow\epsilon_\rel} 
    + \frac{\vec u^2}{c^2}  \frac{\gamma_u n}{\epsilon+p}
    \frac{\dow p}{\dow n_\rel}}{1 - \frac{\vec u^2}{c^2} \frac{\dow p}{\dow\epsilon_\rel} 
    - \frac{\vec u^2}{c^2}  \frac{\gamma_u n}{\epsilon+p}
    \frac{\dow p}{\dow n_\rel}}~, \qquad
  \frac{\dow p}{\dow n}
  = \frac{\frac{1}{\gamma_u}\frac{\dow p}{\dow n_\rel}}{
    1 - \frac{\vec u^2}{c^2} \frac{\dow p}{\dow\epsilon_\rel} 
    - \frac{\vec u^2}{c^2}  \frac{\gamma_u n}{\epsilon+p}
    \frac{\dow p}{\dow n_\rel}
  }~, \nn\\
  \frac{\dow p}{\dow\pi^2}
  = \frac{-\frac{1}{\rho} \lb \frac{\dow p}{\dow\epsilon_\rel} 
    + \frac{\gamma_u n}{2(\epsilon+p)}
      \frac{\dow p}{\dow n_\rel} \rb}{1 - \frac{\vec u^2}{c^2} \frac{\dow p}{\dow\epsilon_\rel} 
    - \frac{\vec u^2}{c^2}  \frac{\gamma_u n}{\epsilon+p}
    \frac{\dow p}{\dow n_\rel}}~, \qquad
  \hat w = \frac{\epsilon+p}{\rho} = c^2~.
\end{gather}

\subsection{Comparison to previous works}

The Landau frame dissipative and non-dissipative non-hydrostatic transport
coefficients appearing above can be related to the ones discussed in the
uncharged case in eq.~(5.6) of~\cite{deBoer:2020xlc} as\footnote{The dissipative
  coefficient $t$ has been called $\ft$ in~\cite{deBoer:2020xlc} and is negative
  semi-definite. We use the notation $t$ to avoid sign confusion with our
  convention of positive semi-definite dissipative coefficients. The mapping of
  transport coefficients with~\cite{deBoer:2020xlc} requires that we flip
  $v^\mu \to -v^\mu$.}
\begin{gather}
  \tilde{\bar\fv}_{12} = - \frac{f^{\text{NHS}}}{|\vec u|}~, \qquad
  \tilde{\bar\fs}_{12} = - \frac{s^{\text{NHS}}_2}{|\vec u|}~, \qquad
  \tilde{\bar\fs}_{13} = - \frac{s^{\text{NHS}}_1}{|\vec u|}~, \qquad
  \tilde{\bar\fs}_{23} = s_3^{\text{NHS}}~, \nn\\
  \tilde\ft = -t~, \qquad
  \tilde\fv_{11} = - \frac{f_1}{\vec u^2}~, \qquad
  \tilde\fv_{12} = - \frac{f_3}{|\vec u|}~, \qquad
  \tilde\fv_{22} = - f_2~, \nn\\
  \tilde\fs_{11} = - \frac{s_1}{\vec u^2}~, \quad
  \tilde\fs_{12} = - \frac{s_4}{|\vec u|}~, \quad
  \tilde\fs_{13} = - \frac{s_5}{|\vec u|}~, \qquad
  \tilde\fs_{22} = - s_2~, \quad
  \tilde\fs_{23} = - s_6~, \quad
  \tilde\fs_{33} = - s_3~.
\end{gather}
9 non-hydrostatic non-dissipative and 17 dissipative coefficients reduce to
4 and 10 respectively in the uncharged case, as reported
by~\cite{deBoer:2020xlc}. In addition, three hydrostatic coefficients
$F_{0,1,2}$ reduce down to just two $F_{1,2}$, as commented upon in
\cref{sec:hydrostatic}.

The comparison of our work to the analysis of~\cite{Novak:2019wqg}, on the other
hand, is considerably more involved.\footnote{We thank the authors
  of~\cite{Novak:2019wqg} for aiding us in this comparison.} Firstly, comparison
with the constitutive relations, given in eq. (2.24) of~\cite{Novak:2019wqg},
can only be done in the limit $c\to\infty$ in the thermodynamic Landau frame
definition in \cref{eq:Landau-frame}, or equivalently $B\to 0$ in eq.~(2.23)
in~\cite{Novak:2019wqg}. For $B\neq 0$, the basis of independent non-hydrostatic
data used in~\cite{Novak:2019wqg} is not compatible with the off-shell formalism
because the resultant dissipation matrices in \cref{eq:Landau-diss} are
asymmetric. Specialising to the $c\to\infty$ case, we can find the mapping of
the transport coefficients $\bar\eta$, $\bar\zeta$, $\bar\sigma$, $\bar\alpha$,
$\bar\gamma$, $\bar\pi$, $\gamma_{1,\ldots,23}$ appearing in eq. (2.24)
of~\cite{Novak:2019wqg} to the ones introduced by us; in the non-hydrostatic
non-dissipative sector we find\footnote{The signs of the coefficients $\gamma_1$,
  $\gamma_3$, $\gamma_4$ in $\Pi^0_{~j}$ in eq. (2.24) of~\cite{Novak:2019wqg}
  are incorrect, as they violate the Landau frame conditions. We are unable to
  reproduce the non-hydrostatic combinations reported in eqs.~(2.48)--(2.56)
  in~\cite{Novak:2019wqg}.}
\begin{gather}
  \tilde{\bar\fs}_{01} = \bar\gamma
  - \lb \frac{\gamma_4}{2T} + \gamma_{18}\rb\vec u^2, \qquad
  \tilde{\bar\fs}_{02} = -|\vec u| \lb
  \frac{\gamma_{12}}{2T}+ \frac{\gamma_{14}}{2T}
  + \gamma_{19} + \frac{\gamma_{20}}{2} \rb, \qquad
  \tilde{\bar\fs}_{03} = - \frac{|\vec u|}{2} \lb \gamma_{20} - \frac{\gamma_{17}}{T}
  \rb, \nn\\
  \tilde{\bar\fs}_{12}
  = |\vec u| \lb \gamma_2 + \frac{\gamma_3}{2}
  - \gamma_{10} - \frac{\gamma_7}{2}
  \rb, \qquad
  \tilde{\bar\fs}_{13}
  = |\vec u| \lb \frac{\gamma_3}{2} + \gamma_{16} \rb, \qquad
  \tilde{\bar\fs}_{23}
  = \vec u^2 \lb \frac{\gamma_{11}}{2} + \gamma_{15} \rb, \nn\\
  \tilde{\bar\fv}_{01} = \bar\gamma, \qquad
  \tilde{\bar\fv}_{02} = -|\vec u| \lb \frac{\gamma_{14}}{2T} + \gamma_{22} \rb,
  \qquad
  \tilde{\bar\fv}_{12} = -|\vec u| \lb \frac{\gamma_7}{2} + \gamma_5 \rb,
  \label{eq:nhsnd-withers}
\end{gather}
and in the dissipative sector
\begin{gather}
  \tilde\fs_{00} = \bar\sigma - \frac{\vec u^2}{T} \gamma_{21}, \qquad
  \tilde\fs_{01} = -\bar\alpha + \lb \frac{\gamma_4}{2T} - \gamma_{18}\rb
  \vec u^2~, \qquad
  \tilde\fs_{11} = \bar\pi + 2\vec u^2\gamma_1~, \nn\\
  \tilde\fs_{02} = |\vec u| \lb
  \frac{\gamma_{12}}{2T}+ \frac{\gamma_{14}}{2T}
  - \gamma_{19} - \frac{\gamma_{20}}{2} \rb, \qquad
  \tilde\fs_{03} = - \frac{|\vec u|}{2} \lb \gamma_{20}
  + \frac{\gamma_{17}}{T} \rb~, \nn\\
  \tilde\fs_{12}
  = |\vec u| \lb \gamma_2 + \frac{\gamma_3}{2}
  + \gamma_{10}+ \frac{\gamma_7}{2}
  \rb, \qquad
  \tilde\fs_{13}
  = |\vec u| \lb \frac{\gamma_3}{2} - \gamma_{16} \rb \nn\\
  \tilde\fs_{23}
  = \bar\zeta + 2\frac{d-1}{d}\bar\eta
    + \vec u^2 \lb \frac{\gamma_{11}}{2} - \gamma_{15} \rb~, \qquad
  \tilde\fs_{33}
  = \bar\zeta + \frac{2}{d(d-1)}\bar\eta~, \nn\\
  \tilde\fv_{00} = \bar\sigma~, \qquad
  \tilde\fv_{01} = -\bar\alpha~, \qquad
  \tilde\fv_{11} = \bar\pi~, \qquad
  \tilde\fv_{02} = |\vec u| \lb \frac{\gamma_{14}}{2T} - \gamma_{22} \rb~, \qquad
  \tilde\fv_{12} = |\vec u| \lb \frac{\gamma_7}{2} - \gamma_5 \rb~, \nn\\
  \tilde\fv_{22} = 2\gamma_8~, \qquad
  \tilde\fs_{22} = \bar\zeta + 2\frac{d-1}{d}\bar\eta
    + \vec u^2 (4\gamma_8+2\gamma_9+\gamma_{11})~, \qquad
    \tilde\ft = \bar\eta~.
    \label{eq:d-withers}
\end{gather}
The three remaining coefficients $\gamma_6 - 2\gamma_5$, $\gamma_{13} - 2\gamma_8$,
$\gamma_{23} - 2\gamma_{22}$ from~\cite{Novak:2019wqg} do not appear in the maps
above. They will, however, get non-trivial contributions in the hydrostatic
sector from $F_{0,1,2}$ in \cref{sec:hydrostatic}. We do not perform this
detailed analysis here.

The authors in~\cite{Novak:2019wqg} introduced a different set of dissipative
coefficients $b_{0,\ldots,21}$ for the entropy-production quadratic form
$\Delta$ in eqs. (2.33)-(2.39), and hydrostatic coefficients
$\tilde c_{1,2,4,8}$ for the non-canonical entropy current $s^\mu_{\text{non-can}}$
in eqs. (2.30)-(2.32). The relation to the aforementioned $\bar\eta$,
$\bar\zeta$, $\bar\sigma$, $\bar\alpha$, $\bar\gamma$, $\bar\pi$,
$\gamma_{1,\ldots,23}$ coefficients is presented by the authors in a companion
notebook. They also find 2 constraints
\begin{equation}
  b_{15} = b_{14}~~, \qquad b_{20} + b_{21} = 2b_{19}~~.
\end{equation}
It should be noted that a complete analysis of the second law constraints is not
provided in~\cite{Novak:2019wqg}. It should also be noted that 9 non-dissipative
non-hydrostatic coefficients, given in eq. (2.48)-(2.56)
of~\cite{Novak:2019wqg}, do not show up in the non-canonical entropy current or
entropy production. To map the $b_{0,\ldots,21}$, $\tilde c_{1,2,4,8}$
coefficients to our formalism, it is easier to work in the thermodynamic density
frame. Mapping the $\Delta$'s in the two frameworks, we find that the 20
independent $b_{0,\ldots,14}$, $b_{16,\ldots,20}$ coefficients map to 17
dissipative coefficients
\begin{align}
  &\frac{1}{\kB T}
  \begin{pmatrix}
    \Lambda^\rmT & 0 \\
    0 & 1
  \end{pmatrix}
  \begin{pmatrix}
    \fs_{00} & \fs_{01} & \fs_{02} & \fs_{03} \\
    \fs_{01} & \fs_{11} & \fs_{12} & \fs_{13} \\
    \fs_{02} & \fs_{12} & \fs_{22} & \fs_{23} \\
    \fs_{03} & \fs_{13} & \fs_{23} & \fs_{33}
  \end{pmatrix}
  \begin{pmatrix}
    \Lambda & 0 \\
    0 & 1
  \end{pmatrix}
                =
  \nn\\
  &
  \begin{pmatrix}
    b_0 + b_1\vec v^2 & b_2 + b_3\vec v^2
    & b_6|\vec v| + b_7|\vec v|^3 + 2b_8|\vec v| & b_6|\vec v| \\
    b_2 + b_3\vec v^2 & b_4 + b_5\vec v^2
    & b_{10}|\vec v| + b_{11}|\vec v|^3 + 2b_{12}|\vec v| & b_{10}|\vec v| \\
    b_6|\vec v| + b_7|\vec v|^3 + 2b_8|\vec v|
    & b_{10}|\vec v| + b_{11}|\vec v|^3 + 2b_{12}|\vec v|
    & 2b_{14} + b_{16} + b_{17} \vec v^4 + 2b_{18}\vec v^2 + 4b_{19}\vec v^2
    & b_{16} + b_{18}\vec v^2 \\
    b_6|\vec v| & b_{10}|\vec v| & b_{16} + b_{18}\vec v^2
    & b_{16} + \frac{2}{d-1}b_{14}
  \end{pmatrix},
      \nn
\end{align}
\begin{gather}
  \frac{1}{\kB T}
  \Lambda^\rmT
  \begin{pmatrix}
    \fv_{00} & \fv_{01} & \fv_{02} \\
    \fv_{01} & \fv_{11} & \fv_{12} \\
    \fv_{02} & \fv_{12} & \fv_{22}
  \end{pmatrix}
  \Lambda
  =
  \begin{pmatrix}
    b_0 & b_2 & b_{8} \\
    b_2 & b_4 & b_{12} \\
    b_8 & b_{12} & b_{14} + b_{19} \vec v^2
  \end{pmatrix}, \nn\\
  \frac{1}{\kB T}\ft = b_{14}~,
\end{gather}
where
\begin{equation}
  \Lambda = 
  \begin{pmatrix}
    - \mu/T & 1 & 0 \\
    1/T & 0 & 0 \\
    - |\vec u|/T & 0 & 1 \\
  \end{pmatrix}~,
\end{equation}
is the transformation matrix arising from converting $\dow_i(\mu/T)$,
$\dow_i(1/T)$, $\dow_i(u_j/T)$ basis to $\dow_iT$, $\dow_i\mu$, $\dow_iu_j$
basis in~\cite{Novak:2019wqg}. The comparison also leads to 3 equality
constraints\footnote{In eq.~(2.33) of~\cite{Novak:2019wqg}, the authors write
   the dissipation matrix in terms of $\dow_i T$, $\dow_i\mu$, and
  $\dow_i u_j$, however, only the symmetric derivatives of $u^i$ are
  non-hydrostatic and can contribute to dissipation. This leads to the said 3
  constraints.}
\begin{equation}
  b_9 = b_8~, \qquad
  b_{13} = b_{12}~, \qquad
  b_{20} = b_{19}~,
  \label{eq:d-constraints}
\end{equation}
which can be thought of as arising from requiring entropy production to be
non-negative. To map the hydrostatic $\tilde c_{1,2,4,8}$ coefficients, we note
that the non-canonical entropy current from eqs. (2.30)-(2.32)
of~\cite{Novak:2019wqg} is given by
\begin{align}
  s^0_{\text{non-can}}
  &= \tilde c_1 u^k\dow_k \vec u^2
    + \tilde c_2 \lb u^k\dow_k \frac{\mu}{T}
    + m_1 \dow_t \vec u^2 \rb
    + \tilde c_4 \dow_k u^k, \nn\\
  s^i_{\text{non-can}}
  &= - u^i \tilde c_1 \dow_t \vec u^2
    - \tilde c_4 \dow_t u^i
    + \tilde c_8 \lb u^k \dow_k u^i - u^i \dow_k u^k \rb \nn\\
  &\qquad
    + m_2 \tilde c_2 \dow_t u^i
    + \tilde c_2 u^i \lb
    m_4 u^k\dow_k \vec u^2
    + u^k\dow_k \frac{\mu}{T}
    + (m_1+m_4) \dow_t \vec u^2
    + m_3 \dow_k u^k
    \rb,
\end{align}
where $m_{1,2,3,4}$ are some known thermodynamic parameters. This can be
compared to our \cref{eq:EC-non-can}, in the absence of background fields, using
the equations of motion, and up to a total-derivative shift of the entropy current
$s^0_{\text{non-can}} \to s^0_{\text{non-can}} + \dow_i X^i$,
$s^i_{\text{non-can}} \to s^i_{\text{non-can}} - \dow_t X^i$ for some $X^i$ that
leaves the divergence of the entropy current invariant. This relates 3 independent
$\tilde c_{1,2,4}$ coefficients to 3 hydrostatic coefficients $F_{0,1,2}$ according to
\begin{align}
  F_0
  &= \tilde c_2
    - Tn X
    - T \frac{\dow Y}{\dow \mu}~~, \nn\\
  F_1
  &= - \frac{\mu}{T} \tilde c_2 
    - Ts X
    - T \frac{\dow Y}{\dow T}~~, \nn\\
  F_2
  &= T\tilde c_1 - Tm_1\tilde c_2
    - \half T\rho X
    - T \frac{\dow Y}{\dow \vec u^2}~,
\end{align}
where
\begin{align}
  X
  &= 2m_1\tilde c_2 \lb
    \frac{1}{\rho} - \frac{\vec u^2}{\rho} \frac{\dow\rho}{\dow\epsilon} 
    - 2 \vec u^2 \frac{\dow\rho}{\dow\vec\pi^2} \rb~, \nn\\
  Y
  &= \tilde c_4 
  - 2 \vec u^2 m_1\tilde c_2 \lb 1- \frac{\epsilon+p}{\rho} 
    \frac{\dow\rho}{\dow\epsilon}
  - \frac{n}{\rho} \frac{\dow \rho}{\dow n}\
  - 2\rho \vec u^2 \frac{\dow\rho}{\dow\vec\pi^2} \rb.
\end{align}
We also get a constraint\footnote{An easy way to understand this constraint is
  by noting that the contribution to entropy-divergence coupled to $\tilde c_8$
  does not vanish in equilibrium, and hence must be set to zero.}
\begin{equation}
  \tilde c_8 = 0.
  \label{eq:hs-constraints}
\end{equation}
To summarise,~\cite{Novak:2019wqg} reports a total of 29 coefficients in the
constitutive relations $\bar\sigma$, $\bar\alpha$, $\bar\gamma$, $\bar\pi$,
$\gamma_{1,\ldots,23}$ (related to our formalism according to
\cref{eq:nhsnd-withers,eq:d-withers}). Ref.~\cite{Novak:2019wqg} also classifies
20 possibly dissipative coefficients $b_{0,\ldots,14}$, $b_{16,\ldots,20}$ in
the entropy production quadratic form, 4 possibly hydrostatic coefficients
$\tilde c_{1,2,4,8}$ in the non-canonical entropy current, along with 9
non-hydrostatic non-dissipative coefficients that do not contribute to the
non-canonical entropy current or to entropy production. The ensuing second law
analysis was not performed in \cite{Novak:2019wqg}. Accounting for the second
law, we find 3 constraints among the coefficients $b$'s, given in
\cref{eq:d-constraints}, and one constraint among the coefficients $\tilde c$'s,
given in \cref{eq:hs-constraints}. Thus, the final number of independent
transport coefficients consists of 17 dissipative, 9 non-dissipative
non-hydrostatic, and 3 hydrostatic.

\section{Interaction vertices}
\label{sec:interactions}

In this appendix, we record the effective Lagrangian that accounts for
interactions between hydrodynamic and stochastic degrees of freedom. Taking into
account only the ideal order part of the Lagrangian from \cref{eq:flat-action},
and expanding to cubic order in fluctuations, we obtain the three-point
interaction Lagrangian given by
\begin{align}
  \mathcal{L}_3
  &= \half\gamma_{nn}^i \delta n^2\dow_i\varphi_a
    + \half\gamma_{\epsilon\epsilon}^i \delta\epsilon^2\dow_i\varphi_a
    + \gamma_{n\epsilon}^i \delta n\,\delta\epsilon\,\dow_i \varphi_a \nn\\
  &\qquad\qquad
    + \gamma_{n\pi}^{ij} \delta n\,\delta\pi_i\dow_j\varphi_a
    + \gamma_{\epsilon\pi}^{ij} \delta\epsilon\,\delta\pi_i\dow_j \varphi_a 
    + \gamma_{\pi\pi}^{ijk}\delta\pi_i\delta\pi_j\dow_k \varphi_a \nn\\
  &\qquad
    - \half\alpha_{nn}^i \delta n^2\dow_iX^t_a
    - \half\alpha_{\epsilon\epsilon}^i \delta \epsilon^2\dow_iX^t_a
    - \alpha_{n\epsilon}^i\delta n\,\delta\epsilon\,\dow_iX^t_a \nn\\
  &\qquad\qquad
    - \alpha_{n\pi}^{ij} \delta n\,\delta\pi_i\dow_jX^t_a
    - \alpha_{\epsilon\pi}^{ij} \delta\epsilon\,\delta\pi_i\dow_jX^t_a
    - \alpha_{\pi\pi}^{ijk} \delta\pi_i\delta\pi_j\dow_k X^t_a \nn\\
  &\qquad
    + \half\beta_{nn}^{ij} \delta n^2 \dow_j X_{ai}
    + \half\beta_{\epsilon\epsilon}^{ij} \delta \epsilon^2 \dow_j X_{ai} 
    + \beta_{n\epsilon}^{ij}\delta n\,\delta\epsilon\, \dow_j X_{ai} \nn\\
  &\qquad\qquad
    + \beta_{n\pi}^{ijk} \delta n\,\delta\pi_i\, \dow_j X_{ak} 
    + \beta_{\epsilon\pi}^{ijk} \delta\epsilon\,\delta\pi_i\, \dow_j X_{ak}
    + \beta_{\pi\pi}^{ijkl} \delta\pi_i \delta\pi_j\, \dow_l X_{ak}~~.
    \label{eq:L3}
\end{align}
Here we have defined the following coupling structures
\begin{gather}
  \gamma^i_{nn} = \rho\frac{\dow^2\hat n}{\dow n^2} u^i~, \qquad
  \gamma^i_{\epsilon\epsilon} = \rho\frac{\dow^2\hat n}{\dow\epsilon^2} u^i~,
  \qquad
  \gamma_{n\epsilon}^i
  = \rho\frac{\dow^2\hat n}{\dow n\dow\epsilon}u^i~, \nn\\
  \gamma_{n\pi}^{ij}
  = 2\rho^2\frac{\dow^2\hat n}{\dow n\dow\vec\pi^2}u^i u^j
  + \frac{\dow\hat n}{\dow n} \delta^{ij}~, \qquad
  \gamma_{\epsilon\pi}^{ij}
  = 2\rho^2 \frac{\dow^2\hat n}{\dow\epsilon\dow\vec\pi^2}u^iu^j
  + \frac{\dow\hat n}{\dow\epsilon} \delta^{ij}~, \nn\\
  \gamma_{\pi\pi}^{ijk}
  = 2\rho^3\frac{\dow^2\hat n}{\dow (\vec\pi^2)^2}u^i u^j u^k
  + \rho\frac{\dow\hat n}{\dow \vec\pi^2}
  (\delta^{ij}u^k+2u^{(i}\delta^{j)k})~, \nn\\
  \alpha^i_{nn} = \rho\frac{\dow^2\hat w}{\dow n^2} u^i~, \qquad
  \alpha^i_{\epsilon\epsilon} = \rho\frac{\dow^2\hat w}{\dow\epsilon^2} u^i~,
  \qquad
  \alpha_{n\epsilon}^i
  = \rho\frac{\dow^2\hat w}{\dow n\dow\epsilon}u^i~, \nn\\
  \alpha_{n\pi}^{ij}
  = 2\rho^2\frac{\dow^2\hat w}{\dow n\dow\vec\pi^2}u^i u^j
  + \frac{\dow\hat w}{\dow n} \delta^{ij}~, \qquad
  \alpha_{\epsilon\pi}^{ij}
  = 2\rho^2 \frac{\dow^2\hat w}{\dow\epsilon\dow\vec\pi^2}u^iu^j
  + \frac{\dow\hat w}{\dow\epsilon} \delta^{ij}~, \nn\\
  \alpha_{\pi\pi}^{ijk}
  = 2\rho^3\frac{\dow^2\hat w}{\dow (\vec\pi^2)^2}u^i u^j u^k
  + \rho\frac{\dow\hat w}{\dow \vec\pi^2}
  (\delta^{ij}u^k+2u^{(i}\delta^{j)k})~, \nn\\
  \beta_{nn}^{ij}
  = \rho^2\frac{\dow^2\rho^{-1}}{\dow n^2}u^iu^j
  + \frac{\dow^2 p}{\dow n^2} \delta^{ij}~, \qquad
  \beta_{\epsilon\epsilon}^{ij}
  = \rho^2\frac{\dow^2\rho^{-1}}{\dow\epsilon^2}u^iu^j
  + \frac{\dow^2 p}{\dow\epsilon^2} \delta^{ij}~, \qquad
  \beta_{n\epsilon}^{ij}
  = \rho^2\frac{\dow^2\rho^{-1}}{\dow n\dow\epsilon}u^iu^j
  + \frac{\dow^2 p}{\dow n\dow\epsilon} \delta^{ij}, \nn\\
  \beta_{n\pi}^{ijk}
  = - \frac{2}{\rho} \frac{\dow\rho}{\dow n} \delta^{i(j}u^{k)}
  + 2\rho^3\frac{\dow^2\rho^{-1}}{\dow n\dow\vec\pi^2} u^iu^ju^k
  + 2\rho\frac{\dow^2 p}{\dow n\dow\vec\pi^2}u^i\delta^{jk}~, \nn\\
  \beta_{\epsilon\pi}^{ijk}
  = - \frac{2}{\rho} \frac{\dow\rho}{\dow\epsilon} \delta^{i(j}u^{k)}
  + 2\rho^3\frac{\dow^2\rho^{-1}}{\dow\epsilon\dow\vec\pi^2}u^iu^ju^k
  + 2\rho\frac{\dow^2 p}{\dow\epsilon\dow\vec\pi^2}u^i\delta^{jk}, \nn\\
  \beta_{\pi\pi}^{ijkl}
  = \frac{1}{\rho} \delta^{ik} \delta^{jl}
  + \frac{\dow p}{\dow\vec\pi^2} \delta^{ij} \delta^{kl} 
  - \frac{\dow\rho}{\dow\vec\pi^2}
  (4u^{(i}\delta^{j)(k}u^{l)} + \delta^{ij}u^ku^l)
  + 2\rho^2\frac{\dow^2 p}{\dow (\vec\pi^2)^2} u^iu^j \delta^{kl}
  + 2\rho^4\frac{\dow^2\rho^{-1}}{\dow (\vec\pi^2)^2}
  u^iu^ju^ku^l~.
\end{gather}
This procedure can analogously be iterated to obtain higher derivative and
higher-point interactions (see~\cite{Jain:2020vgc} for the discussion in
Galilean case). We leave the analysis of the effects of \eqref{eq:L3} on 
hydrodynamic equations of motion and correlation functions to 
future work.

\makereferences

\end{document}